\documentclass[aps,prx,twocolumn,preprintnumbers,showpacs,amsmath,amssymb,superscriptaddress]{revtex4-1}
\pdfoutput=1

\usepackage{graphicx}
\usepackage{dcolumn}
\usepackage{color}
\usepackage{amsmath}
\usepackage[breaklinks,colorlinks,bookmarks=false,citecolor=blue,linkcolor=red,urlcolor=blue]{hyperref}

\newcommand {\JR}{J_\perp}
\newcommand {\JRc}{J_{\perp,c}}
\newcommand {\JL}{J_\|}
\newcommand {\JD}{J_\times}

\begin{document}

\preprint{Phys. Rev. B {\bfseries 94}, 094402  (2016)}

\title{Multi-triplet bound states and finite-temperature dynamics in highly frustrated quantum spin ladders}

\author{Andreas Honecker}
\affiliation{Laboratoire de Physique Th\'eorique et Mod\'elisation, CNRS 
UMR 8089, Universit\'e de Cergy-Pontoise, % Site de Saint-Martin,
F-95302 Cergy-Pontoise Cedex, France}
%\affiliation{Institut f\"ur Theoretische Physik, Georg-August-Universit\"at 
%G\"ottingen, D-37077 G\"ottingen, Germany}

\author{Fr\'ed\'eric Mila}
\affiliation{Institute of Physics, Ecole Polytechnique F\'ed\'erale
Lausanne (EPFL), CH-1015 Lausanne, Switzerland}

\author{B. Normand}
\affiliation{Department of Physics, Renmin University of China, Beijing
100872, P.~R.~China}

\date{May 30, 2016; revised July 29, 2016}

\begin{abstract}

Low-dimensional quantum magnets at finite temperatures present a complex 
interplay of quantum and thermal fluctuation effects in a restricted phase 
space. While some information about dynamical response functions is available 
from theoretical studies of the one-triplet dispersion in unfrustrated chains 
and ladders, little is known about the finite-temperature dynamics of 
frustrated systems. Experimentally, inelastic neutron scattering studies 
of the highly frustrated two-dimensional material SrCu$_2$(BO$_3$)$_2$ show 
an almost complete destruction of the one-triplet excitation band at a 
temperature only 1/3 of its gap energy, accompanied by strong scattering 
intensities for apparent multi-triplet excitations. We investigate these 
questions in the frustrated spin ladder and present numerical results from 
exact diagonalization for the dynamical structure factor as a function of 
temperature. We find anomalously rapid transfer of spectral weight out of 
the one-triplet band and into both broad and sharp spectral features at a 
wide range of energies, including below the zero-temperature gap of this 
excitation. These features are multi-triplet bound states, which develop 
particularly strongly near the quantum phase transition, fall to particularly 
low energies there, and persist to all the way to infinite temperature. Our 
results offer valuable insight into the physics of finite-temperature spectral 
functions in SrCu$_2$(BO$_3$)$_2$ and many other highly frustrated spin systems.

\end{abstract}

\pacs{75.10.Jm, 75.40.Gb, 75.40.Mg}

\maketitle

\section{Introduction}
\label{sec:Intro}

Quantum mechanics mandates a degree of uncertainty in the properties of 
a physical system. In many-body systems, this uncertainty is manifest as 
quantum fluctuations between different, and often classically inspired, 
states of the system or its subcomponents. One-dimensional (1D) quantum 
antiferromagnets provide an excellent example of a situation where the 
classical state, which would be N\'eel order, is destroyed completely by 
quantum fluctuations, replaced by gapped or gapless states with complex 
correlations (or ``entanglement'') but no magnetic order \cite{rrt}. 
The canonical gapless 1D quantum magnet, the $S = 1/2$ Heisenberg chain, 
has quasi-long-ranged spin correlations and massless, fractionalized 
``spinon'' excitations \cite{FaTa81}; its gapped counterparts, including 
the two-leg $S = 1/2$ spin ladder \cite{DRS92,DaRiRev96,Dagotto99}, the 
$S = 1$ ``Haldane'' chain \cite{haldane83}, and the frustrated $S = 1/2$ 
($J_1$--$J_2$, ``Majumdar-Ghosh'') chain \cite{MG69,Ma69,ShaSu81j1j2}, all 
show exponentially decaying spin correlations, accompanied in the first two 
cases by robust triplet excitations and in the third by massive spinons.  
These properties are consequences purely of quantum spin fluctuations in the 
restricted phase space available in one spatial dimension, and significant 
progress has been made over the last two decades both in their theoretical 
description and in their experimental observation \cite{rre}.

Far less well understood is the additional effect of finite temperatures 
on these systems. The problem of describing the combination of quantum and 
thermal fluctuations within the same restricted phase space, to deduce 
the consequences for the ground and excited states, has proven difficult 
to address. In the most straightforward description of the triplet 
excitations of a two-leg spin ladder as robust $\delta$-function peaks 
in energy \cite{rnr}, a significant band-narrowing effect is obtained 
as the temperature is increased, and this was in fact observed in the 3D 
coupled dimer system TlCuCl$_3$ \cite{rrnmnfkgbsm}. In lower dimensions, 
however, more sophisticated means of modeling thermal effects in 
unfrustrated spin chains suggest rather a systematic and asymmetric 
broadening of peak widths \cite{rml,re,JamesPhD,rfsu,Fauseweh2015}, 
driven by a mixing of available states within the triplet band, and this 
is confirmed by recent observations in the alternating spin-chain materials 
Cu(NO$_3$)$_2\cdot$2.5D$_2$O \cite{rtljenmfkct} and BaCu$_2$V$_2$O$_8$ 
\cite{BaCu2V2O8}. By contrast, the situation in frustrated 1D systems at 
finite temperatures remains essentially unexplored. 

While systematic experimental data for frustrated spin chains are not yet 
available, the 2D material SrCu$_2$(BO$_3$)$_2$ offers some crucial insight 
into the effects of competing interactions. The $S = 1/2$ Cu$^{2+}$ ions in 
SrCu$_2$(BO$_3$)$_2$ form the ``Shastry-Sutherland'' geometry 
\cite{ShaSu81} of ``rung'' bonds bridging the opposite corners of a 
square lattice, and at large values of the rung coupling there is an exact 
ground state of dimer singlets on these bonds. A triplet excitation of any 
single dimer experiences complete frustration (up to sixth order in a 
perturbative expansion \cite{rkbmhu,MiUe03}) and therefore the excitation 
band measured at low temperatures is almost completely flat, at an energy 
of approximately $3.0$~meV ($\simeq 35$~K) \cite{rknaoynkku}. Remarkably, if 
the temperature is increased to only $6$~K, this mode suffers a serious loss 
of intensity, and by 12 K it is essentially indiscernible, its spectral 
weight spread over the entire energy spectrum \cite{rglhcbdqc,rzrr}. Some 
phenomenological proposals \cite{rglhcbdqc,rzrr} and one numerical study 
\cite{rebs} have tried to account for this anomalously strong decay of the 
one-triplet excitation, but with limited success. 

We begin with the hypothesis that this behavior is the consequence of 
bound-state formation, which is strongly enhanced by the presence of 
frustration. We comment that the Hamiltonian for SrCu$_2$(BO$_3$)$_2$ does 
contain nontrivial additional terms, specifically Dzyaloshinskii-Moriya 
interactions, whose effects would need to be computed in a quantitative 
treatment of triplet decay. However, our aim is to test the concept that 
the formation of (many) strongly bound states and the resulting strong 
enhancement of one-triplet decay may be generic to the finite-temperature 
dynamical response of frustrated quantum magnets. As a candidate system 
for the demonstration of a nontrivial thermal redistribution of spectral 
weight, we investigate the fully frustrated $S = 1/2$ two-chain spin ladder, 
introduced in Sec.~\ref{sec:FFL}; this model is well suited for our study 
because of the existence of exact multi-triplet bound states over the full 
range of parameters giving a rung-singlet ground state. 

Computing the dynamical spectral function of a low-dimensional quantum spin 
system at finite temperatures is a challenging problem, requiring knowledge 
both of the full excitation spectrum and of all matrix elements. Although some 
analytical progress has been possible by Bethe-Ansatz techniques \cite{re} and 
by a recent diagrammatic approach \cite{rfsu,Fauseweh2015}, the applicability 
of these methods to frustrated systems is limited. Among the available 
numerical techniques, exact diagonalization (ED) has in the past \cite{rml,
rtljenmfkct,rebs} been the only fully systematic approach for all models, 
parameters, and temperatures. Quantum Monte Carlo (QMC) methods are able 
to deliver dynamical response functions for the lowest excitations of 
unfrustrated systems, but are in general unsuitable for frustrated ones 
because of the sign problem. Density-matrix renormalization-group (DMRG) 
techniques have previously been of limited value in dealing with the 
combination of time (dynamics) and temperature, but we comment that recent 
advances in methodology have the potential to expand very significantly the 
study of finite-temperature response functions and we review these briefly 
in Sec.~\ref{sec:FTspec} for their pedagogical value. For the present study, 
the intrinsic advantages of ED far outweigh its traditional limitation to 
small system sizes, as we discuss in detail in Sec.~\ref{sec:FTspec}.

Significant information in support of bound-state effects in frustrated 
systems can be obtained from their thermodynamic response. The calculation 
of quantities such as the specific heat and susceptibility of a quantum system 
is also a hard problem, requiring again the full excitation spectrum. Once 
again, analytical assistance is limited to the spin-1/2 Heisenberg chain 
\cite{Takahashi99,Gaudin71,EAT94,Kl98,KlJ00,THKO10}, also one of the first 
systems to which ED was applied \cite{rbf}, and beyond this model a variety 
of numerical methods have been employed for different 1D systems. In parallel 
with the present study, we have performed a systematic analysis of the 
magnetic specific heat and susceptibility of the frustrated spin ladder in 
Ref.~\cite{rus}, to which we refer the reader for full details. Here we draw 
attention only to the use of ED methods to obtain the thermodynamic response of 
frustrated spin chains, specifically $J_1$--$J_2$ models motivated by certain 
$S = 1/2$ and $S = 1$ materials \cite{rhhv,rnbsylkdnghmrplj}. QMC techniques 
have been exploited widely for unfrustrated ladders \cite{Frischmuth96,
Johnston00b}, but for frustrated systems are stymied by the sign problem 
except in special cases \cite{rus,nakamura98,radp}. Quantum Transfer Matrix 
(QTM)-DMRG is particularly well suited to the extraction of thermodynamic 
information for 1D systems and has been applied both to unfrustrated 
\cite{WangXiang97,rtx} and to frustrated ($J_1$--$J_2$) spin chains, where 
two characteristic energy scales are found \cite{rms,LWQT06,rtf,Sirker10}. 
However, these studies have largely been restricted to methods and to 
specific parameter choices, and we are unaware of any systematic 
investigation of frustration effects. In Ref.~\cite{rus} we provide such 
a study for the fully frustrated ladder, which demonstrates clearly the 
role of bound states in determining the characteristic evolution of the 
thermodynamic response and the dominant effect of extended multi-triplet 
bound states close to a quantum critical point; a summary of these results 
is presented in Sec.~\ref{sec:FFL}.

The structure and primary results of this article are as follows. In 
Sec.~\ref{sec:FFL}, we present the model of the fully frustrated two-leg 
spin ladder, summarize its phase diagram and the nature of its multi-triplet 
bound states, and review analytical and numerical results obtained for the 
thermodynamics of the system. In Sec.~\ref{sec:FTspec}, we show the results 
of ED calculations of the dynamical spectral function at all temperatures, 
for a selection of ladders with different coupling ratios and degrees of 
frustration. Figures \ref{dsf211}-\ref{dsfinf} provide a complete overview 
of the momentum dependence of the dynamical structure factor and 
Fig.~\ref{dsf2o3} in particular demonstrates a very rapid redistribution 
of spectral weight with increasing temperature. In Sec.~\ref{sec:Analysis}, 
we analyze our results by considering the bound-state spectra, scattering 
matrix elements, and transfer of spectral weight from the one-triplet sector 
to the multi-triplet bound states that emerge particularly strongly close to 
the quantum phase transition. Figures \ref{dsfkp211} to \ref{dsfkp21p9} 
quantify the temperature dependence of the dynamical structure factor and in 
Fig.~\ref{dsfkpinf} we demonstrate that remarkably sharp spectral features 
survive up to infinite temperatures. Figure \ref{ftwmtm} shows the thermal 
evolution of the leading peaks in the spectral function for one value
of the exchange ratio, Fig.~\ref{ftwj} 
analyzes the temperature dependence of the one-triplon line for several 
values of the exchange ratio, demonstrating accelerated thermal suppression 
close to the quantum phase transition. Figure \ref{fth} interprets these 
findings in terms of emergent, effective temperature scales of the system. 
Section \ref{sec:Summary} presents a brief summary, while three appendices 
contain complementary details concerning the analytical treatment of 
few-triplon bound states (Apps.~\ref{appa} and \ref{appb}) and finite-size 
effects in the one-triplon spectral weight (App.~\ref{app:FS}).

\section{Fully Frustrated Ladder}
\label{sec:FFL}

\begin{figure}[t!]
\centering\includegraphics[width=0.8\columnwidth]{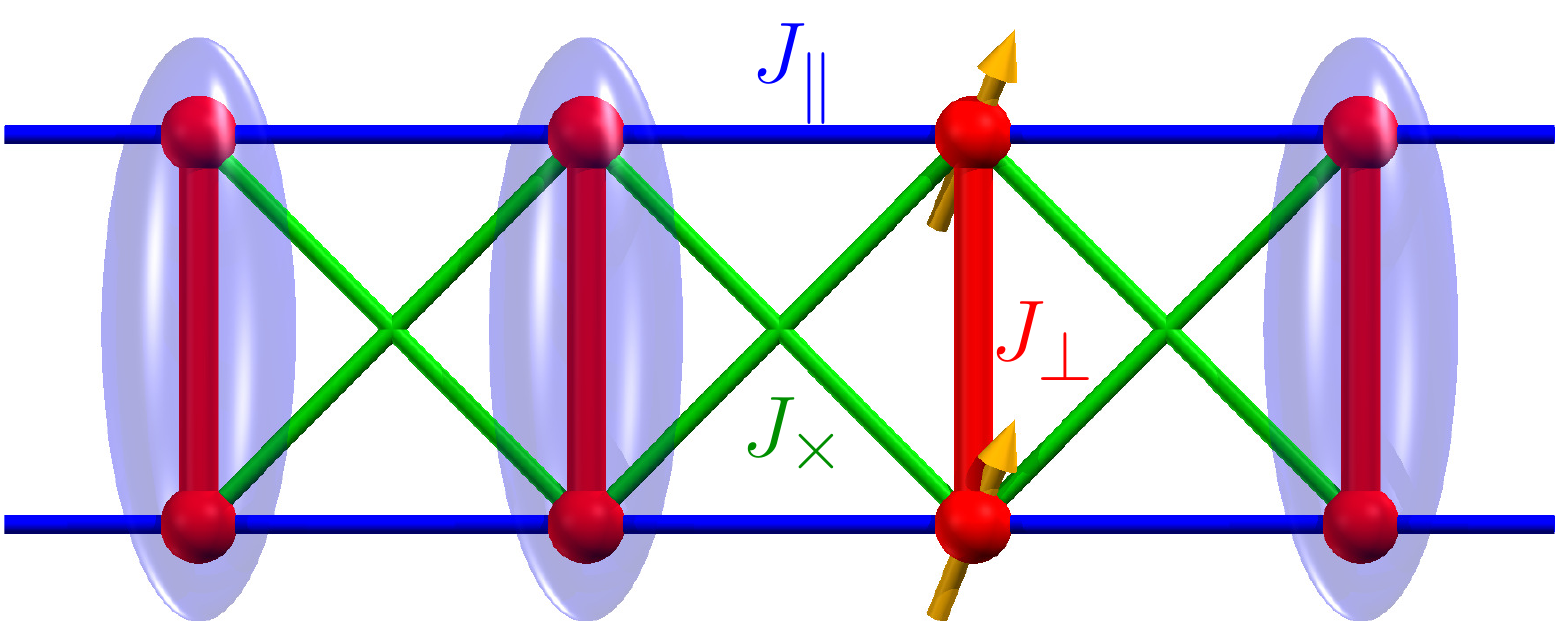}
\caption{Representation of superexchange interactions in a frustrated spin 
ladder (reproduced from Ref.~\cite{rus}). Each ladder site (spheres) hosts 
an $S = 1/2$ quantum spin and the Heisenberg couplings between spins are 
specified by the parameters $\JR$ for the ladder rungs, $\JL$ for the ladder 
legs, and $\JD$ for the cross-plaquette couplings, which we take to be 
symmetrical. Purple ellipses represent singlet spin states on the rungs 
and their absence a rung triplet (center).}
\label{fig:ladder}
\end{figure}

The Hamiltonian of the frustrated $S = 1/2$ Heisenberg spin ladder 
represented in Fig.~\ref{fig:ladder} is  
\begin{equation} 
\!\!\!\! H \! = \! \sum_{i} \! \JR {\vec S}_{i}^1 \cdot {\vec S}_{i}^2 + \!\!
\sum_{i,m=1,2} \!\! \left[ \JL {\vec S}_{i}^m \cdot {\vec S}_{i+1}^m
 + \JD {\vec S}_{i}^m \cdot {\vec S}_{i+1}^{\bar m} \right] ,
\label{essh} 
\end{equation} 
where $i$ the rung index, $m = 1$ and $2$ denote the two chains of the 
ladder, and ${\bar m}$ is the chain opposite to $m$. The spectrum of 
the ladder is gapped for all finite values of $\JR$, and for all values 
$\JR > 2 \{\JL,\JD\}$ the spin correlations are dominated by singlets 
occupying all of the rung bonds. In this regime, a triplet excitation on 
one of the rungs (to which, for reasons of clarity, we refer henceforth as 
a ``triplon''), as represented in Fig.~\ref{fig:ladder}, is a propagating 
quasiparticle with hard-core-boson nature. However, as the inter-rung 
frustration, $\JD$, approaches $\JL$, the nature of the ground state can 
change (below) and at the fully frustrated point, $\JD = \JL$, the triplon 
band is completely flat, with energy $\omega_k = \JR$ for all wave vectors 
$k_\|$. 

The fully frustrated $S = 1/2$ ladder has been considered in a number 
of studies. It was first shown in Ref.~\cite{rg} that, for a range of 
weak inter-rung couplings, all (one-, two-, three-triplon, \dots) excited 
states are strictly localized objects, and the critical coupling for this 
range was found to be $j' = \JR/\JL = \JR/\JD \ge j'_c \simeq 1.401$. 
In Ref.~\cite{rx} it was shown that, up to an additive constant, the 
Hamiltonian (\ref{essh}) may be expressed as 
\begin{equation} 
H/\JL = \sum_{i} \left[ j' ({\vec S}_{i}^1 \cdot {\vec S}_{i}^2 - 1/4) + 
{\vec P}_i \cdot {\vec P}_{i+1}\right], 
\label{exeh} 
\end{equation}
where the first term is finite only for singlet rung states and the second 
is a Heisenberg interaction between $S = 1$ spin operators, ${\vec P}_i$, 
describing the triplon states. A similar expression was obtained in 
Ref.~\cite{rhmt} for the general case of the fully frustrated $m$-leg ladder, 
where the total spin of each rung remains a good quantum number and thus 
gives an infinite number of conserved quantities in the thermodynamic limit. 

Focusing on the two-leg case, when the inter-rung coupling ratio, $1/j'$, 
becomes a significant fraction of 1, it is more favorable for the fully 
frustrated ladder to abandon the rung-singlet state in favor of a rung-triplet 
one, thereby satisfying all the inter-rung bonds. The phase diagram of the 
model has been found \cite{rx,rhmt,rccbz} to have only this one, 
first-order transition. Because the rung-triplet state has many properties 
in common with the Haldane chain, one may use Eq.~(\ref{exeh}) together 
with the very accurately known ground-state energy of the Haldane chain
\cite{white93,golinelli94} to deduce \cite{rx} that $j'_c = 1.401 \, 484 \, 
038 \, 971(4)$; the nature of the excited states of the system for $j' < j'_c$ 
may also be understood in detail from this parallel \cite{rx,rus}. The phase 
diagram of the frustrated ladder has been discussed in detail in
Refs.~\cite{rg,rx,rhmt,BoGa93,rzko,KSE99,rw,rccbz} 
and has also been considered for related models including tetrahedral cluster 
chains \cite{rbb2,rtm,rvh,Kim08,rhs,Poilblanc10,rlm}.

Returning to the nature of the fully localized excited states, it was 
shown for the two-leg ladder \cite{rg} that these are exact bound states. 
In the rung-singlet regime at $j' > j'_c$, the properties of an $n$-triplon 
ladder segment may be deduced by considering the $n$-site Haldane chain 
with open boundary conditions \cite{kennedy90,NUZ97}. As an example, the 
excitation with two triplons on neighboring rungs forms an exact bound state 
consisting of a singlet with energy $E_{2s}/\JL = 2(j' - 1)$, a triplet with 
energy $E_{2t}/\JL = 2 j' - 1$, and a quintet with $E_{2q}/\JL = 2 j' + 1$. 
The spectrum of the analogous 27-branch multiplet for the case $n$ = 3, which 
along with $n = 2$ will turn out to be most important for our analysis, is 
presented in App.~\ref{appa}. 
 
The existence of the exact bound-state regime is a strong asset for the 
purpose of assessing whether bound or scattering states arising due to 
frustration effects may contribute to a complex evolution of the dynamical 
spectral function at finite temperatures, of the type observed in 2D in 
SrCu$_2$(BO$_3$)$_2$. It is clear by inspection of the energy $E_{2s}$, the 
singlet level of the two-triplon bound state, that the gap to the lowest 
excitation of the system changes from $\JR$ to $E_{2s}$ in the parameter 
region $j'_c < j' < 2$. Similarly, the lowest branch of the three-triplon 
state is a net triplet whose energy, $E_{3tc}$, becomes the lowest triplet 
excitation when $j'$ falls below 1.5. Thus one may expect immediate and 
potentially dramatic effects in both the thermodynamic and dynamical response 
functions of the system in the vicinity of the quantum phase transition. 

\begin{figure}[t!]
\centering\includegraphics[width=0.98\columnwidth]{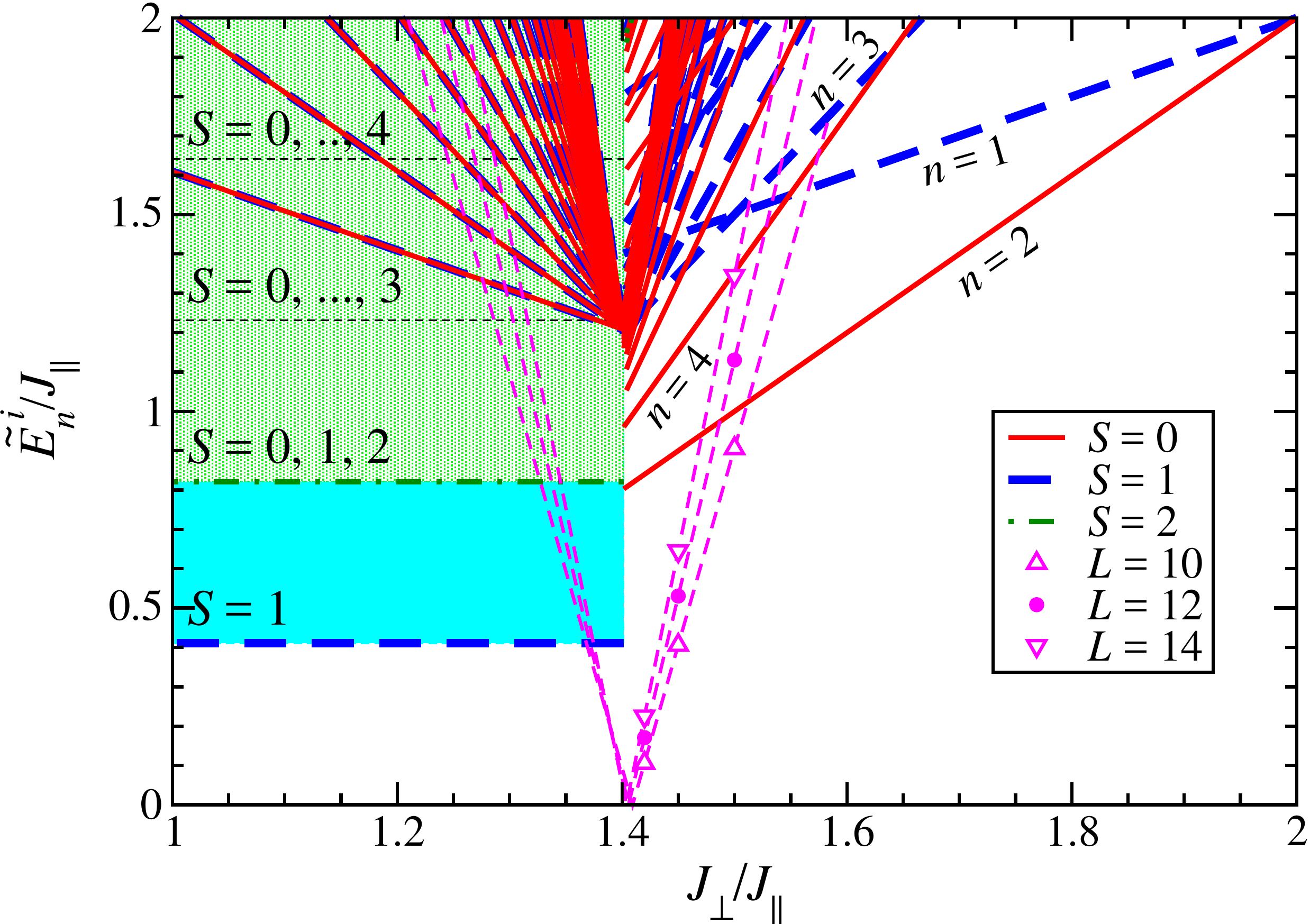}
\caption{Schematic representation of the low-lying energy levels, 
${\tilde E}_n^i$, of the infinite, fully frustrated spin-1/2 ladder, shown 
as a function of $\JR/\JL$ for a broad region around the quantum phase 
transition at $\JRc = 1.410484\,\JL$. Excitations are classified according to 
their total spin quantum number, $S$. The additional label $n$ for the 
rung-singlet phase specifies the origin of the excitations in the bound 
states of $n$ rung triplons. Shaded regions in the rung-triplet phase denote 
the one-magnon band and multiple-magnon continua. We indicate in addition 
the lowest excited energy state appearing close to the transition in our 
calculations for ladders of $L = 10$, $12$, and $14$ rungs; these ``intruder 
states,'' marked as thin dashed lines, and with their positions at $j' = 
1.42$, $1.45$, and $1.5$ marked by symbols, are finite-size features whose 
effects are discussed in the text and in App.~\ref{app:FS}.}
\label{fig:LinfAllExec}
\end{figure}

The energies of the lowest singlet and triplet levels of the exact $n$-triplon 
bound states at $j' = j'_c$ are shown in Ref.~\cite{rus} up to $n = 20$. 
Their key property is illustrated in Fig.~\ref{fig:LinfAllExec}, which we 
adapt from Fig.~4 of Ref.~\cite{rus}. As $j'$ approaches $j'_c$, a very 
large number of excitation branches falls to low (but finite) energy 
values below the one-triplon gap. The low-lying levels at $j' > j'_c$ are 
mostly the lowest branches of the $n$-triplon multiplets for all $n$ 
[Fig.~\ref{fig:LinfAllExec}], which are respectively singlets or triplets 
for even or odd $n$ \cite{rus}. We will show in Secs.~\ref{sec:FTspec} and 
\ref{sec:Analysis} how these states are responsible for anomalous 
redistribution of spectral weight at finite temperatures. For $j' < j'_c$, 
the discrete lines are bound states of singlet excitations within the 
rung-triplet background, while the shaded areas of Fig.~\ref{fig:LinfAllExec} 
represent the continuous spectrum arising from the magnon excitations of the 
Haldane chain, with different sectors for different magnon numbers.

We stress that Fig.~\ref{fig:LinfAllExec} represents the spectrum of the 
system in the thermodynamic limit, except for the low-lying levels very 
close to the phase transition. We include as thin, dashed lines the levels 
on both sides of the transition that lie lower than any others in our 
finite-size calculations. As discussed in Sec.~\ref{sec:FTspec}, these are 
performed by ED of the Hamiltonians (\ref{essh}) and (\ref{exeh}) for ladders 
with a finite number, $L$, of rungs and periodic boundary conditions. The 
levels appearing in Fig.~\ref{fig:LinfAllExec} for ladders of $L = 10$, 12, 
and 14 rungs are in fact the ground states of the system on the opposite 
side of the transition and their energies are infinite for systems of 
infinite length. However, in finite-system calculations they remain present 
as spurious ``intruder'' states, which are low-lying when $|j' - j'_c| 
\lesssim 0.05$ (for $L = 14$). These intruder states are the primary
source of finite-size effects in our calculations and their consequences 
for the physical quantities we compute are discussed in detail in 
Sec.~\ref{sec:Analysis} and App.~\ref{app:FS}.

Regarding thermodynamic quantities, in Ref.~\cite{rus} we computed the magnetic 
specific heat, $C (T)$, and susceptibility, $\chi (T)$, over the entire range 
of fully frustrated coupling ratios in Eqs.~(\ref{essh}) and (\ref{exeh}), 
spanning the rung-singlet and -triplet regimes. ED methods are appropriate 
for all values of $j'$ away from the phase transition, where the correlation 
lengths are far shorter than the accessible system size ($L = 14$ rungs). 
Close to $j' = j'_c$ we used a quantum Monte Carlo technique, which we adapted 
to be free of the sign problem affecting most frustrated systems, to access 
sizes up to $L = 200$ and thereby obtain numerically exact results. 
Qualitatively, the effect of frustration is to redistribute the thermodynamic 
response to both lower and higher energies, despite the one-triplon band 
remaining entirely flat, and near the transition a sharp, low-energy peak 
develops in $C (T)$. In a straightforward description based on 2-, 3-, $\dots$, 
$n$-triplon bound states within the rung-singlet chain, multi-triplon effects 
are observable over a wide range of $j'$ values, but become critically 
important in the regime $j' \rightarrow j'_c$, where many high-$n$ states 
approach the gap energy (Fig.~\ref{fig:LinfAllExec}) and the peak of $C (T)$ 
[the upturn in $\chi(T)$] is pushed to very low temperatures. We now begin 
to investigate the consequences of these bound-state effects for the 
finite-temperature dynamics of frustrated spin ladders, presenting our 
methods and results in Sec.~\ref{sec:FTspec} and a detailed analysis in 
Sec.~\ref{sec:Analysis}. 

\section{Finite-Temperature Dynamical Spectral Functions}

\label{sec:FTspec}

\subsection{Calculating Spectral Functions at Finite Temperatures}

\label{sec:FTspecA}

We calculate the dynamical structure factor 
\begin{equation}
S^{zz} ({\vec k}, \omega) = \frac{1}{\pi Z} \sum_{i,j} {\rm Im} \frac{ 
e^{-E_i/T} \big| \langle j | {S}^z ({\vec k}) | i \rangle \big|^2}
{\omega - (E_j - E_i + i \eta)}, 
\label{edsfah}
\end{equation}
where $Z$ is the partition function and we set $\hbar$ and $k_{\rm B}$ to 1. 
The Fourier-transformed spin operators are defined by
\begin{equation}
\vec{S} (k_\|,k_{\perp}) \! = \! \frac{1}{\sqrt{2L}} \! \sum_{\ell=1}^L \sum_{m=1}^2
\exp [i(k_\| \ell + k_\perp m)] \vec{S}_{\ell}^m ,
\label{eq:defSk}
\end{equation}
where the ``transverse momentum,'' $k_\perp$, takes the values $0$ (symmetric 
channel) and $\pi$ (antisymmetric channel). The quantity $\eta$ is a small 
imaginary part, which appears as a line-broadening in the spectral function; 
in the limit $\eta \to 0$, one recovers a representation in terms of
$\delta$-functions. As discussed in App.~\ref{appb}, the full rotational 
symmetry of the problem in spin space means that it is sufficient to compute 
$S^{zz} ({\vec k},\omega)$.

For the calculation of Eq.~(\ref{edsfah}), we employ ED in order to perform 
a complete investigation of two-chain spin ladder over a broad range of 
temperature. The high symmetries and exact conservation laws in the fully 
frustrated case extend the accessible system lengths to $L = 16$ ladder rungs 
and the extremely short spin-spin correlation length of the model makes this an 
eminently viable size. ED has the further significant advantages \cite{rhea} 
of providing spectral functions with arbitrarily fine energy resolution and 
of working with periodic boundary conditions in Eqs.~(\ref{essh}) and 
(\ref{exeh}). Our calculations exploit the conservation of $S^z$ and the 
full translational symmetry, as well as certain aspects of the reflection 
and spin-inversion symmetries of the ladder. 

As noted in Sec.~\ref{sec:Intro}, recent advances in DMRG techniques
\cite{White_1992,White_1993,Schollwoeck_2011} raise the possibility 
of qualitative improvements in the computation of finite-temperature spectral 
functions. Dynamical correlation functions were originally obtained at finite 
temperatures by applying DMRG to QTMs \cite{rnwzl}, including the computation 
of their full real-time evolution \cite{rsk}, and later by the use of 
minimally entangled typical thermal states (METTS) \cite{rbb1,rbvdw}.
While most modern DMRG approaches retain the basic structure of purifying the  
mixed density operator and applying real-time evolution \cite{rvgc,rzv,rfw,
rbsw,rkbm1,rb,rkbm2,rpeat}, and have been used with some success for real 
materials \cite{rltcbsnf}, a further recent development \cite{rtmph,rhea} is 
to work directly in frequency space at finite temperature. However, the high 
degeneracies and many conservation laws of the fully frustrated ladder provide 
an additional challenge to METTS-based DMRG techniques in accessing the full 
space of available states. For the present purposes, ED therefore provides 
the optimal approach for computing the spectrum of exact multi-triplon bound 
states, which are the key to the unconventional thermodynamic and dynamical 
response of the frustrated ladder and allow a complete analytical 
interpretation of our numerical results.

A direct approach to the evaluation of Eq.~(\ref{edsfah}) is based on full 
diagonalization, i.e., on the computation of all eigenvalues, $E_i$, and the
corresponding eigenstates, $|i \rangle$. Although our initial computations 
for ladders of $10 \times 2$ spins were performed on the high-performance 
computers of HLRN II, this calculation is actually feasible on a modern 
desktop. The primary bottleneck is the requirement for approximately 470~GBytes 
of hard-disk space to store all the necessary matrix elements, $\langle j| 
{S}^z ({\vec k}) |i \rangle$, for each set of exchange constants.

To proceed to larger systems, and here we present data for ladders up 
to $16 \times 2$ spin-$1/2$ sites, it is no longer possible to obtain 
the full spectrum. However, one may truncate the sum over $i$ in 
Eq.~(\ref{edsfah}) to low energies, $E_i$, and use the Lanczos algorithm 
\cite{Lanczos,DagottoRevModPhys66} to compute these low-lying states; the 
spectral sum over $j$ is then evaluated by a continued-fraction expansion for 
each eigenstate, $|i \rangle$ \cite{DagottoRevModPhys66,HHK72,GBPhysRevLett59}.
%Typically 
We retain at least 20 initial states, $|i \rangle$, in each symmetry 
sector for ladders of $L = 14$ rungs, pushing this up to several hundred per 
symmetry sector for $L = 12$. The truncation to low energies restricts the 
validity of this approximation to low temperatures. Thus, although we have 
access only to smaller system sizes ($L = 10$ rungs) at higher temperatures, 
the finite-size effects we wish to gauge are important primarily at lower 
temperatures. For this reason, we have pushed our calculations to the largest 
combinations of system size and temperature allowed by our computational 
resources. 

One specific advantage of ED is that we have access to the weight of 
individual poles of the spectral function, i.e., the coefficients of the 
individual $\delta$-functions in the limit $\eta = 0$. In the context of 
full diagonalization, this is contained in the spectral representation 
[Eq.~(\ref{edsfah})], while in the continued-fraction expansion these 
weights are obtained from the eigenvectors of the associated tridiagonal 
Lanczos matrix \cite{DagottoRevModPhys66}.

The effects of the finite temperature are incorporated through Boltzmann 
weighting factors, $e^{-E_i/T}$, on every energy level, $E_i$, in the 
spectrum. We have performed calculations for ladders of $L = 6$, $8$, $10$,
$12$, $14$, and $16$ rungs at zero temperature, and up to $L = 14$ at finite 
temperatures, in order to analyze the effects of finite system size. We will 
demonstrate that these effects are in general very limited, because of the 
extremely short correlation length of the maximally frustrated system, and 
it is only arbitrarily close to $j' = j'_c$ (in fact where multi-triplon 
bound states with $n \approx L$ become relevant) that they become visible 
in the dynamical response.

The two-leg ladder has two types of spectral function, those symmetric or 
antisymmetric between the legs, corresponding as above to the quantum 
numbers $k_{\perp} = 0$ and $\pi$, and we will show both. At $j' > j'_c$ 
(rung-singlet ground state), processes changing the number of triplons, $l$, 
by an odd number appear in the antisymmetric channel, while those involving 
even numbers (including zero, as in intra-multiplet transitions) are 
symmetric. The selection rules of a neutron-scattering process, and indeed 
of the matrix elements in Eq.~(\ref{edsfah}), are that the change in total 
spin may be only $\Delta S = 0, \pm 1$. 

For the fully frustrated ladder, these rules combine with the conservation 
law on the total spin of each rung \cite{rhmt} to dictate that direct 
processes are allowed only from sectors with $l$ excited rung triplons to 
sectors with $l$ or $l \pm 1$. At zero temperature, the rung-singlet ground 
state may therefore be excited by the antisymmetric operators of 
Eq.~(\ref{edsfah}) only into the antisymmetric channel, and we stress that 
it is not possible to couple directly to the triplet branch of an $n$-triplon 
bound state with $n \ge 2$. Because the rung spin is a good quantum number 
for all parameters, the same is true for the rung-triplet ground state. This 
result is a consequence of the perfect frustration and can be relaxed if the 
frustration is incomplete, albeit with small matrix elements, as we will 
illustrate in Secs.~\ref{sec:FTspecC} and \ref{sec:Analysis}. 

As a consequence, all excitation processes of the fully frustrated ladder 
to sectors of higher $l$ take place only at finite temperatures, where the 
probability for multi-triplon states to be populated is non-vanishing. 
However, the transitions between these states and others within the thermal 
population remain restricted to those obeying the selection rules $\Delta S
 = 0, \pm 1$ and $\Delta l = 0, \pm 1$. The effect of the temperature is to 
alter the relative weights of the different terms in Eq.~(\ref{edsfah}), 
establishing the thermal evolution of $S ({\vec k}, \omega)$. 

\subsection{Fully Frustrated Ladders}

\label{sec:FTspecB}

We first provide results to illustrate the evolution of the dynamical spectral 
function with temperature. Our primary focus is on the fully frustrated ladder 
($\JL = \JD = 1$), for which we compare the situations within and beyond the 
exact bound-state regime ($j' > j'_c$ and $j' < j'_c$), with particular 
attention paid to the critical region around $j' = j'_c$. To gain more 
perspective on these results, we also compare the fully frustrated case 
with unfrustrated ($\JD = 0$) and less frustrated ladders ($\JD = 0.9 \JL$).

\begin{figure}[t!]
\centering\includegraphics[width=0.98\columnwidth]{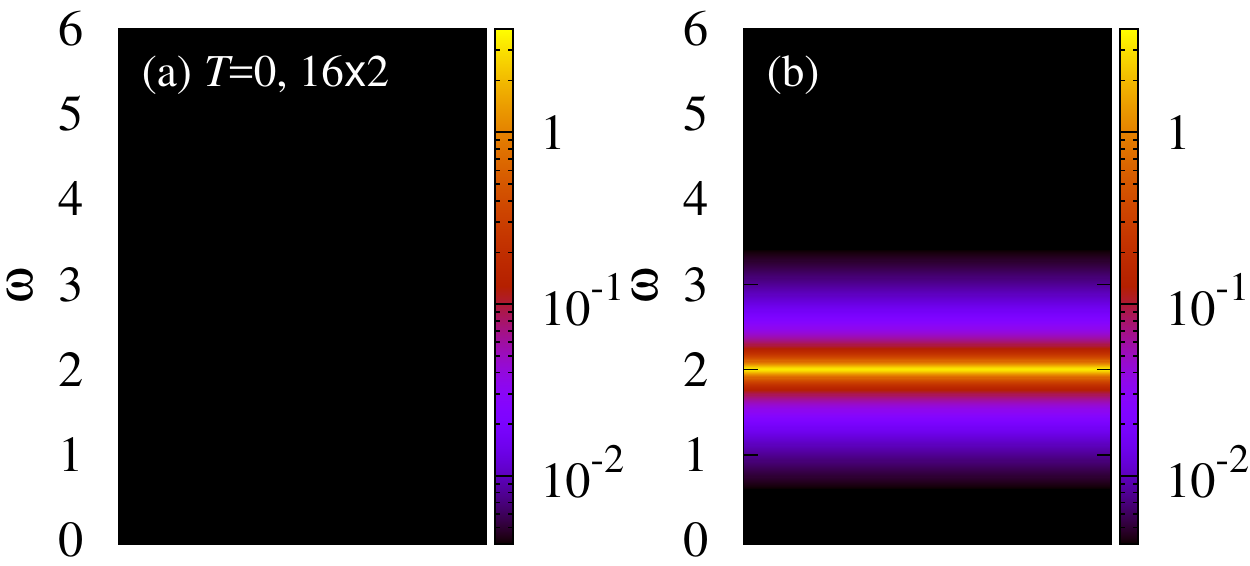}
\centering\includegraphics[width=0.98\columnwidth]{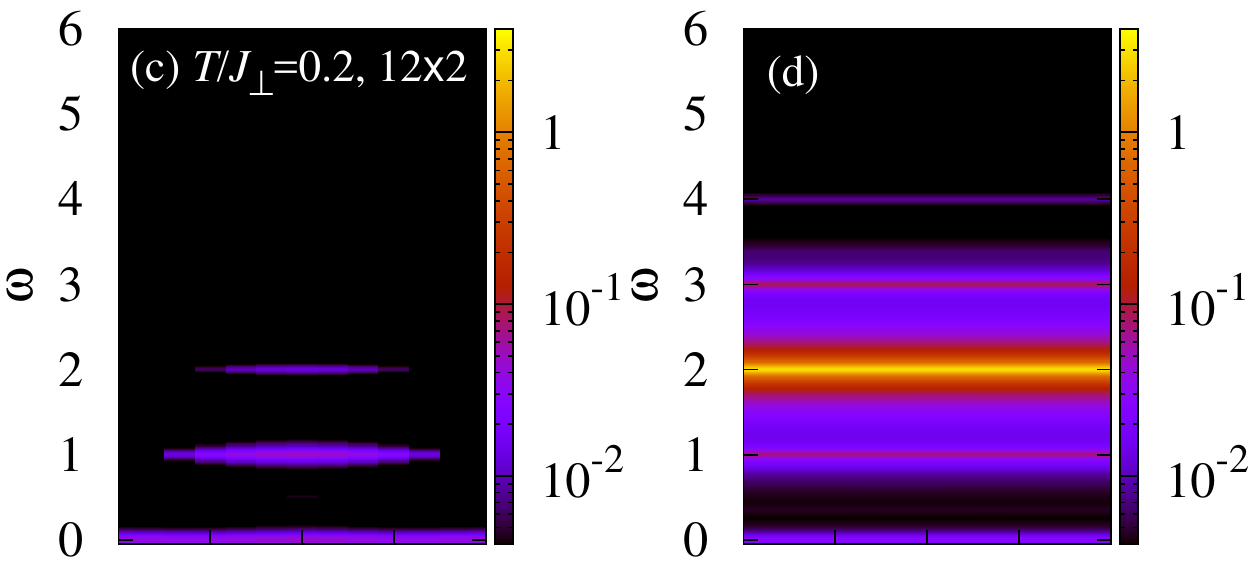}
\centering\includegraphics[width=0.98\columnwidth]{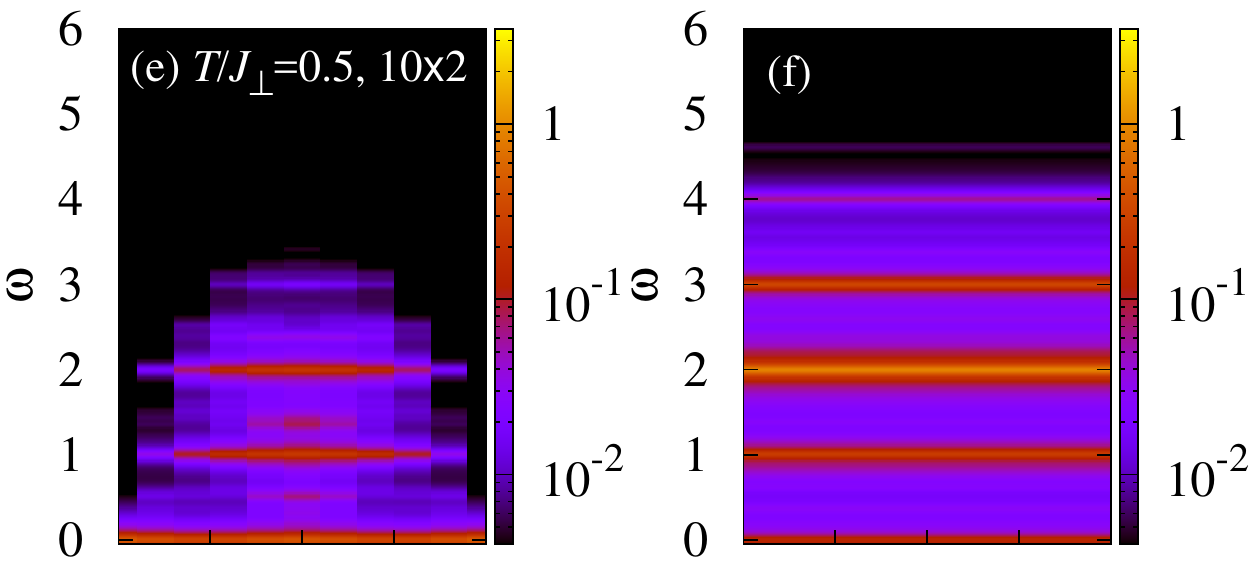}
\centering\includegraphics[width=0.98\columnwidth]{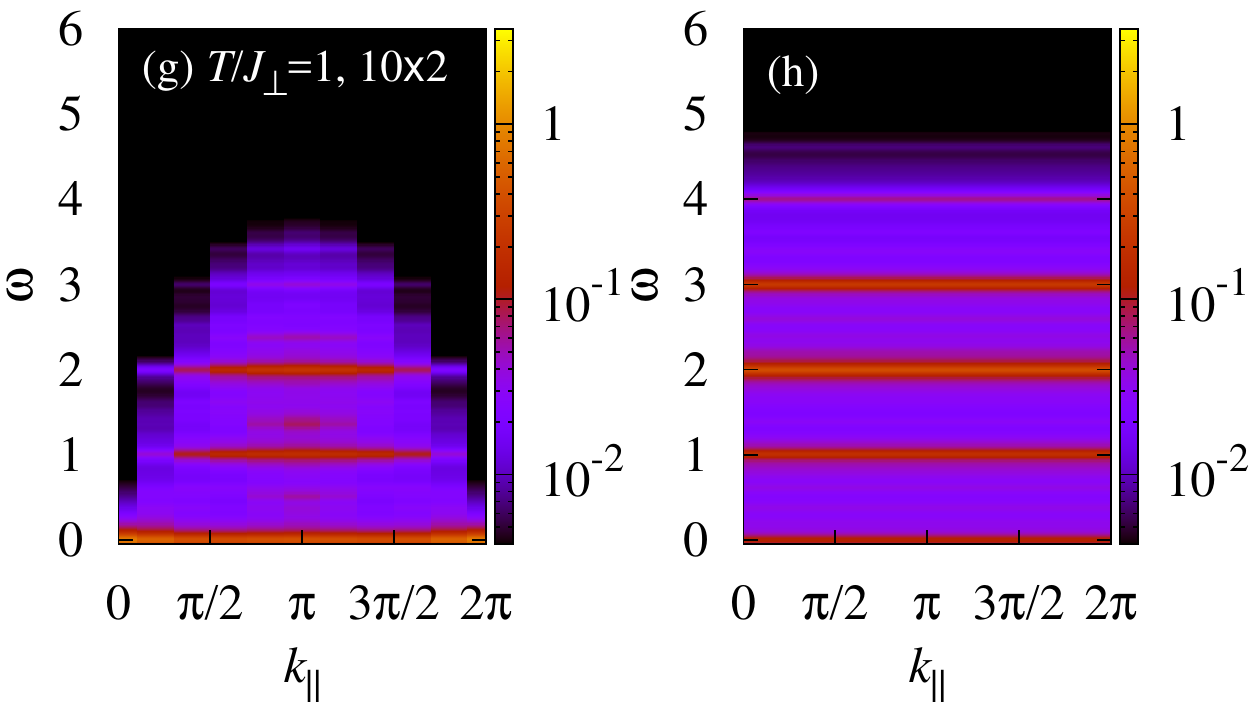}
\caption{Dynamical structure factor in (a,c,e,g) the symmetric and (b,d,f,h)
the antisymmetric channel for the ladder of Fig.~\ref{fig:ladder} with 
coupling parameters $\JR = 2$ and $\JL = \JD = 1$, for temperatures of (a,b) 
$T/\JR = 0$ ($16 \times 2$ spins), (c,d) $0.2$ ($12 \times 2$ spins), (e,f) 
$0.5$, and (g,h) $1$. The system size for $T = 0.5$ and 1 is $10 \times 2$ 
spins and a Lorentzian broadening $\eta = \JL/20$ is applied to the data.}
\label{dsf211}
\end{figure}

\subsubsection{$j' > j'_c$}

We begin by presenting the calculated dynamical structure factor for a fully 
frustrated ladder with $\JR = 2$ and $\JL = \JD = 1$. Figure \ref{dsf211} 
shows the spectral weights for a range of temperatures between 0 and $\JR$, 
with the symmetric channel in the left panels and the antisymmetric channel 
on the right. We draw attention to the fact that the color contours in all 
the figures in this section represent logarithmic intensity scales, and 
thus the strong branches are indeed features of very high relative weight. 
At low temperatures, only one feature is visible, the one-triplon 
excitation appearing in the antisymmetric channel. Its band dispersion, 
$\omega(k) = \JR$, is completely flat as a consequence of the perfect 
frustration. As the temperature increases, increasing spectral weight 
may be found in this channel at a significant number of different, 
discrete, and non-dispersive energy levels, which must arise from two- or 
more-triplon excitations. The number of visible levels continues to increase 
with $T$, as, with one obvious exception, does the weight they contain. The 
one-triplon mode has clearly lost appreciable spectral weight once $T = 0.5 
\JR$, and this loss continues with $T$. 

In the symmetric channel, we also find a range of discrete, non-dispersive 
excitations, but these have zero spectral weight at $k_\| = 0$ and a maximum 
at $k_\| = \pi$, causing a modulation of the scattered intensity. Again, both 
the number of visible levels and their weight increase with temperature. An 
explanation for the origin, possible degeneracy, and intensity evolution of 
the strongest energy levels is deferred to Sec.~\ref{sec:Analysis}. 

\begin{figure}[t!]
\centering\includegraphics[width=0.98\columnwidth]{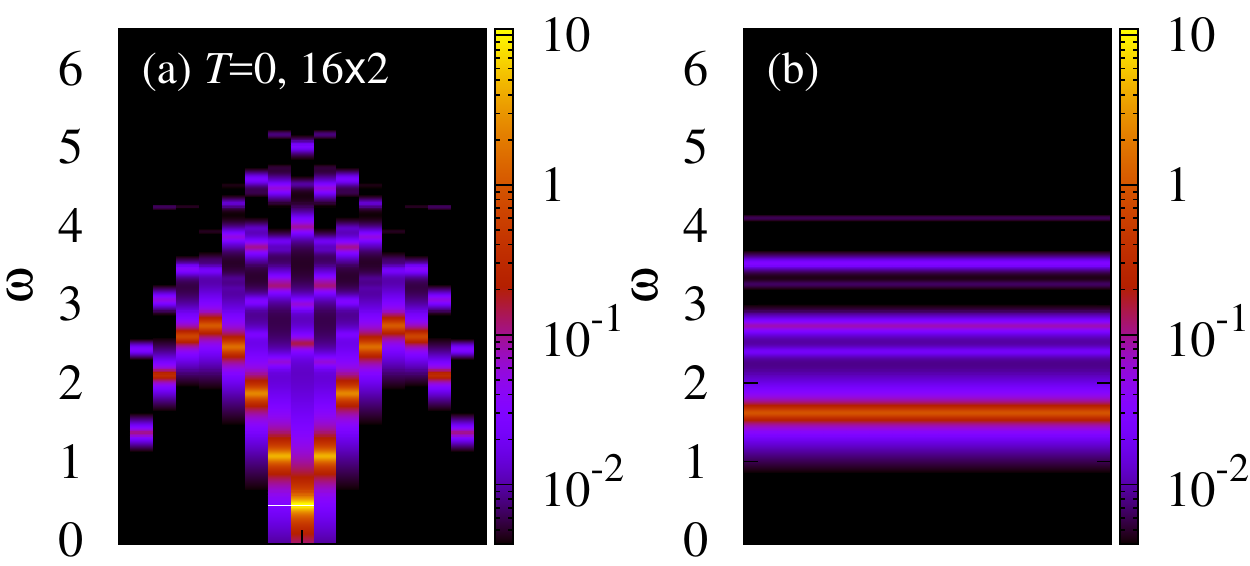}
\centering\includegraphics[width=0.98\columnwidth]{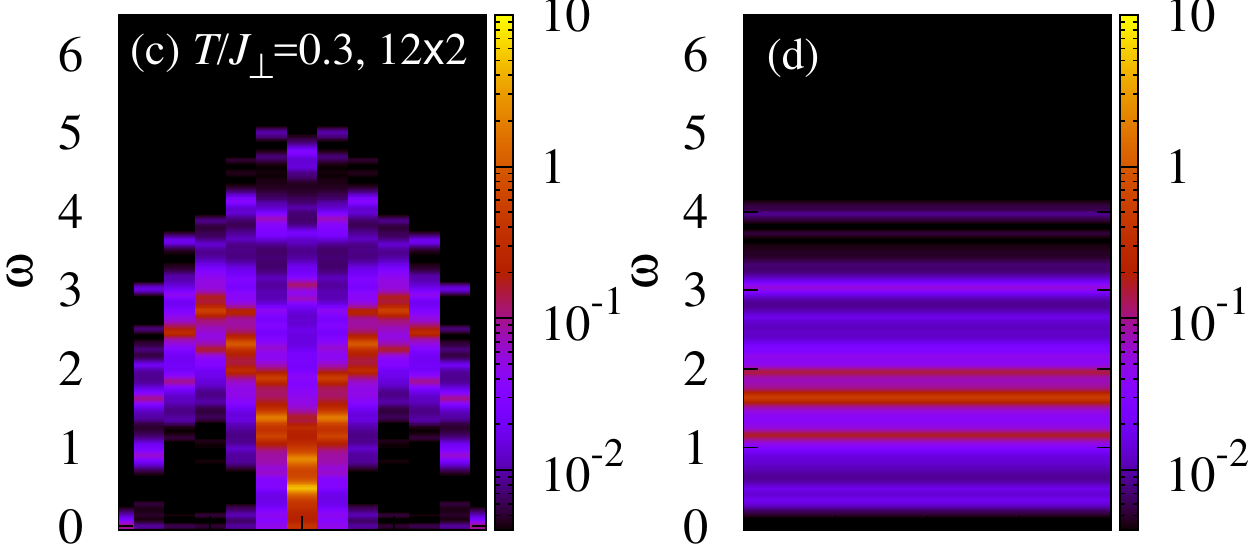}
\centering\includegraphics[width=0.98\columnwidth]{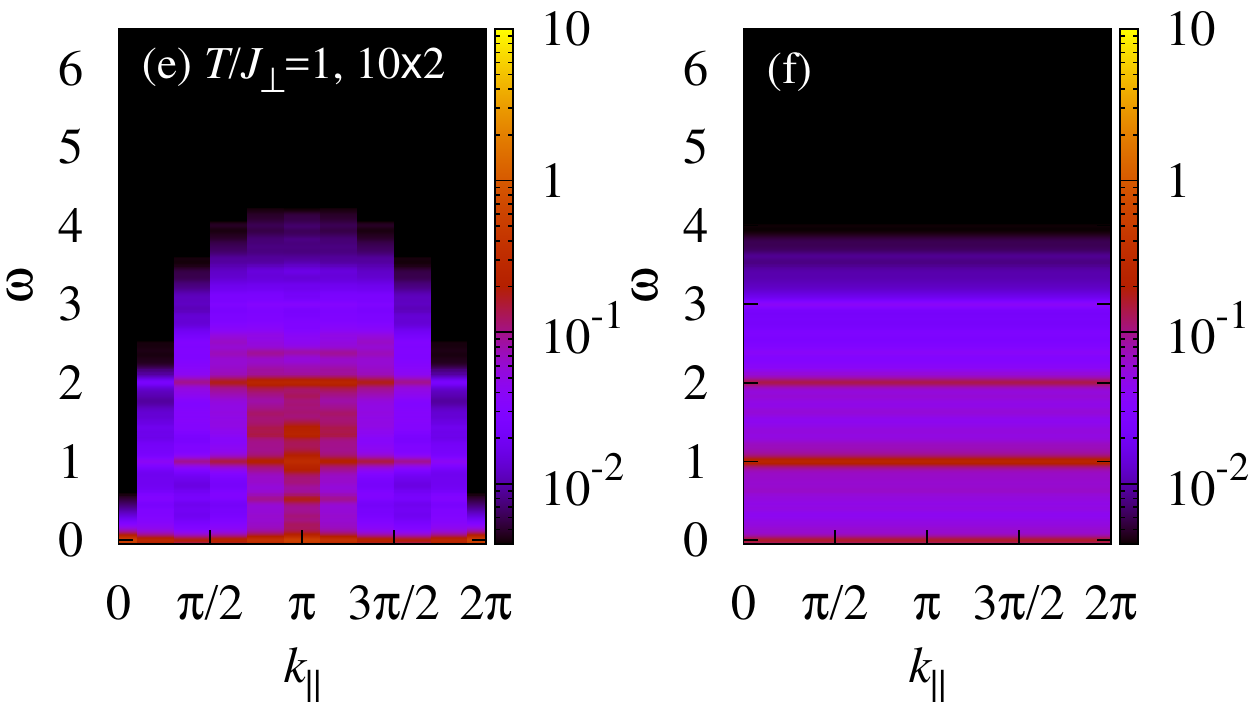}
\caption{Dynamical structure factor in (a,c,e) the symmetric and (b,d,f) 
the antisymmetric channel for the ladder of Fig.~\ref{fig:ladder} with 
coupling parameters $\JR = \JL = \JD = 1$, for temperatures of (a,b) $T/\JR
 = 0$ ($16 \times 2$ spins), (c,d) $0.3$ ($12 \times 2$ spins), and (e,f) $1$ 
($10 \times 2$ spins). A Lorentzian broadening $\eta = \JL/20$ is applied to 
the data.}
\label{dsf111}
\end{figure}

\subsubsection{$j' < j'_c$}

Remaining with fully frustrated ladders, in Fig.~\ref{dsf111} we show similar 
results for a ladder with $\JR = \JL = \JD = 1$, values for which the system 
is located in the rung-triplet (Haldane-type) phase. The antisymmetric channel 
again shows completely non-dispersive triplet bands at zero temperature, but 
in this case more than one. As the temperature is increased, these are joined 
by a larger number of weaker levels, which have almost merged into an 
intensity continuum when $T = \JR$. We observe that the one-triplon level 
dominating the low-$T$ spectrum has vanished almost completely at this 
temperature. 

In the symmetric channel, at low temperatures one may recognize the 
characteristic triplet (``magnon'') dispersion relation of the Haldane 
chain, whose dynamical structure factor at $T = 0$ is shown in Fig.~4(b) 
of Ref.~\cite{rSMKYM98} for similar system sizes. More specifically, in  
Fig.~\ref{dsf111}(a) we compute a band minimum at $k_\| = \pi$ of $\Delta/\JL 
\simeq 0.443$ for the $L = 16$ case \cite{SaTa90,golinelli94}, compared 
with the true Haldane gap of $\Delta/\JL = 0.4105$, and find the band maximum 
where it is cut off by the descent of the two-magnon continuum towards $k_\|
 = 0$ and $2\pi$ \cite{rwa}. At intermediate temperatures, more flat-band 
features begin to appear in the spectral weight, and at high temperatures 
the weight distribution is quite similar to the case with $\JR = 2$, shown 
in Fig.~\ref{dsf211}, suggesting the predominance of local physics. 

\begin{figure}[t!]
\centering\includegraphics[width=0.98\columnwidth]{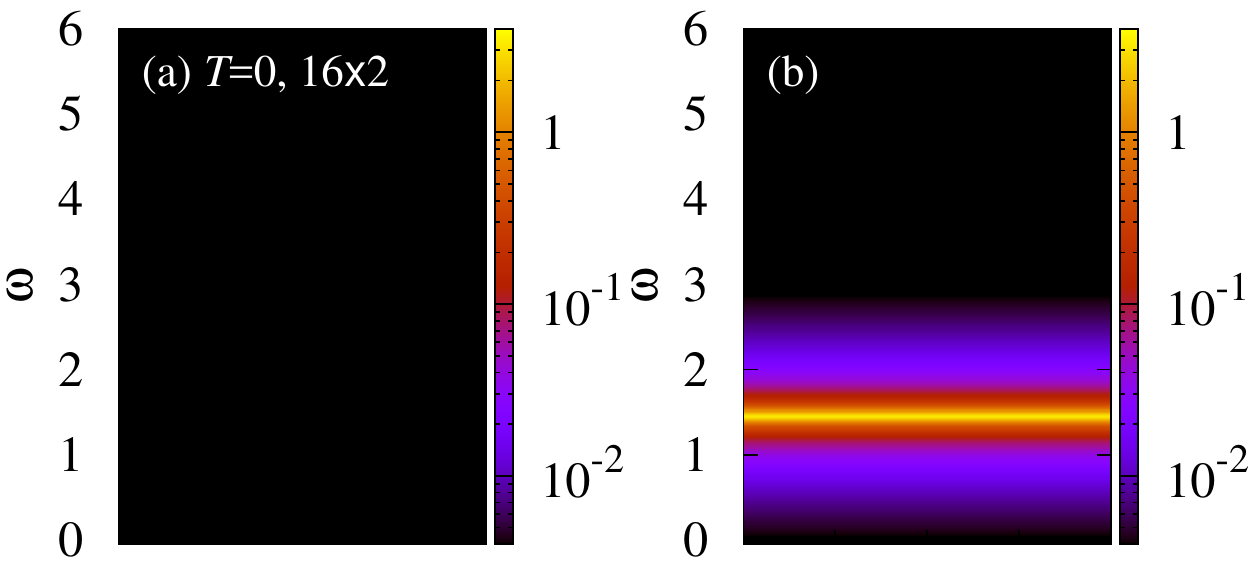}
\centering\includegraphics[width=0.98\columnwidth]{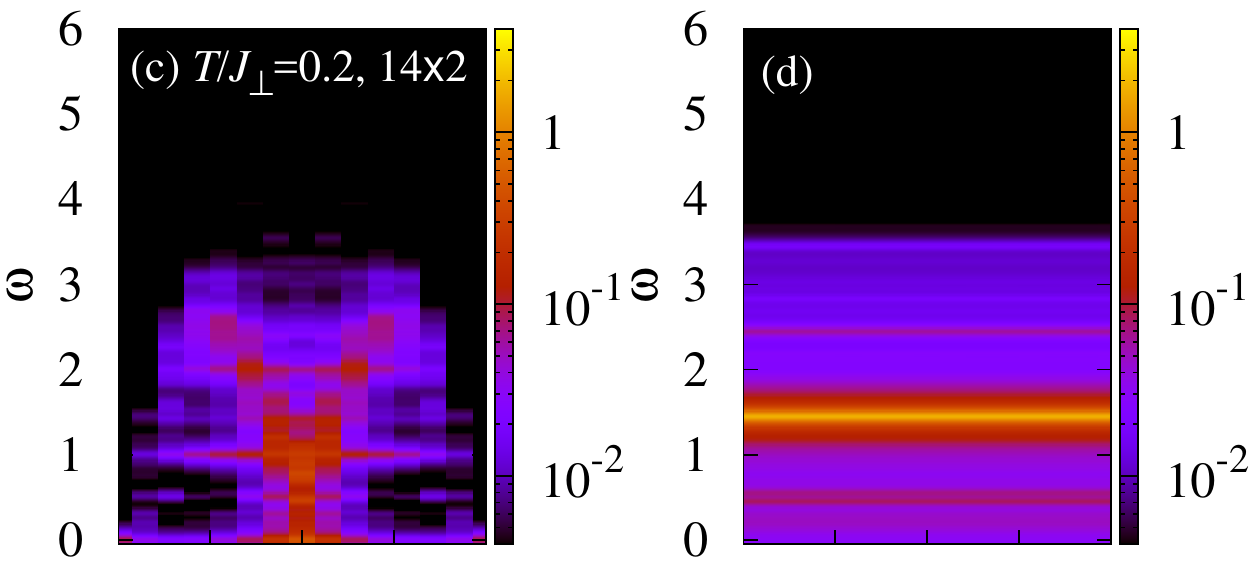}
\centering\includegraphics[width=0.98\columnwidth]{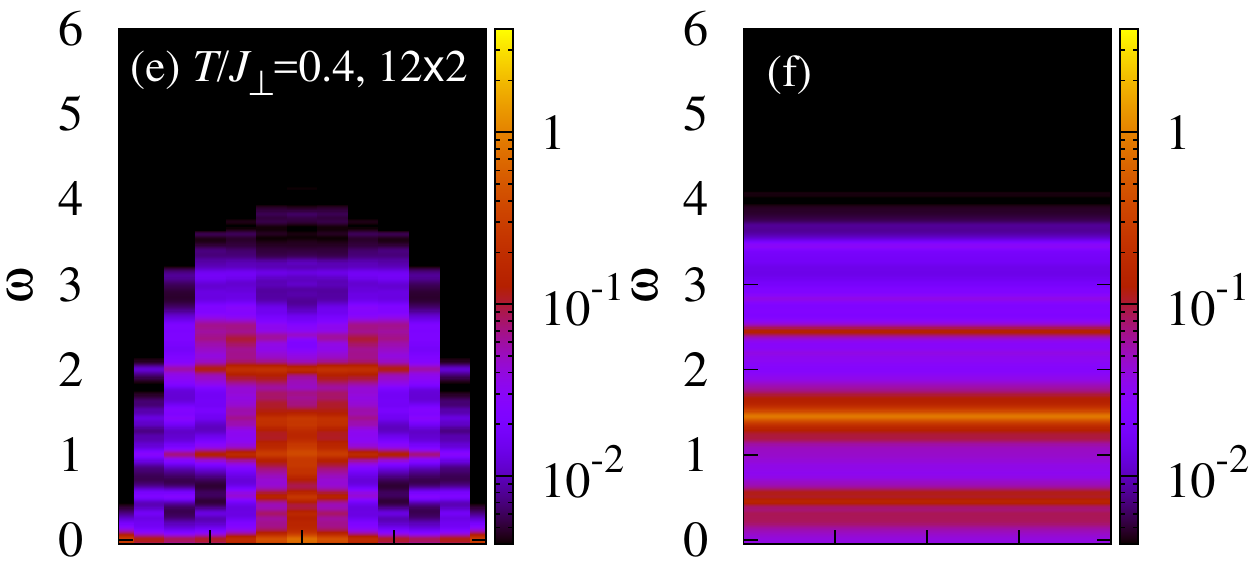}
\centering\includegraphics[width=0.98\columnwidth]{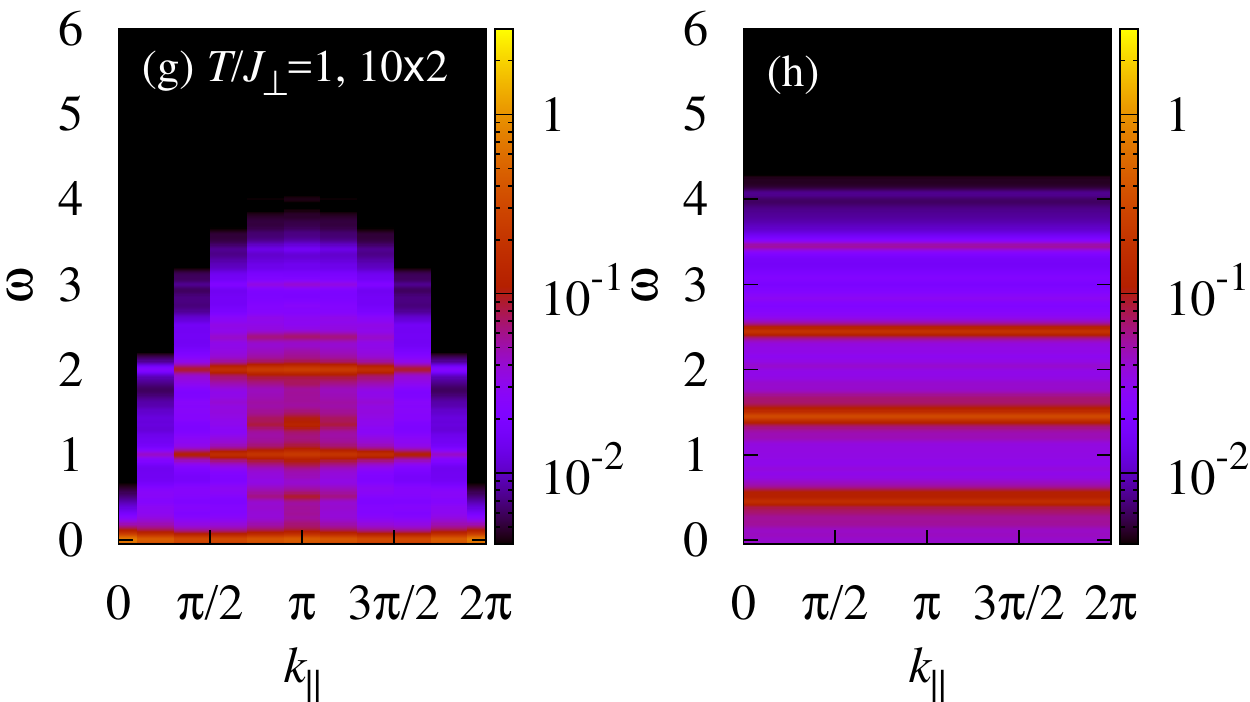}
\caption{Dynamical structure factor in (a,c,e,g) the symmetric and (b,d,f,h)
the antisymmetric channel for the ladder of Fig.~\ref{fig:ladder} with 
coupling parameters $\JR = 1.45$ and $\JL = \JD = 1$, for temperatures of 
(a,b) $T/\JR = 0$ ($16 \times 2$ spins), (c,d) $0.2$ ($14 \times 2$ spins),
(e,f) $0.4$ ($12 \times 2$ spins), and (g,h) $1$ ($10 \times 2$ spins).
A Lorentzian broadening $\eta = \JL/20$ is applied to the data.}
\label{dsf2o3}
\end{figure}

\subsubsection{$j' \rightarrow j'_c$}

As discussed in Sec.~\ref{sec:FFL} and demonstrated in the thermodynamic 
properties of the system \cite{rus}, the most dramatic phenomena in the 
fully frustrated ladder occur close to the quantum phase transition ($j'_c$), 
where large numbers of multi-triplet bound states appear at very low 
energies. However, these are of necessity extended objects, and therefore 
our calculations for small systems are no longer perfectly representative 
of the thermodynamic limit as $j' \rightarrow j'_c$. Nevertheless, in 
Fig.~\ref{dsf2o3} we show $S^{zz} (k,\omega,T)$ calculated for a ladder 
with $\JR = 1.45$ and $\JL = \JD = 1$, where we find our results to be 
largely converged at $L = 14$; thus the critical regime dominated by truly 
high-$n$ bound states remains rather narrow \cite{rus}. It is clear that 
the spectral weight evolves with temperature in a manner very similar to 
the $\JR = 2$ case (Fig.~\ref{dsf211}), which may be regarded as representing 
generic and non-critical fully frustrated behavior, but in an accelerated 
manner. The weight growth visible in Fig.~\ref{dsf2o3} is equivalent to an 
effective rescaling of the temperature by a factor of approximately 2.5 
compared with Fig.~\ref{dsf211}, and yet was obtained with no rescaling of 
energies (although we note that $T$ is normalized to $\JR$). 

To summarize our observations for fully frustrated ladders in the rung-singlet 
regime, the dynamical spectral function is dominated by discrete and 
non-dispersive excitations, whose origin lies in the existence of exact 
bound states. As a function of temperature, all of the bound-state levels 
are populated increasingly at the cost of the intensity in the one-triplon 
band. The intensity distribution is $k_\|$-independent in the antisymmetric 
channel, but peaked at $k_\| = \pi$ and vanishing at $k_\| = 0$ in the symmetric 
channel. The dependence of this thermal evolution effect on the inter-rung 
coupling ratio may be broadly characterized, from studies we do not show 
here for reasons of space, as follows. The dynamical spectral function for 
ladders with values of $j' \gtrsim 3$ varies little from the case $j' = \infty$ 
(isolated rung singlets) for any realistic values of the temperature. For 
values $1.5 \lesssim j' \lesssim 3$, we observe the essential phenomenon 
of thermal evolution in a frustrated system, that spectral weight from the 
one-triplon band is redistributed to lower and higher energies by the 
formation of few-triplon bound states. For values $j'_c \lesssim j \le 1.5$, 
we find highly anomalous versions of this effect, with extremely rapid 
spectral-weight transfer out of the one-triplon band and into bound states 
of many different triplons. For values $j < j'_c$, we find a composite spectral 
function containing some features of the Haldane chain \cite{rwa} and some of 
local excitations (rung-singlet bound states), a topic we discuss in more 
detail in Sec.~\ref{sec:Analysis}. 

\subsection{Unfrustrated and Partially Frustrated Ladders}

\label{sec:FTspecC}

\begin{figure}[t!]
\centering\includegraphics[width=0.98\columnwidth]{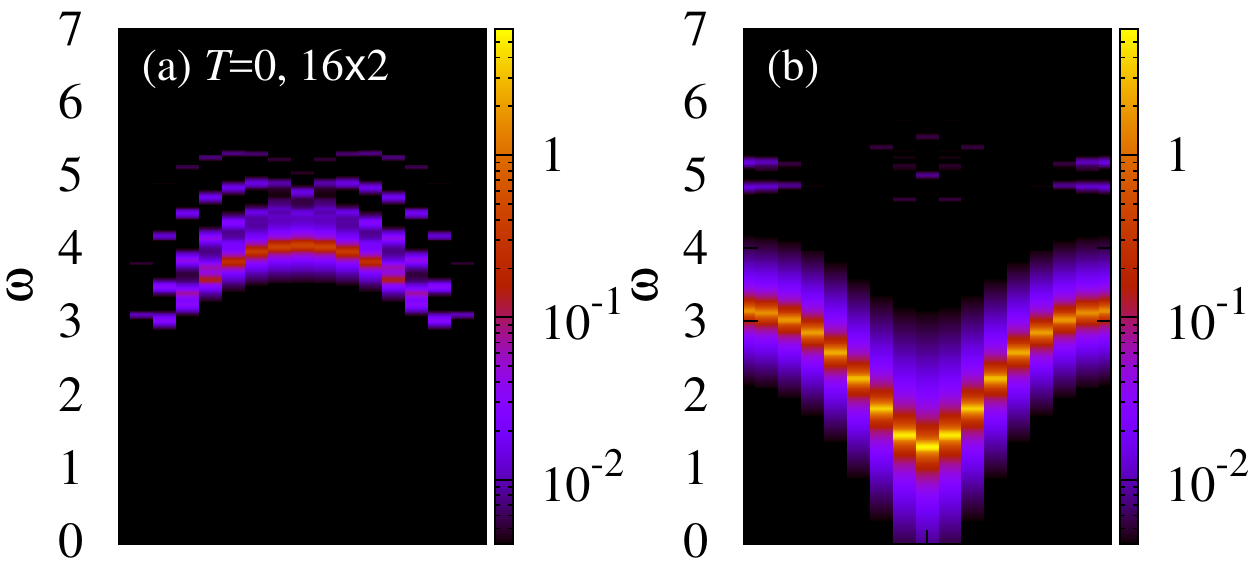}
\centering\includegraphics[width=0.98\columnwidth]{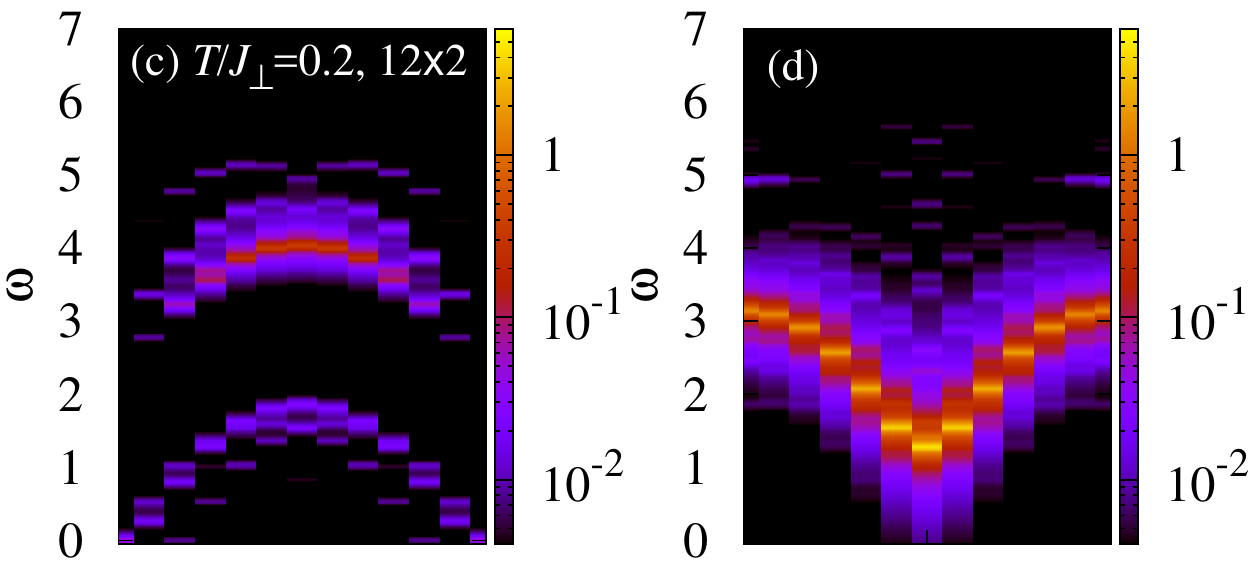}
\centering\includegraphics[width=0.98\columnwidth]{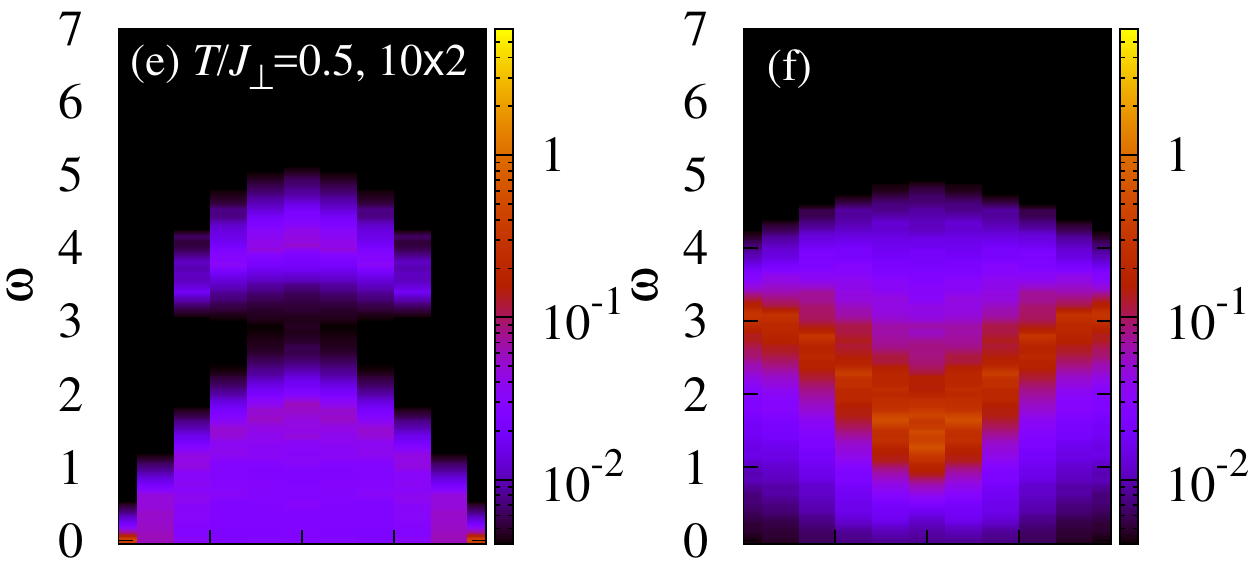}
\centering\includegraphics[width=0.98\columnwidth]{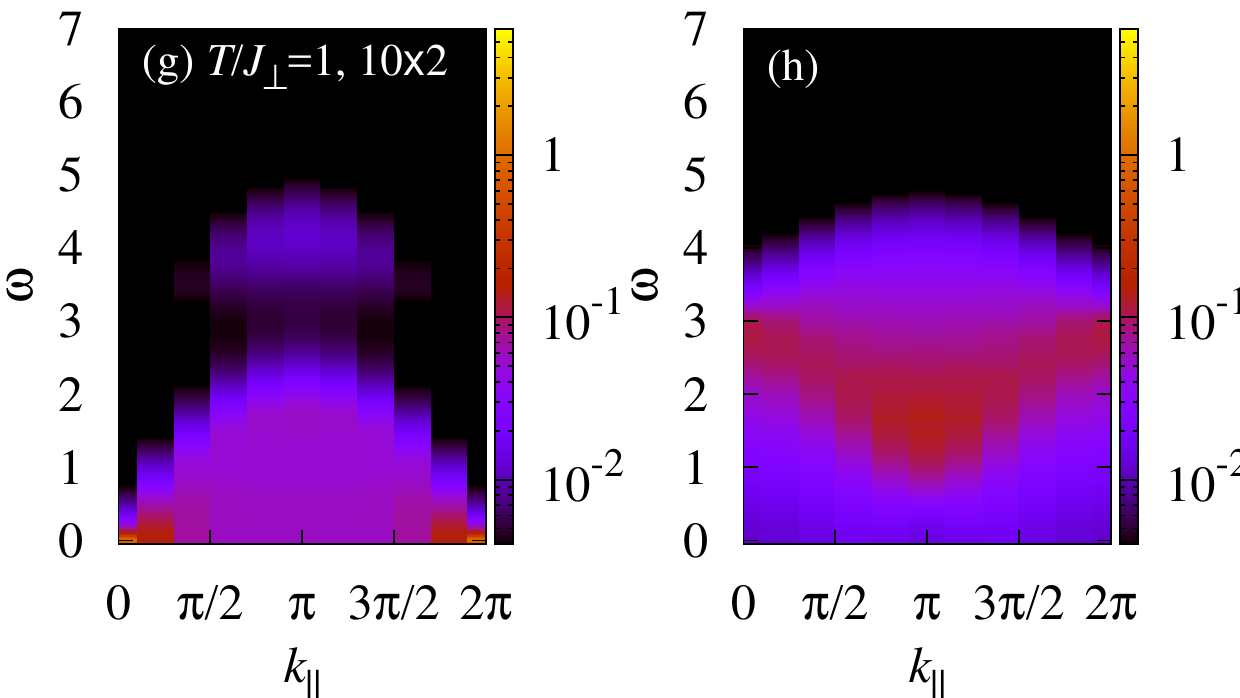}
\caption{Dynamical structure factor in (a,c,e,g) the symmetric and (b,d,f,h) 
the antisymmetric channel for the ladder of Fig.~\ref{fig:ladder} with coupling 
parameters $\JR = 2$, $\JL = 1$, and $\JD = 0$, for temperatures of (a,b) 
$T/\JR = 0$ ($16 \times 2$ spins), (c,d) $0.2$ ($12 \times 2$ spins),
(e,f) $0.5$, and (g,h) $1$ (both $10 \times 2$ spins). A Lorentzian 
broadening $\eta = \JL/20$ is applied to the data.}
\label{dsf210}
\end{figure}

To benchmark the effects of frustration on the temperature dependence of 
the dynamical spectral function, we switch them off in whole or in part. In 
Fig.~\ref{dsf210} we show results analogous to those of Fig.~\ref{dsf211}, but 
for an unfrustrated 14-rung ladder with $\JR = 2$, $\JL = 1$, and $\JD = 0$. In 
this case, the one-triplon band visible in the low-temperature antisymmetric 
response has a clear dispersion, with its minimum ($\omega/\JL \approx 1$) at 
$k_\| = \pi$ and a band width of approximately $2\JL$. This band is broadened 
into a continuum of available states as a function of increasing temperature, 
and at $T \approx \JR$ it has a width of order $\JR$, as expected for 
conventional thermal broadening. The low-temperature spectrum in the 
symmetric channel is that of a dispersive two-triplon bound state
\cite{rtmhzs,rKSGU01,rgea,rSU05}, which has a strongly $k_\|$-dependent 
intensity and a loss of weight from the high-energy feature reappearing at 
low energies for higher temperatures. In the thermodynamic limit, these bands 
are continuous, and simply broaden at finite temperatures into a continuum of 
available states with equal weights. 

\begin{figure}[t!]
\centering\includegraphics[width=0.98\columnwidth]{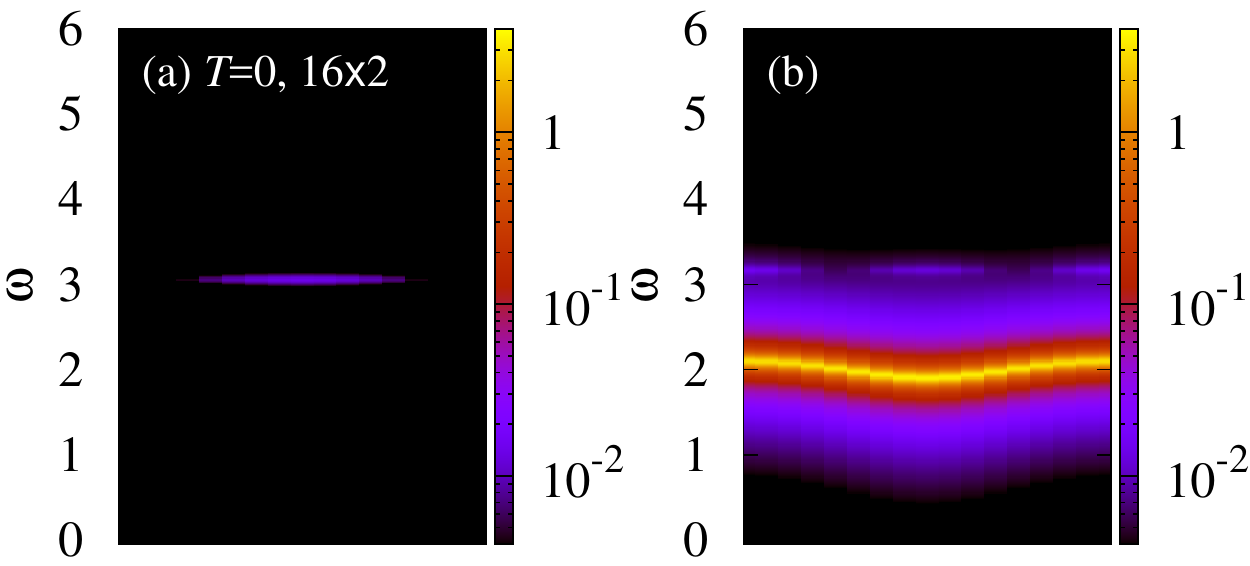}
\centering\includegraphics[width=0.98\columnwidth]{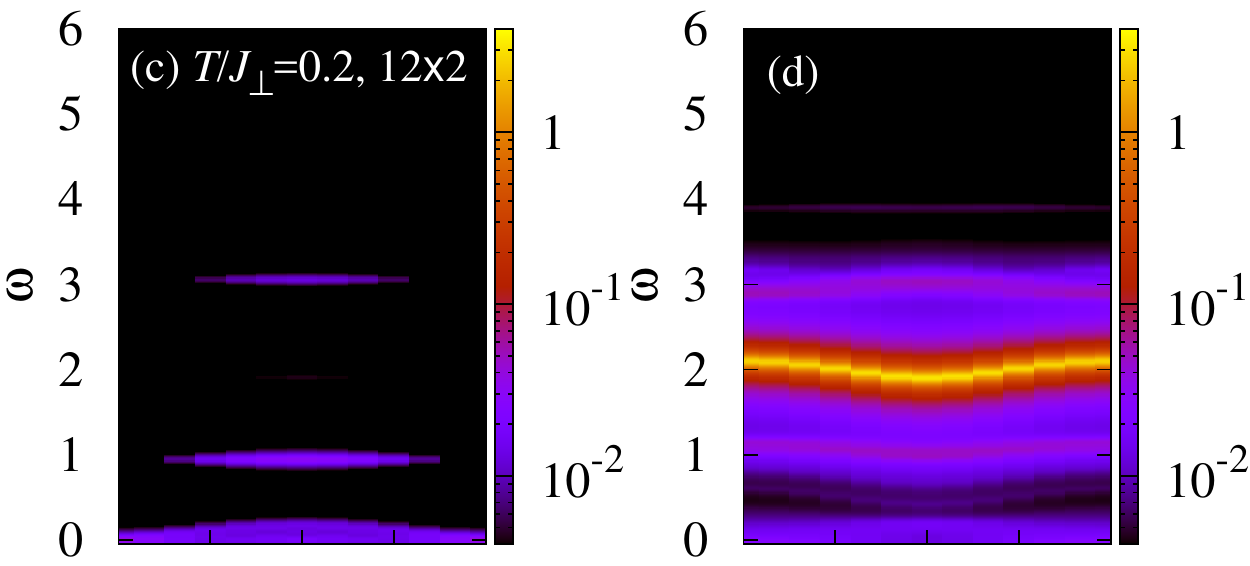}
\centering\includegraphics[width=0.98\columnwidth]{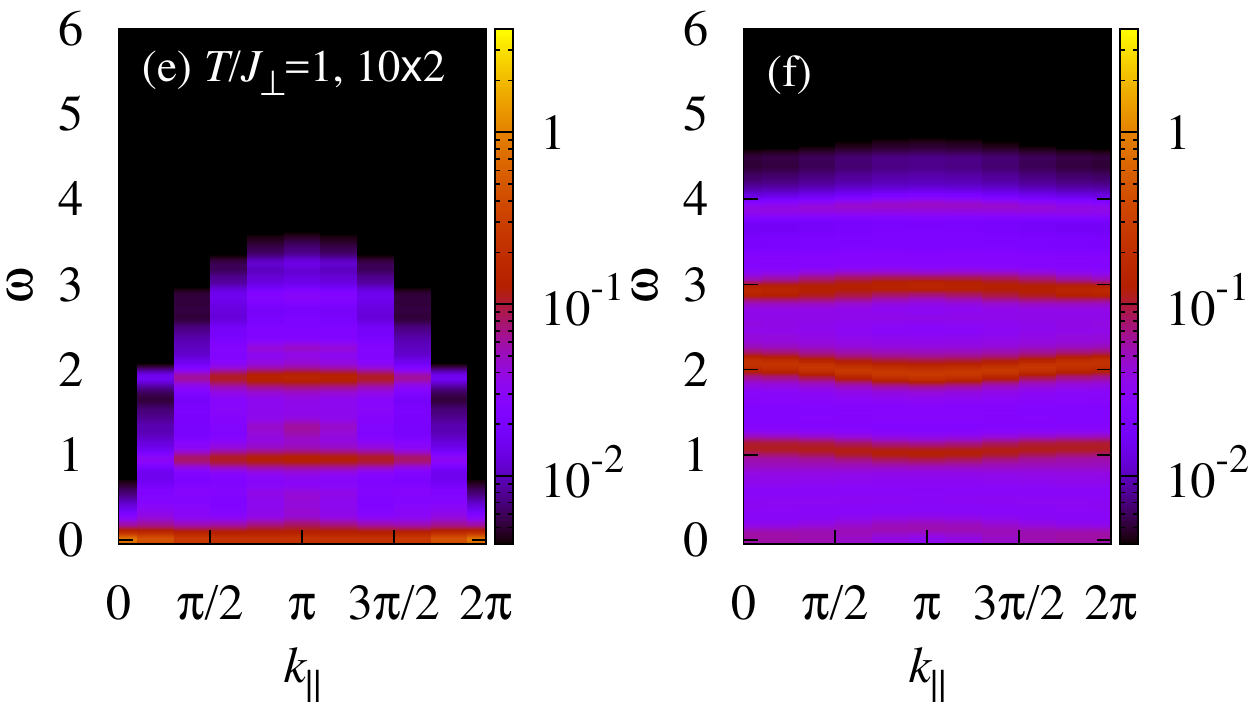}
\caption{Dynamical structure factor in (a,c,e) the symmetric and (b,d,f) the 
antisymmetric channel for the ladder of Fig.~\ref{fig:ladder} with coupling 
parameters $\JR = 2$, $\JL = 1$, and $\JD = 0.9$, for temperatures of (a,b) 
$T/\JR = 0$ ($16 \times 2$ spins), (c,d) $0.2$ ($12 \times 2$ spins), and 
(e,f) $1$ ($10 \times 2$ spins). A Lorentzian broadening $\eta = \JL/20$ is 
applied to the data.}
\label{dsf21p9}
\end{figure}

The characteristic features arising from the discrete energy spectrum of the 
fully frustrated case can be recovered if the frustration parameter, $\JD$, 
is increased continuously from 0 to 1. The conventional spectral features of 
Fig.~\ref{dsf210} are changed smoothly into the very different form of 
Fig.~\ref{dsf211} by the emergence of preferred, discrete levels from the 
continua, whose widths decrease systematically to very narrow values, and 
whose dispersion decreases until the bands become completely flat. This 
evolution is represented in Fig.~\ref{dsf21p9} for the case $\JD = 0.9$. The 
frustration in this case is strong, and the bands only weakly dispersive, such 
that the fingerprints of multi-triplon bound states emerge with increasing 
temperature. However, the continuum nature of the bands remains evident in 
the two-magnon response at $T = 0$ in the symmetric channel, as well as in the 
continuous nature of the thermal broadening in the antisymmetric channel at 
$T/\JR = 1$, which on close inspection lacks the weakly varying intensity 
bands visible between the primary modes in Figs.~\ref{dsf211} and \ref{dsf2o3}. 

\subsection{Infinite-Temperature Spectral Functions}

\label{sec:FTspecD}

\begin{figure}[t!]
\centering\includegraphics[width=0.94\columnwidth]{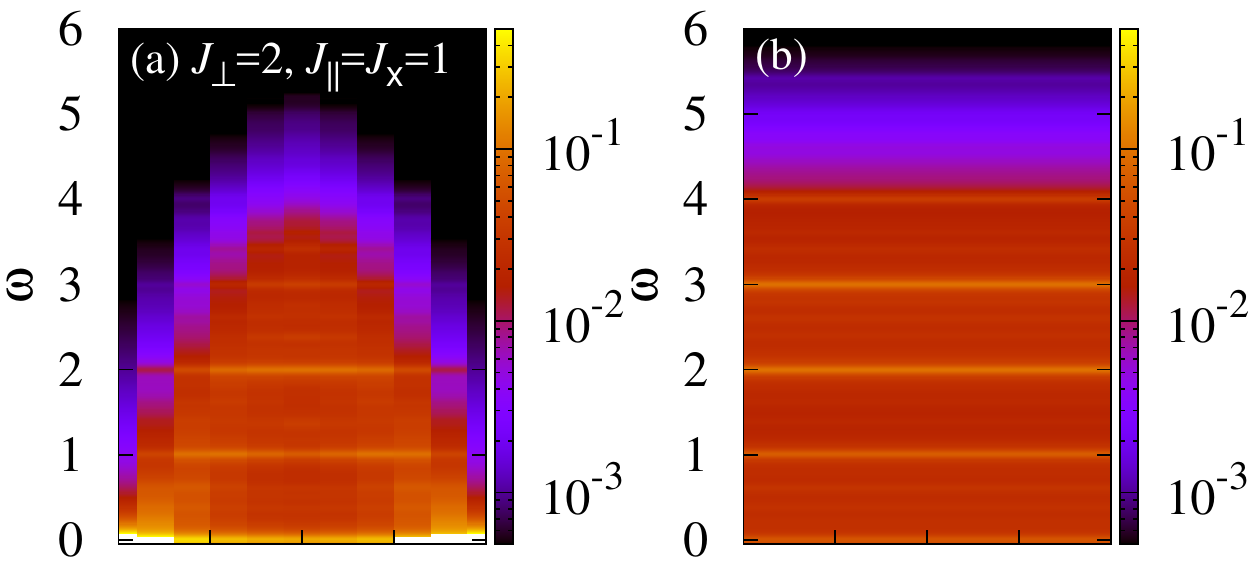}
\centering\includegraphics[width=0.94\columnwidth]{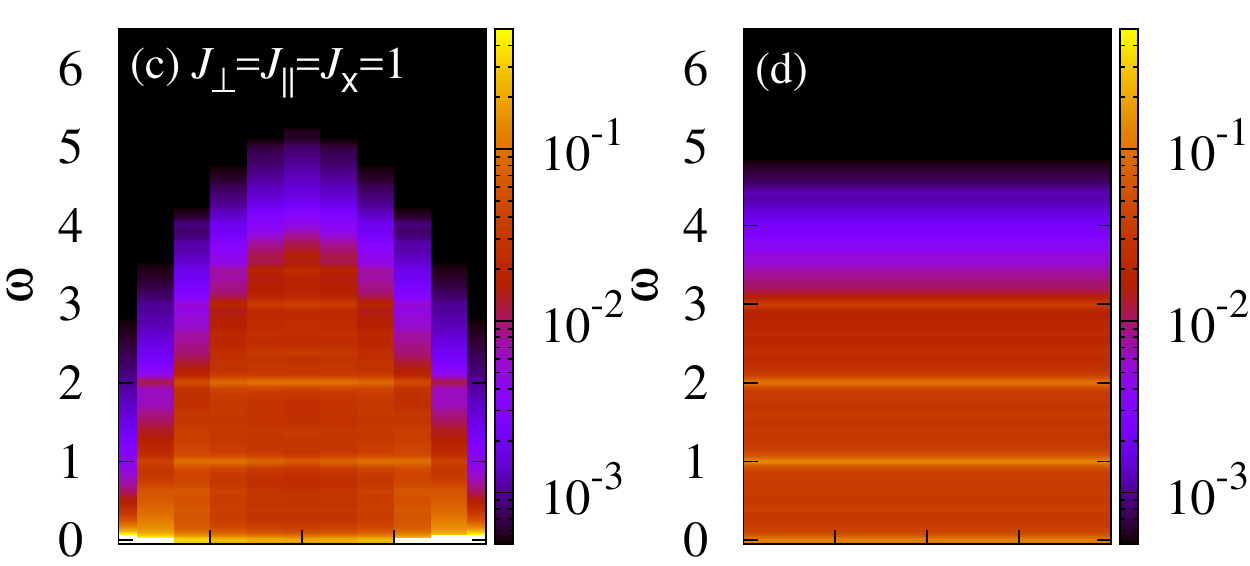}
\centering\includegraphics[width=0.94\columnwidth]{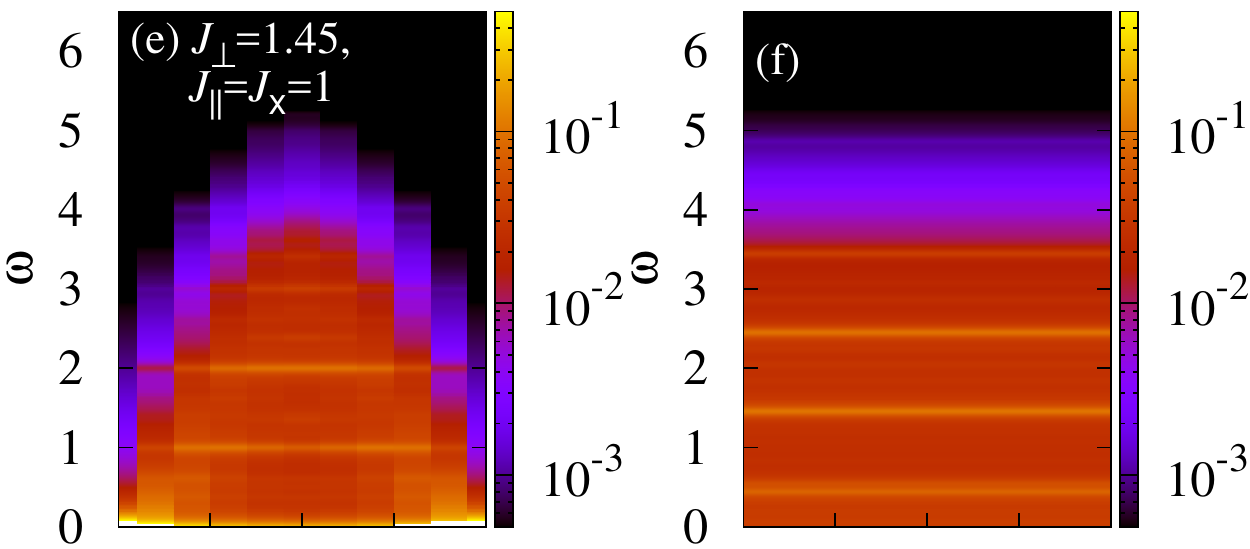}
\centering\includegraphics[width=0.94\columnwidth]{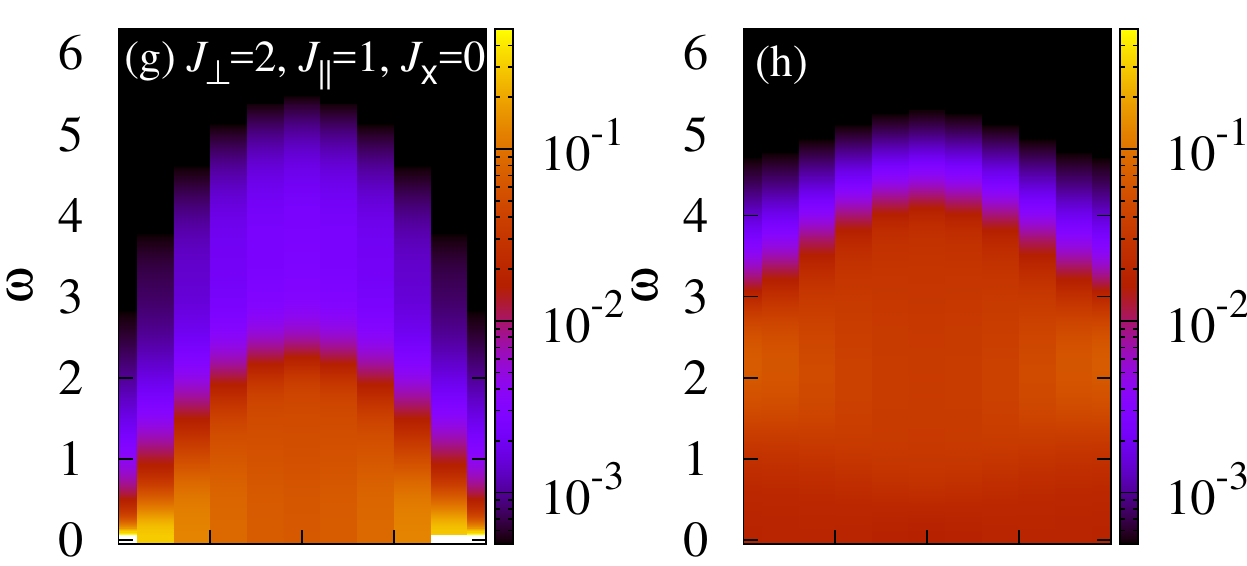}
\centering\includegraphics[width=0.94\columnwidth]{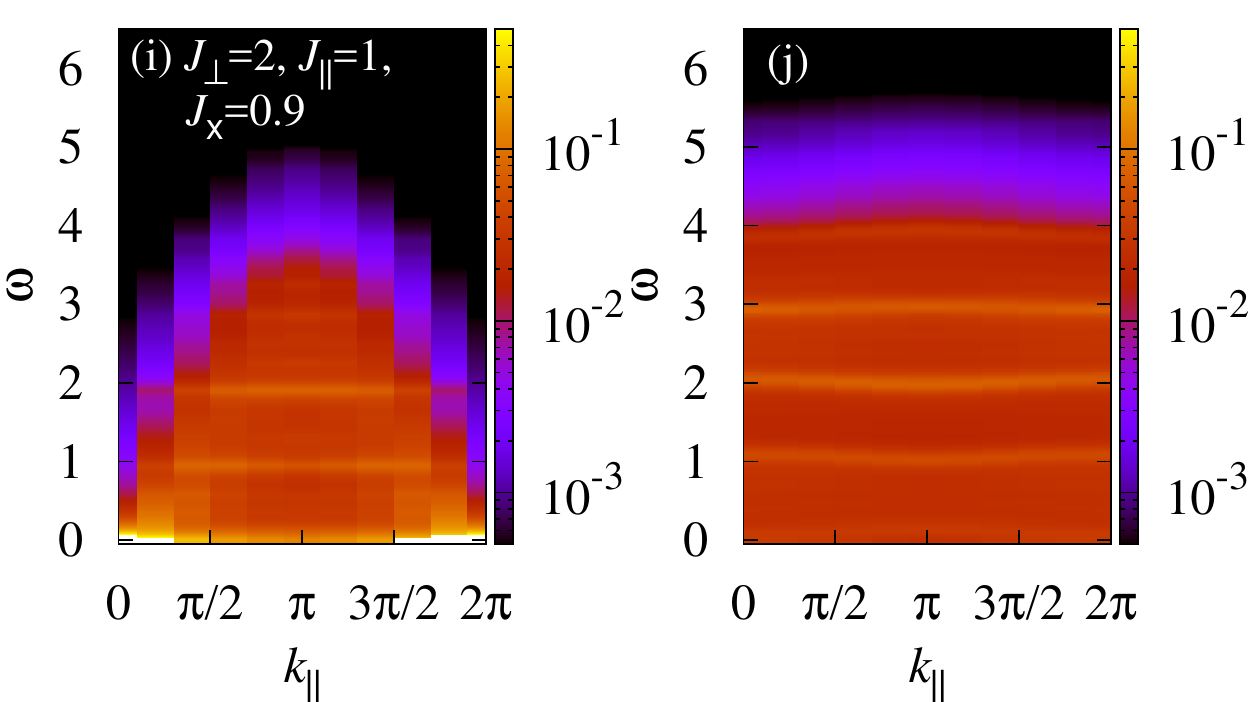}
\caption{Dynamical structure factor in (a,c,e,g,i) the symmetric and 
(b,d,f,h,j) the antisymmetric channel at infinite temperature for ladders with 
coupling parameters $\JR = 2$ and $\JL = \JD = 1$ (a,b); $\JR = \JL = \JD = 1$ 
(c,d);$\JR = 1.45$ and  $\JL = \JD = 1$ (e,f); $\JR = 2$, $\JL = 1$, and $\JD
 = 0$ (g,h); $\JR = 2$, $\JL = 1$, and $\JD = 0.9$ (i,j). The system size is 
$10 \times 2$ spins throughout this figure and a Lorentzian broadening $\eta
 = \JL/20$ is applied to the data.}
\label{dsfinf}
\end{figure}

One of the most striking features of the spectra we calculate is their 
behavior at very high temperatures. Figure \ref{dsfinf} shows the dynamical 
spectral functions for all the ladders of Figs.~\ref{dsf211} to \ref{dsf21p9} 
as $T \to \infty$. The conventional expectation is the situation illustrated 
in Figs.~\ref{dsfinf}(g,h) for the unfrustrated ladder, where the spectral 
weight is spread uniformly to all available energies. By contrast, the primary 
spectral features of the fully frustrated ladders [Figs.~\ref{dsfinf}(a,b), 
\ref{dsfinf}(c,d), and \ref{dsfinf}(e,f)] not only persist at infinite 
temperature but remain completely sharp. It is clear that the discrete 
support, by which is meant the $T = 0$ spectrum of non-dispersive bound 
states, leads to preferred scattering processes at these specific energies at 
all temperatures. This perfect discreteness is the consequence of the perfect 
frustration, and in the nearly-frustrated ladder [Figs.~\ref{dsfinf}(i,j)] one 
observes that the sharpness begins to be lost, although the levels do persist; 
we examine both effects in more detail in Sec.~\ref{sec:Analysis}. 
    
\section{Analysis of Spectral Weights}

\label{sec:Analysis}

\begin{figure}[t!]
\centering\includegraphics[width=0.98\columnwidth]{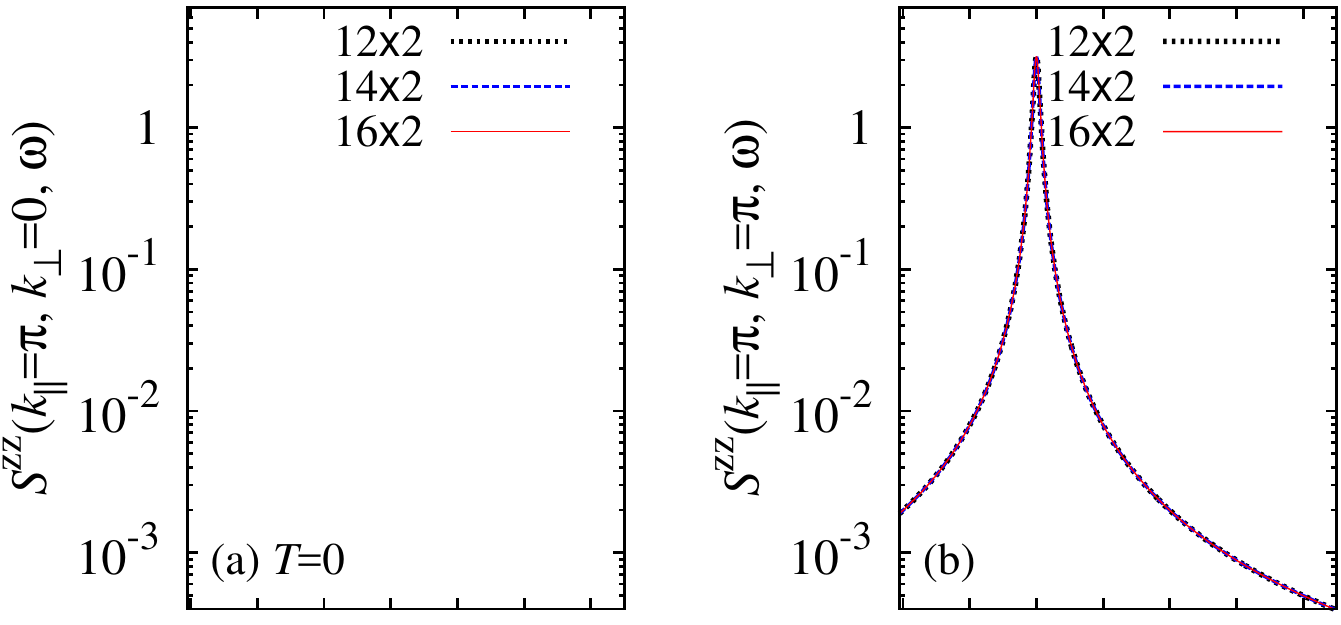}
\centering\includegraphics[width=0.98\columnwidth]{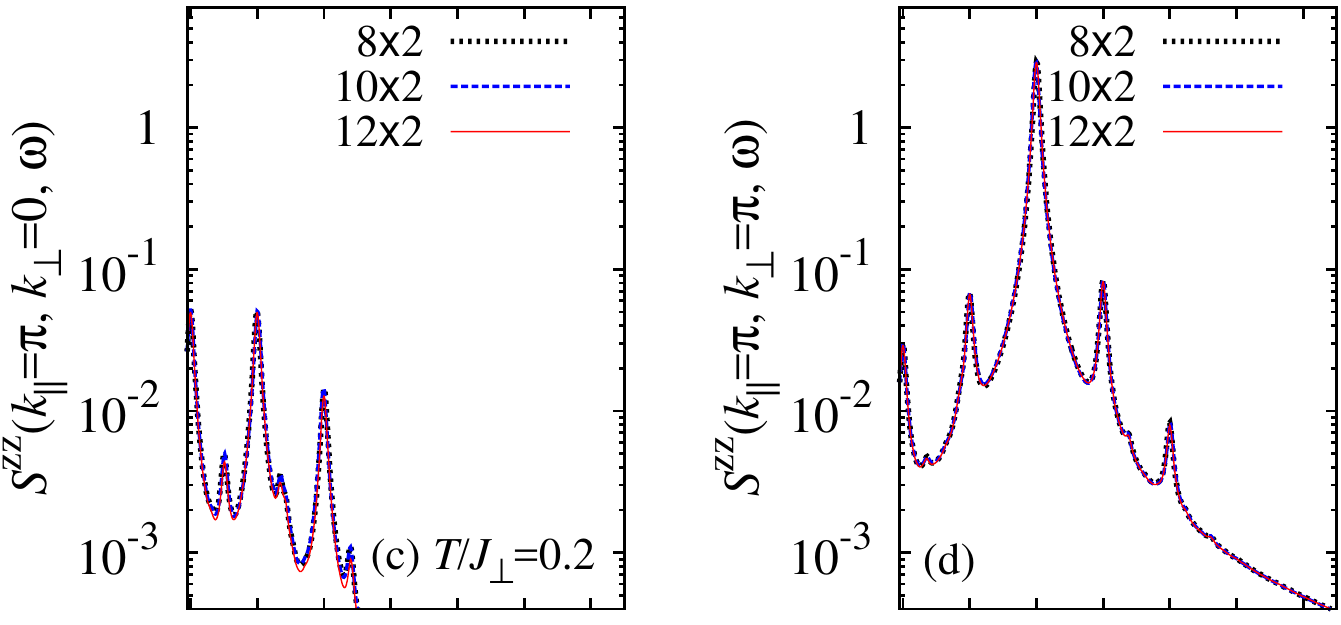}
\centering\includegraphics[width=0.98\columnwidth]{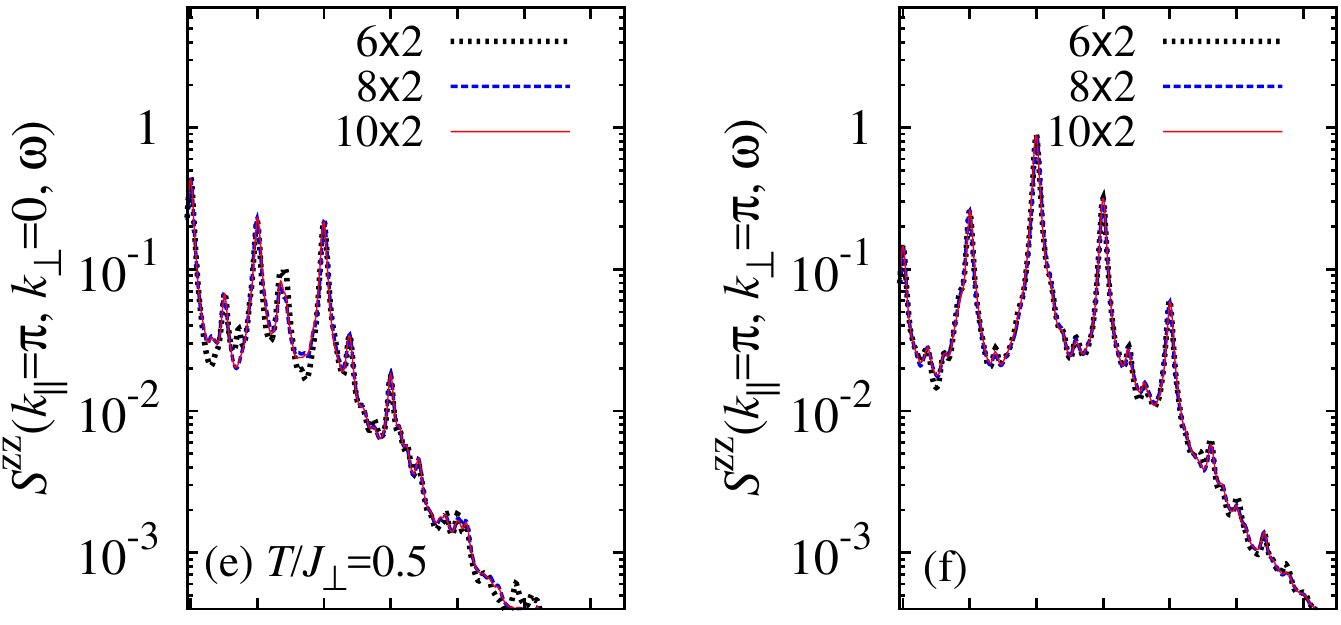}
\centering\includegraphics[width=0.98\columnwidth]{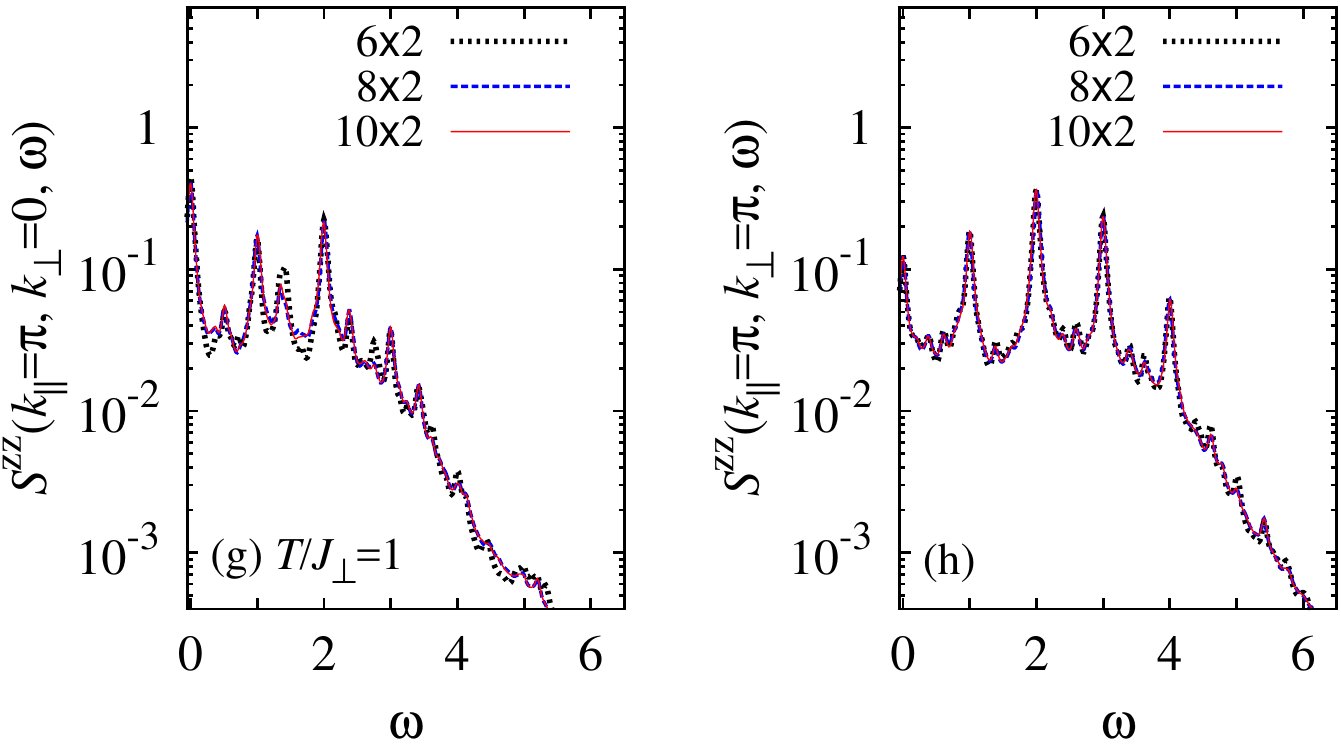}
\caption{Intensity at $k_{||} = \pi$ in (a,c,e,g) the symmetric and (b,d,f,h) 
the antisymmetric channel, shown as a function of energy for the ladder of 
Fig.~\ref{fig:ladder} with the system size specified, coupling parameters 
$\JR = 2$ and $\JL = \JD = 1$, and $\eta = \JL/20$, for temperatures of (a,b) 
$T/\JR = 0$, (c,d) $0.2$, (e,f) $0.5$, and (g,h) $1$.}
\label{dsfkp211}
\end{figure}

\begin{figure}[t!]
\centering\includegraphics[width=0.98\columnwidth]{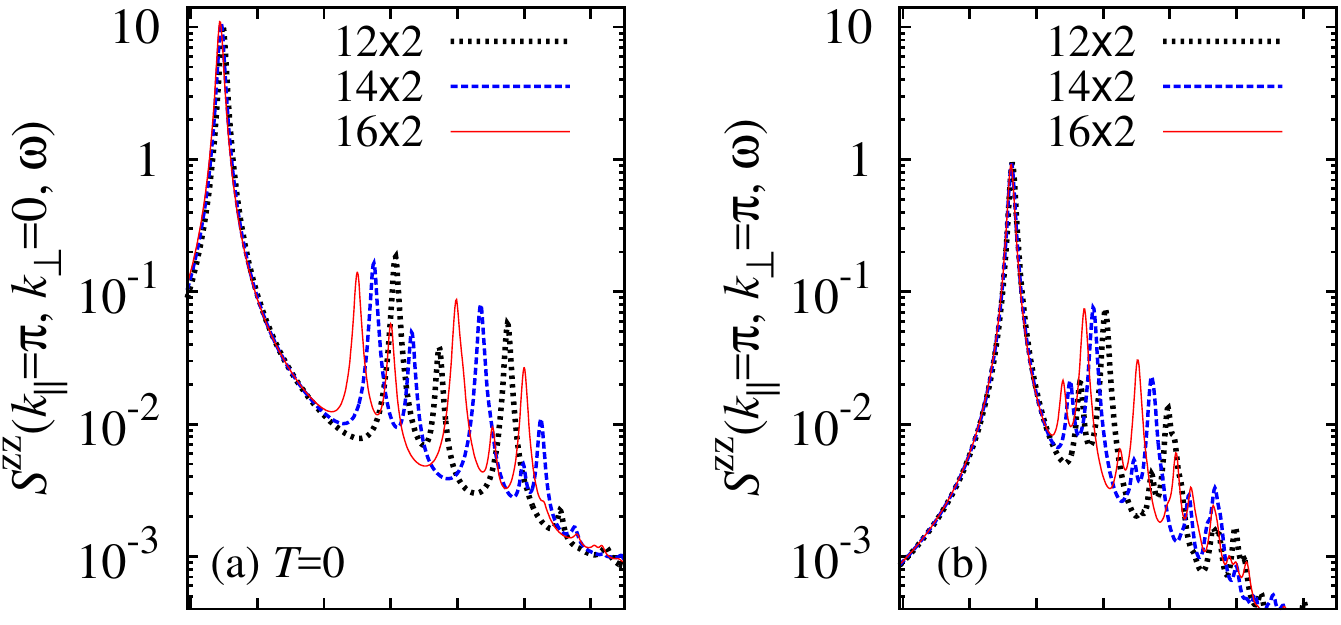}
\centering\includegraphics[width=0.98\columnwidth]{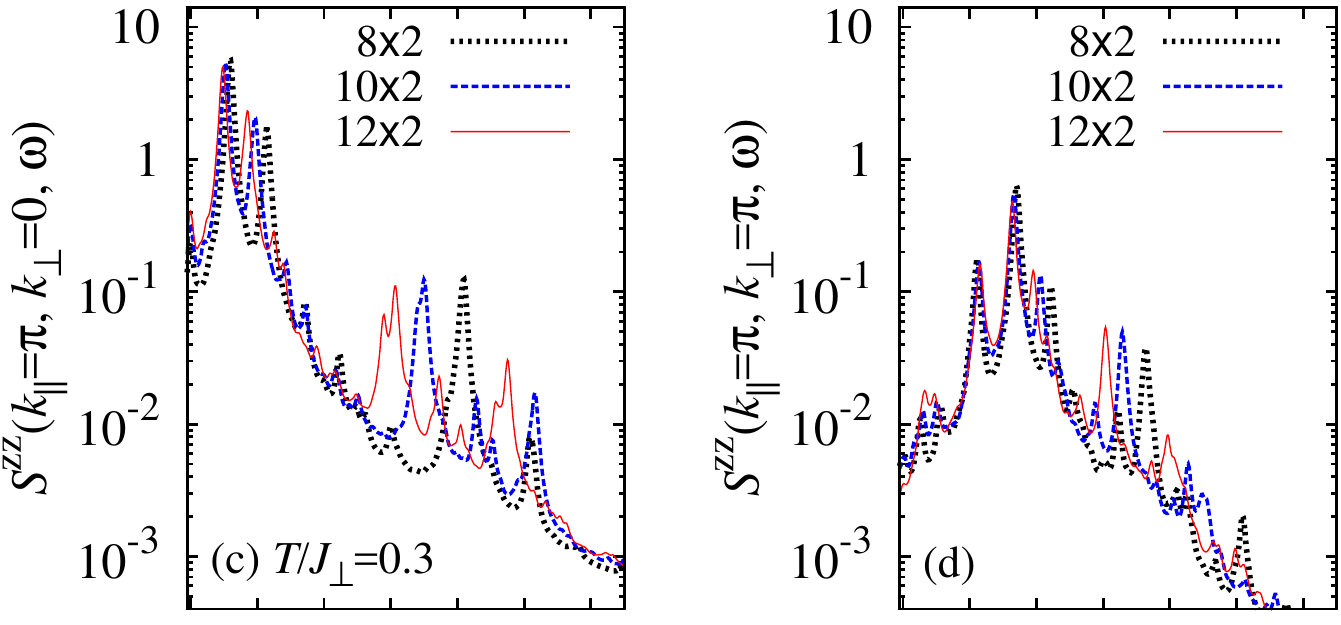}
\centering\includegraphics[width=0.98\columnwidth]{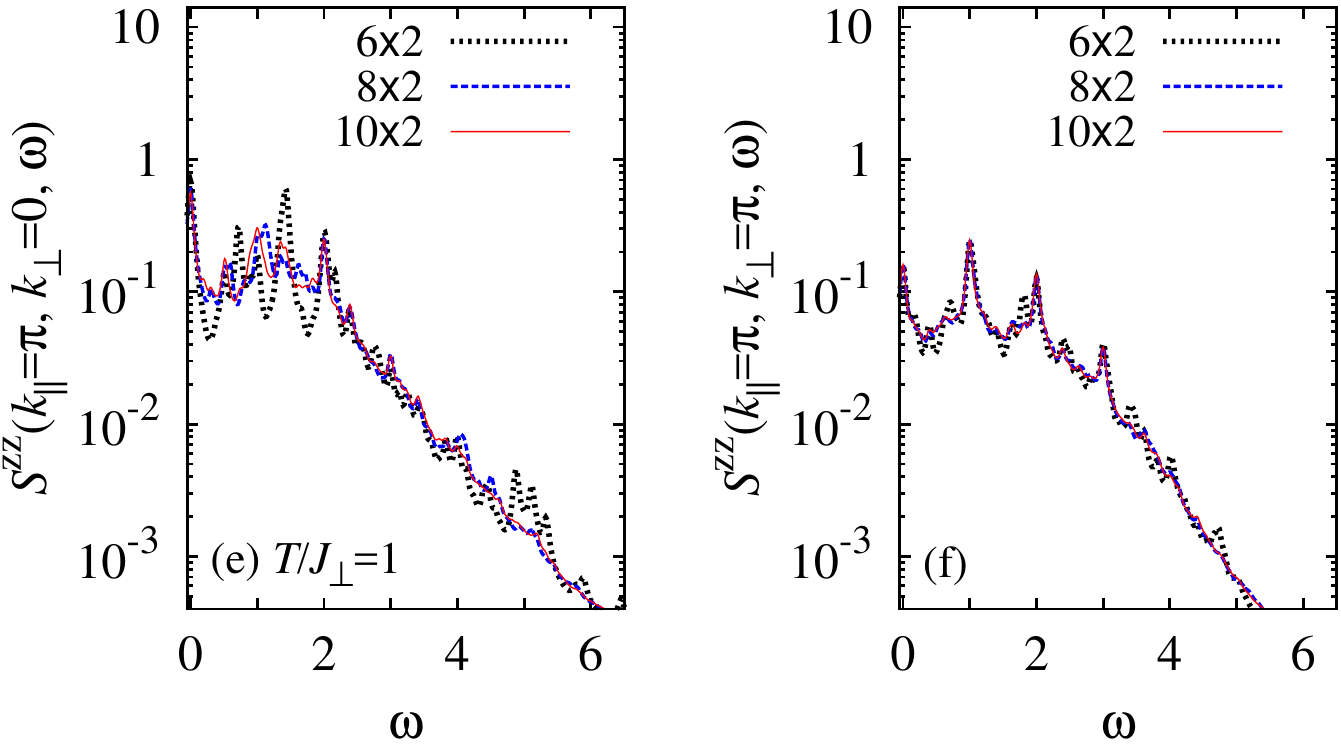}
\caption{Intensity at $k_{||} = \pi$ in (a,c,e) the symmetric and (b,d,f) 
the antisymmetric channel, shown as a function of energy for the ladder of 
Fig.~\ref{fig:ladder} with the system size specified, coupling parameters 
$\JR = \JL = \JD = 1$, and $\eta = \JL/20$, for temperatures of (a,b) 
$T/\JR = 0$, (c,d) $0.3$, and (e,f) $1$.}
\label{dsfkp111}
\end{figure}

\begin{figure}[t!]
\centering\includegraphics[width=0.98\columnwidth]{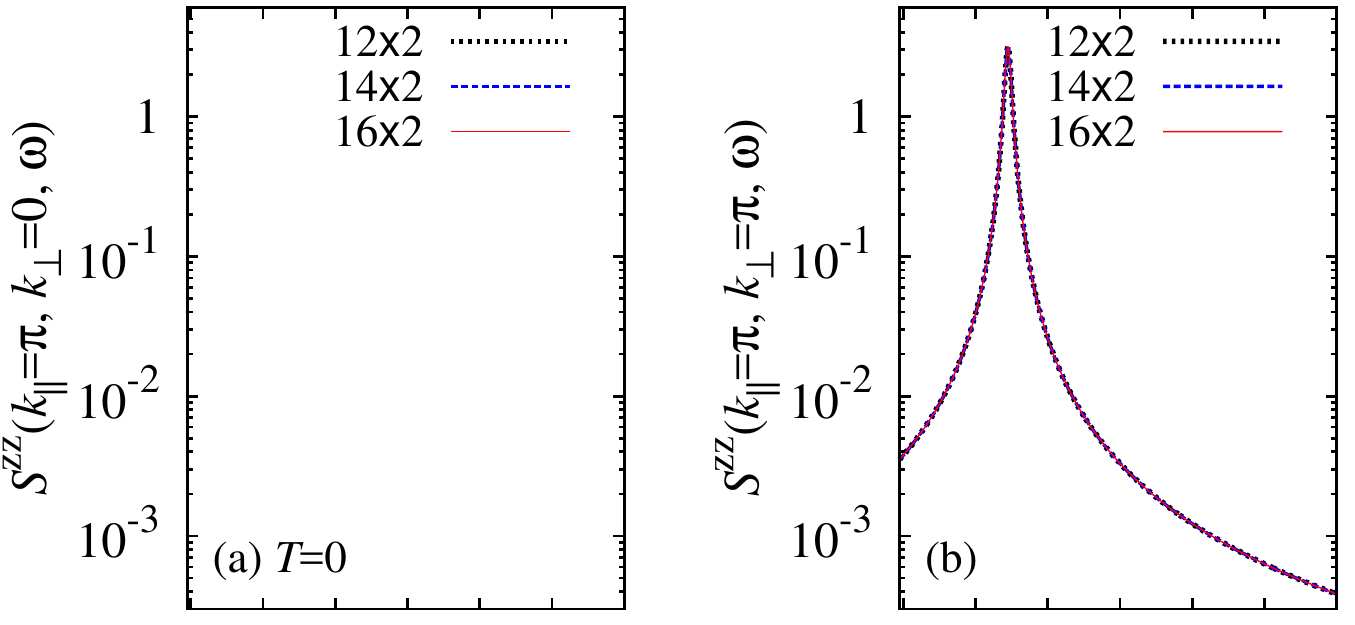}
\centering\includegraphics[width=0.98\columnwidth]{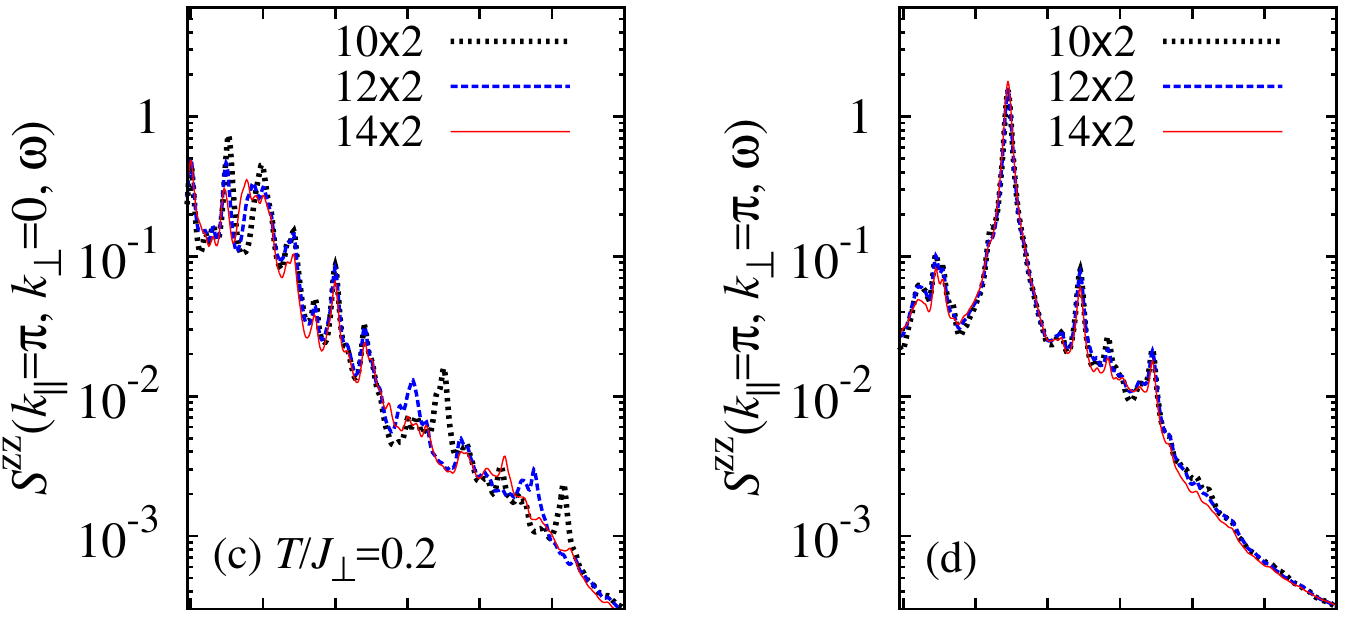}
\centering\includegraphics[width=0.98\columnwidth]{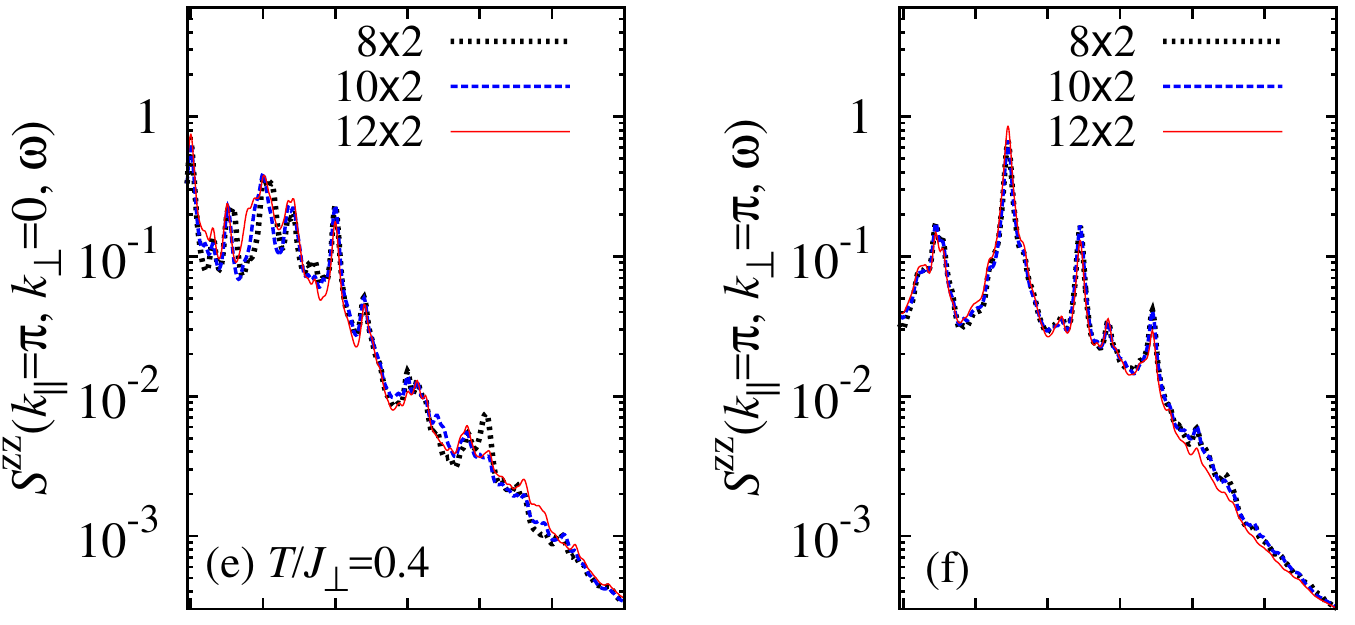}
\centering\includegraphics[width=0.98\columnwidth]{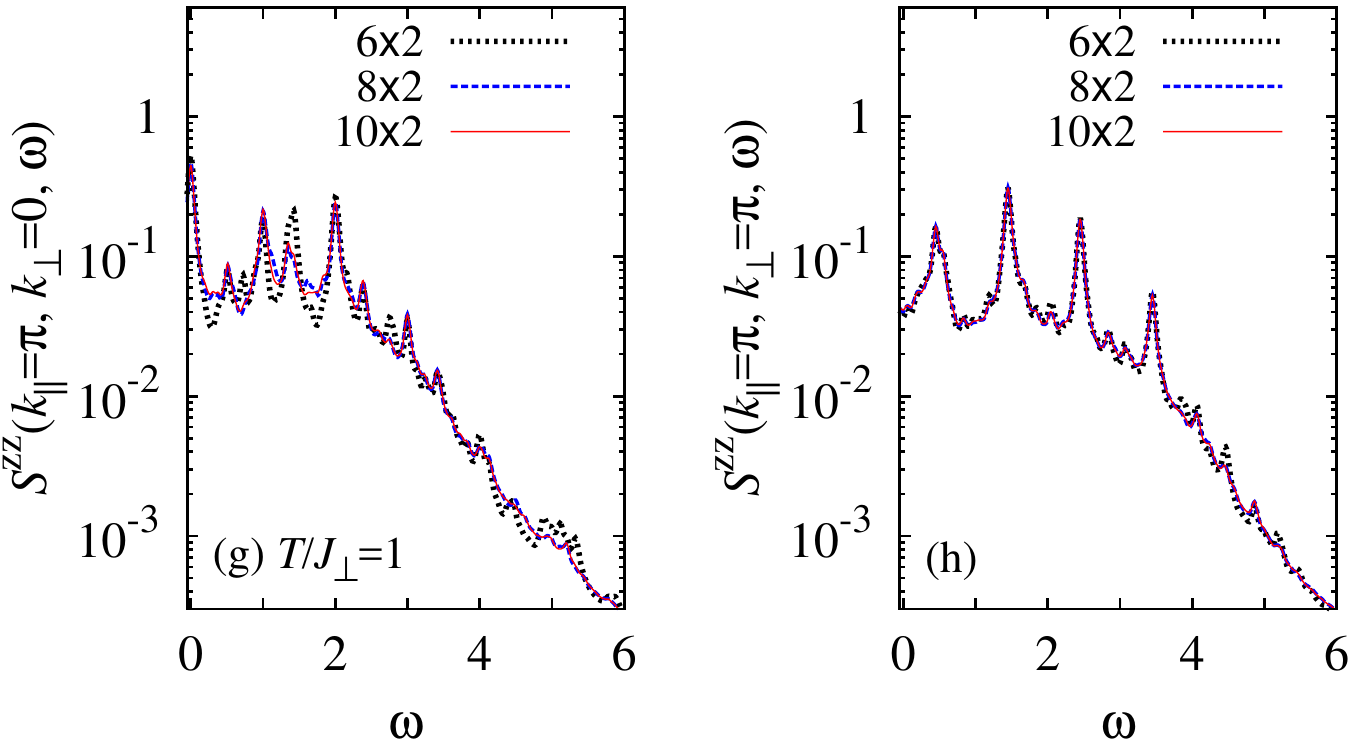}
\caption{Intensity at $k_{||} = \pi$ in (a,c,e,g) the symmetric and (b,d,f,h) 
the antisymmetric channel, shown as a function of energy for the ladder of 
Fig.~\ref{fig:ladder} with the system size specified, coupling parameters 
$\JR = 1.45$ and $\JL = \JD = 1$, and $\eta = \JL/20$, for temperatures of 
(a,b) $T/\JR = 0$, (c,d) $0.2$, (e,f) $0.4$, and (g,h) $1$.}
\label{dsfkp2o3}
\end{figure}

\begin{figure}[t!]
\centering\includegraphics[width=0.98\columnwidth]{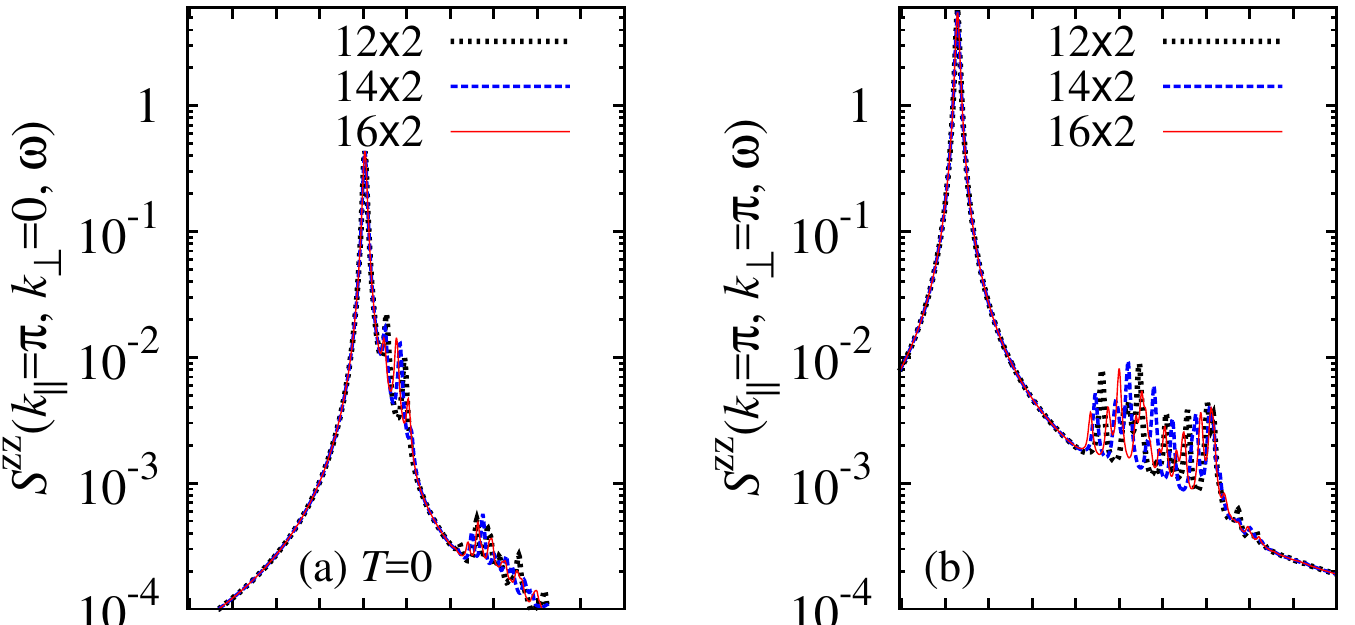}
\centering\includegraphics[width=0.98\columnwidth]{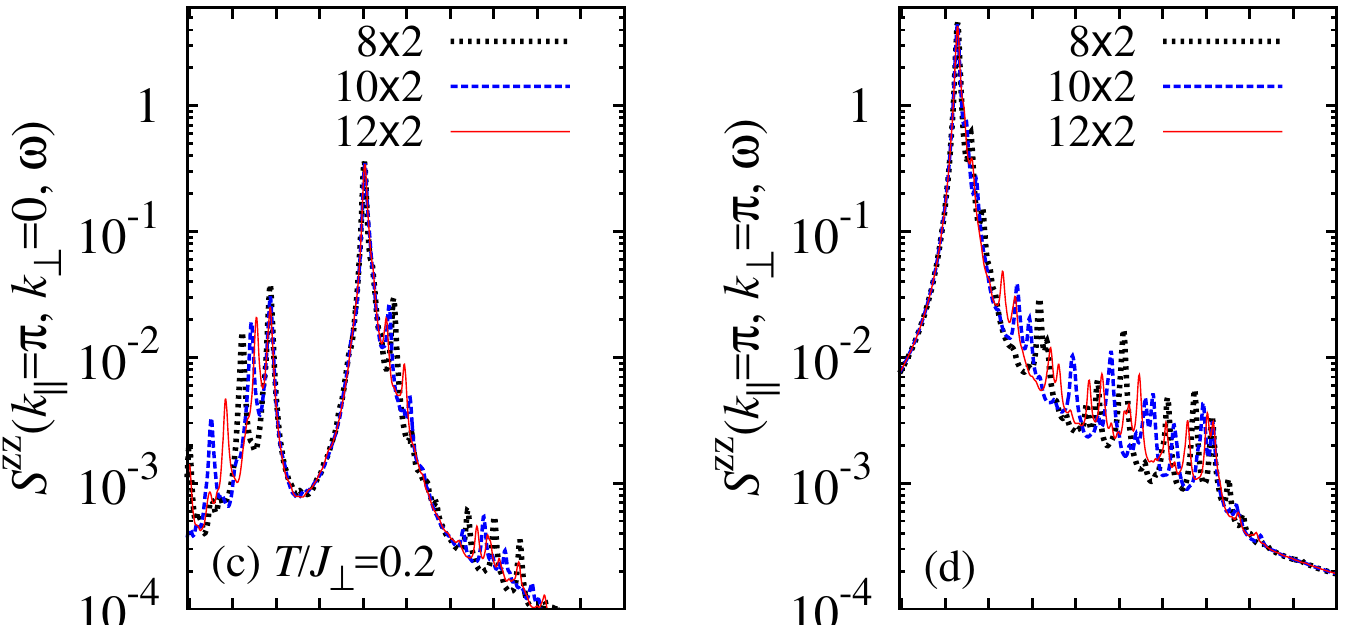}
\centering\includegraphics[width=0.98\columnwidth]{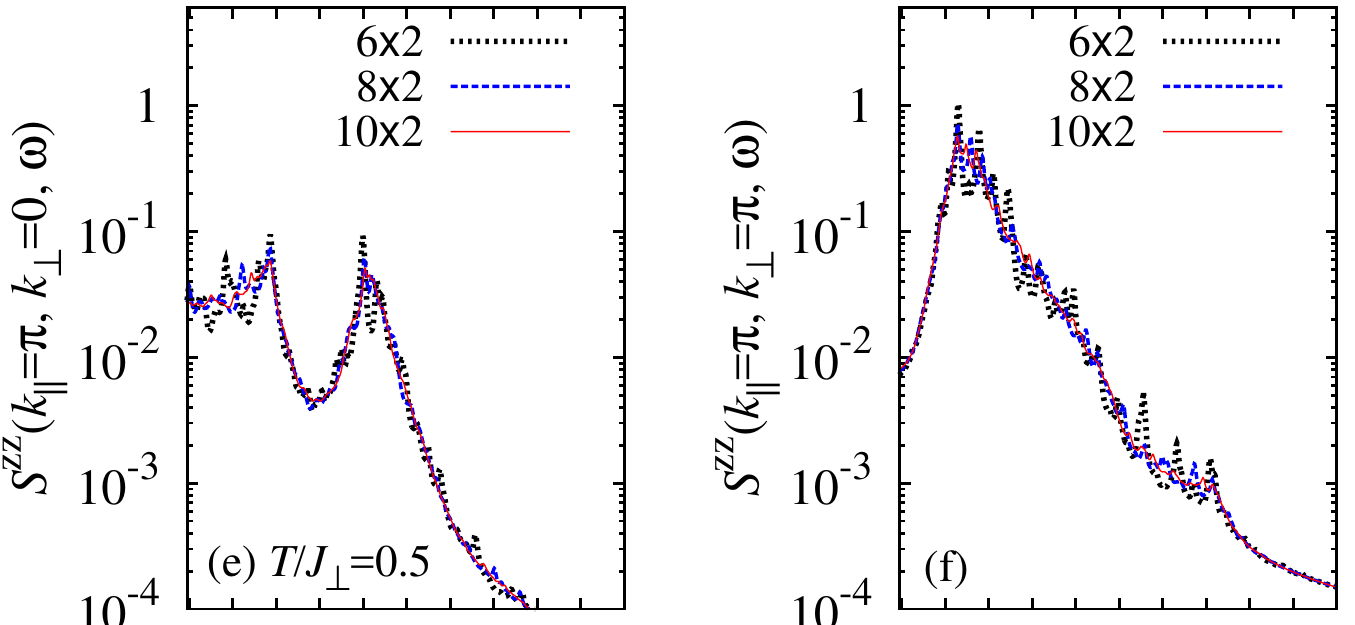}
\centering\includegraphics[width=0.98\columnwidth]{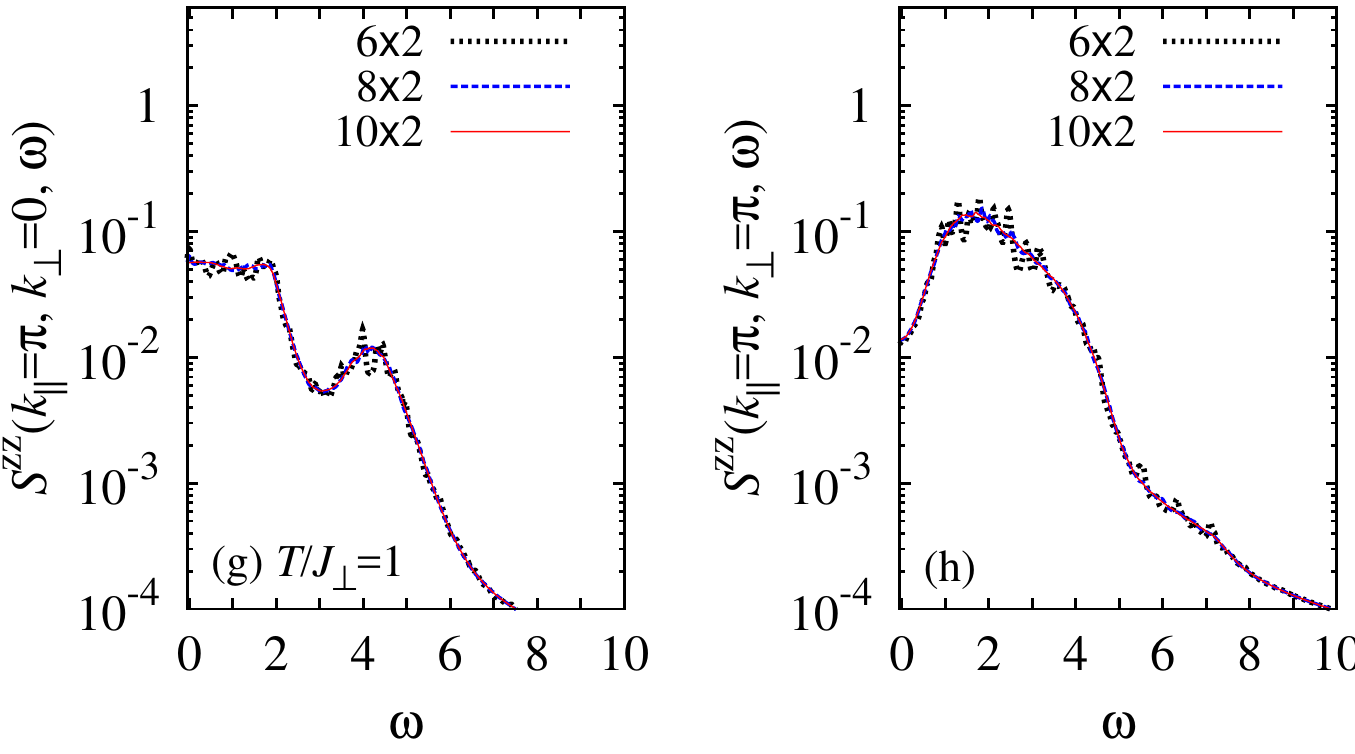}
\caption{Intensity at $k_{||} = \pi$ in (a,c,e,g) the symmetric and (b,d,f,h)
the antisymmetric channel, shown as a function of energy for the ladder of 
Fig.~\ref{fig:ladder} with the system size specified, coupling parameters 
$\JR = 2$, $\JL = 1$, and $\JD = 0$, and $\eta = \JL/20$, for temperatures 
of (a,b) $T/\JR = 0$, (c,d) $0.2$, (e,f) $0.5$, and (g,h) $1$.}
\label{dsfkp210}
\end{figure}

\begin{figure}[t!]
\centering\includegraphics[width=0.98\columnwidth]{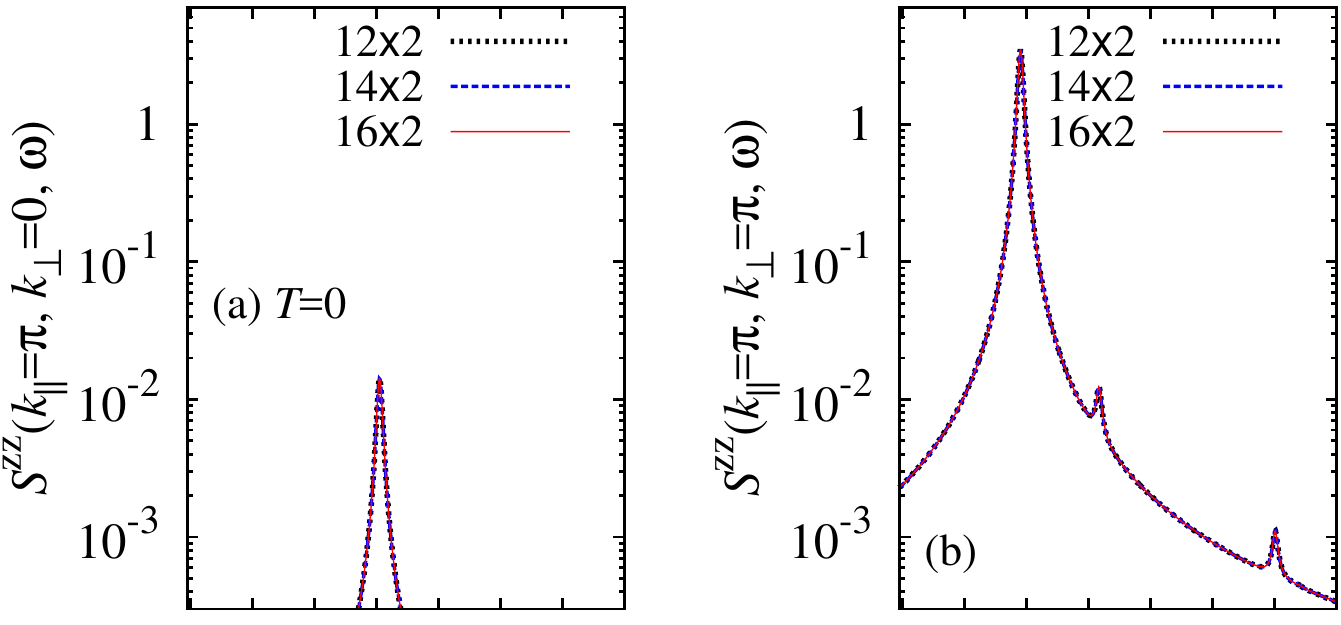}
\centering\includegraphics[width=0.98\columnwidth]{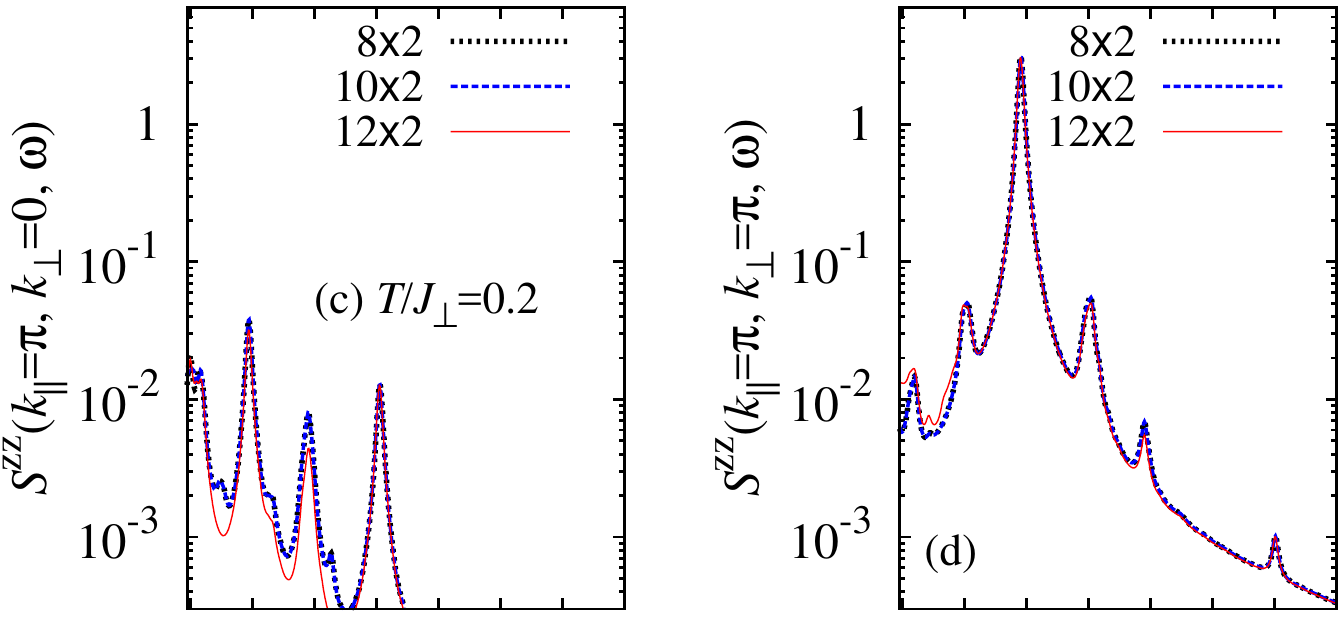}
\centering\includegraphics[width=0.98\columnwidth]{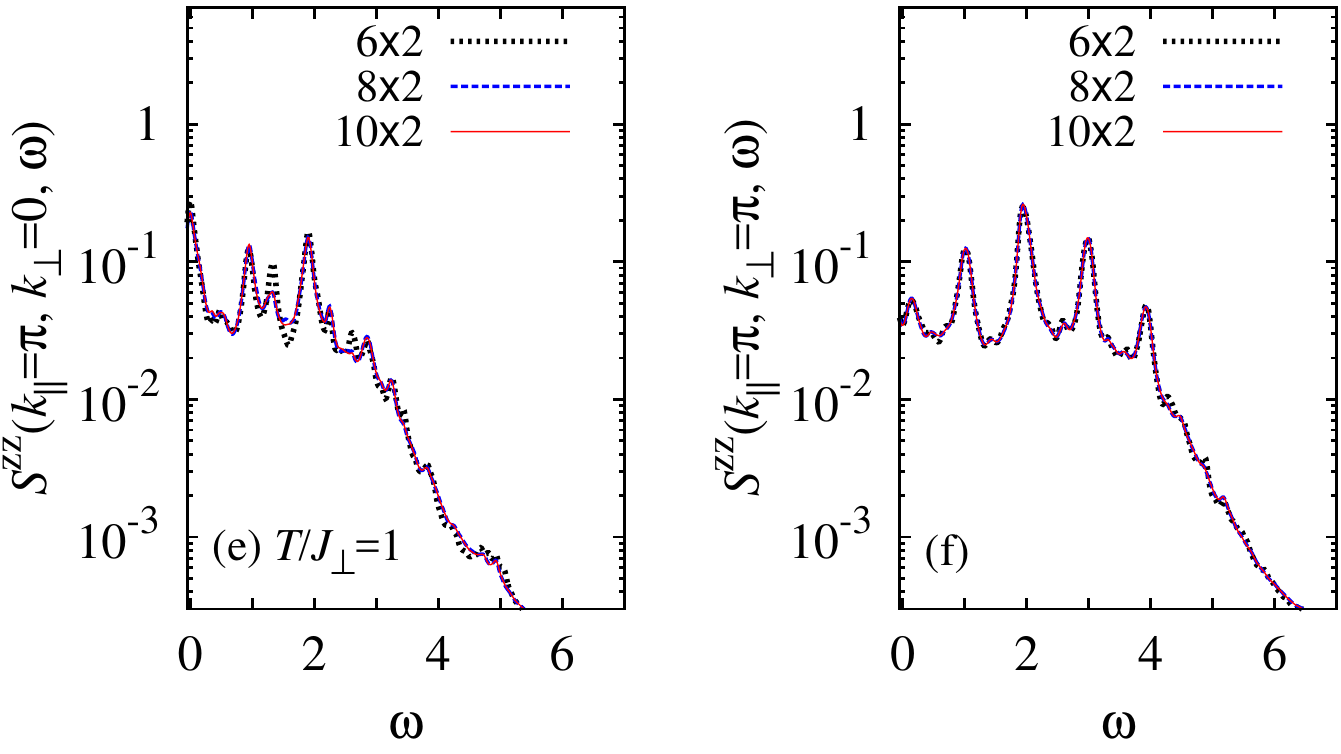}
\caption{Intensity at $k_{||} = \pi$ in (a,c,e) the symmetric and (b,d,f) the 
antisymmetric channel, shown as a function of energy for the ladder of 
Fig.~\ref{fig:ladder} with the system size specified, coupling parameters 
$\JR = 2$, $\JL = 1$, and $\JD = 0.9$, and $\eta = \JL/20$, for temperatures 
of (a,b) $T/\JR = 0$, (c,d) $0.2$, and (e,f) $1$.}
\label{dsfkp21p9}
\end{figure}

We begin a quantitative discussion of our results for the dynamical structure 
factor by displaying our intensity data (Figs.~\ref{dsf211} to \ref{dsfinf}) 
as functions of energy for one specific wave vector, taken to be $k_\| = \pi$; 
although the intensities are identical for all wave vectors in the 
antisymmetric channel, this is not the case in the symmetric channel. Again 
these data are most transparent on a logarithmic scale and Figs.~\ref{dsfkp211} 
to \ref{dsfkp21p9} demonstrate clearly the increasing importance of increasing 
numbers of bound-state energy levels as the temperature is raised. In every 
panel we show data obtained for three successive system sizes in order to 
gauge their finite-size effects. The lines display data obtained with a 
broadening factor of $\eta = 0.05 \JL$, and we comment that, as a result of 
the logarithmic intensity scale, the entire line shape visible in the figures 
of this section is a consequence of $\eta$.

The primary features of the dynamical spectral functions for which we 
seek a quantitative explanation are (i) the nature of the processes generating 
the discrete spectra of the fully frustrated ladder, which we address in 
Sec.~\ref{sec:AnalysisA}, and (ii) the origin of the rapid transfer of spectral 
weight, which we discuss in Secs.~\ref{sec:AnalysisB} and \ref{sec:AnalysisC}. 
We first summarize the qualitative features of the results we show for the two 
different types of ground state, namely the rung-singlet phase of 
Figs.~\ref{dsfkp211} and \ref{dsfkp2o3} and the rung-triplet (Haldane) phase 
of Fig.~\ref{dsfkp111}. Although it stands to reason that there are different 
types of excited state for $j' > j'_c$ and $j' < j'_c$, the lower-energy states 
of one phase are higher-energy excitations of the other, and thus both sides of 
the transition contain messages for interpreting the opposite one. 

In the rung-singlet phase shown in Fig.~\ref{dsfkp211}, we comment first that 
very few finite-size effects (discrepancies between the curves obtained for 
the three system sizes shown) are visible, indicating that even these short 
systems capture all of the physics away from $j'_c$. The one-triplon peak in 
the antisymmetric channel loses weight systematically as a function of 
temperature to a small number of dominant and, for this parameter choice, 
symmetrically distributed states lying above a rising background. The response 
in the symmetric channel is strongest at zero energy, and shows weaker 
finite-temperature contributions from a greater number of levels. Figure 
\ref{dsfkp2o3} shows the same physics developing at lower temperatures, as 
well as additional peaks appearing due to the deliberate asymmetry of the 
parameters (Sec.~\ref{sec:AnalysisA}). 

The spectrum of the rung-triplet phase ($j' < j'_c$, Fig.~\ref{dsfkp111}) 
shows the predominance of weight in the symmetric channel, and displays 
significant finite-size effects. These are expected in general if the origin 
of the features lies in the ``Haldane'' part of the spectrum, which contains 
long-wavelength excitations, but not from the localized (bound-state) 
features. However, we comment that the strongest finite-size effects in 
Fig.~\ref{dsfkp111}(a) actually appear in the three-particle continuum 
\cite{rwa}, whereas those in the one-triplon line are small. The loss of 
strong finite-size features at higher temperatures implies that the primary 
contributions in this regime are in fact from few-rung processes; here one 
may borrow from the $j' > j'_c$ phase to interpret these as isolated rung 
singlets, and (non-exact) bound states of rung singlets, in a background of 
rung triplets. Similarly, as noted in Sec.~\ref{sec:FFL}, the interpretation 
of the spectral function at $j' > j'_c$ can also be borrowed from the 
$j < j'_c$ side in the form of open $n$-site Haldane chains separated by 
singlet rungs. The presence of such segmented Haldane chains is clearest 
in the symmetric channel, in the shape of the envelope of spectral weight 
in Figs.~\ref{dsf211} to \ref{dsf2o3}.

The physics of the spectral-weight distribution, and of its thermal 
redistribution, is shown most clearly in Fig.~\ref{dsf2o3} for a system 
close to $j'_c$. The intensities of the discrete energy levels over the 
entire spectrum rise rapidly with temperature, and the concomitant fall 
in the intensity of the one-triplon band is quantified in Fig.~\ref{dsfkp2o3}. 
The redistribution effect among discrete levels shown in this figure provides 
a stark contrast with the unfrustrated ladder, whose intensity at $k_\| = \pi$ 
is shown in Fig.~\ref{dsfkp210}. For the minimum of the dispersive ladder 
bands, one observes a dominant low-lying peak in the antisymmetric channel 
spreading non-uniformly to all energies with increasing temperature, whereas 
the symmetric channel preserves distinct but dispersive low- and high-energy 
contributions to significant temperatures; the dominant low-temperature feature 
is the two-triplon bound state mentioned in Sec.~\ref{sec:FTspecC} 
\cite{rtmhzs,rKSGU01,rgea,rSU05}.

Figure \ref{dsfkp21p9} provides the imperfectly frustrated hybrid of the two 
different types of response. The perfect frustration therefore produces a 
very unusual but characteristic spectrum of discrete, flat bands, which are 
determined by the bound states discussed in Sec.~\ref{sec:FFL}; the fact that 
they appear in the spectral function with a very specific hierarchy of spectral 
weights indicates that their ratios are determined not only by the temperature 
but also by specific matrix elements.

\subsection{Excitation Processes}

\label{sec:AnalysisA}

To identify the excitation processes present in the calculated spectra, 
we use the energies of the known bound states and begin with the smallest 
clusters. The results for the spectra of the $n = 2$ and 3 multiplets are 
given in App.~\ref{appa}. Our choice of parameters in all the figures of 
Secs.~\ref{sec:FTspec} and \ref{sec:Analysis} for fully frustrated ladders 
is such that $\JL = \JD = 1$, 
with only $\JR$ changing the ratio $j'$. Because $\JL$ determines the multiplet 
splitting of the bound states, and all level separations are integral for 
$n = 2$ and 3, any excitations with irrational energies must automatically 
be a consequence of transitions involving clusters with $n = 4$ or higher. 
However, this situation has the twin disadvantages that primary $n = 2$ and 
3 levels are always degenerate and that ``commensurate'' values of $\JR/\JL$ 
can cause additional energetic degeneracies between different excitation 
processes. 

Considering only the antisymmetric channel, the one-triplon excitation is 
located at an energy of $j'$ (in units of $\JL$) and the leading bound-state 
energies in Figs.~\ref{dsf211} (\ref{dsfkp211}) and \ref{dsf2o3} 
(\ref{dsfkp2o3}) are located at $|j' - 2|$, $j' - 1$, and $j' + 1$. These 
all arise from transitions between the one-triplon state and the two-triplon 
bound state (for which we adopt the notation $1 \rightarrow 2$), respectively 
its singlet, triplet, and quintet components. We note (App.~\ref{appa}) that 
there is no selection rule prohibiting excitation processes from a triplet 
to a triplet in the multiplet basis. We find that the first process is 
degenerate with the one-triplon band for the case $j' = 2$ and thus we use 
``incommensurate'' values of $j'$ to isolate the individual contributions 
from every process where possible. For the choice $j' = 1.45$, we find the 
dominant excitations at $\omega/\JL = 0.55$, $0.45$, and $2.45$ in 
Figs.~\ref{dsf2o3} and \ref{dsfkp2o3}.

The next higher energy levels, and presumably next-strongest intensities, 
will arise from the $2 \rightarrow 3$ transitions, which are expected at 
the energies $|j' - 3|$, $|j' - 2|$, $j' - 1$, and so on. Many of these 
are not detectable because they contribute at exactly the same energies as 
the stronger $1 \rightarrow 2$ signal, and cannot be separated by using 
different values of $j'$. The first non-degenerate option, the $j' + 2$ 
signal, is clearly visible in all of Figures \ref{dsf211} (\ref{dsfkp211}), 
\ref{dsf2o3} (\ref{dsfkp2o3}), and \ref{dsf21p9} (\ref{dsfkp21p9}), the 
last with an inverted dispersion, but is definitely weaker than the $1 
\rightarrow 2$ peaks at all temperatures. As noted above, processes involving 
four-triplet bound states appear at irrational energies and these form the 
leading contributions to the ``background'' of many discrete energies visible 
in all of the spectra shown in Figs.~\ref{dsf211} to \ref{dsf2o3} and 
\ref{dsf21p9} even at moderate $T$.

The logical extension of this discussion is the obvious importance of 
``multiparticle'' excitations, by which is meant the low-lying bound states 
of high-$n$ triplon clusters. Particularly for values of $j'$ just above 
$j'_c$, as in Fig.~\ref{dsf2o3}, these cause the very rapid filling of the 
spectrum as the temperature is raised, with thermal excitations populating 
bound states over a wide range of $n$. These multi-triplon excitations are 
in fact rather spatially extended objects and as such constitute exceptions 
to the notion that all physics tends to be extremely local in highly 
frustrated and gapped systems. This is also the origin of problems with 
finite-size effects entering our calculations as $j' \rightarrow j'_c$,
some of which are visible in Fig.~\ref{dsfkp2o3}.

\subsection{Spectral-Weight Redistribution}

\label{sec:AnalysisB}

\begin{figure}[t!]
\centering\includegraphics[width=0.98\columnwidth]{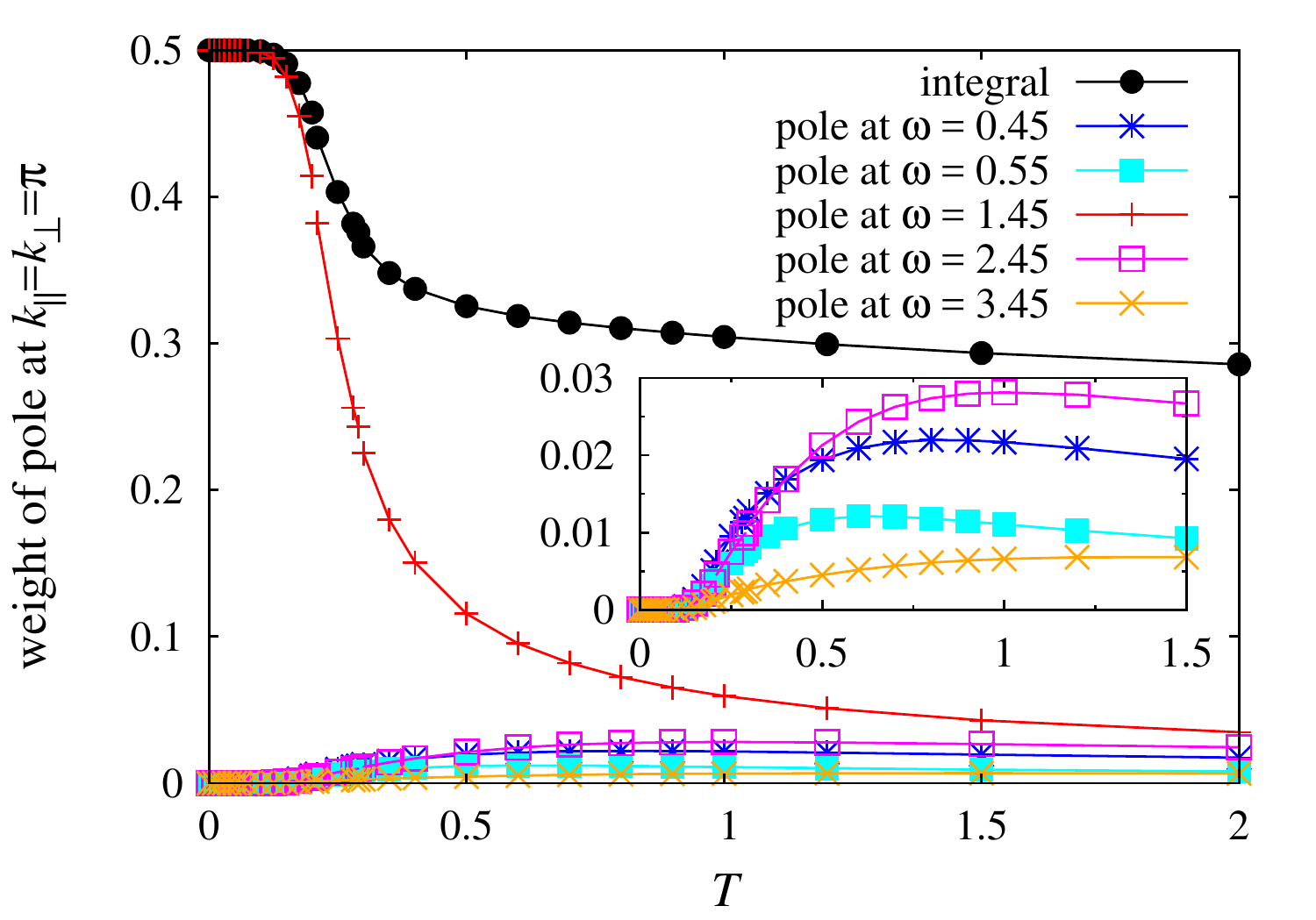}
\caption{Weights of selected poles in the dynamical structure factor in 
the antisymmetric channel for a fully frustrated ladder with $j' = 1.45$, 
shown as a function of temperature for the total spectral weight, the 
one-triplon mode, and the four most intense multi-triplon bound states. 
The thermal evolution of the bound-state intensities (inset) reflects the 
common origin of the three leading lines, which differs from that of the 
fourth (see text). System sizes are $L = 14$ rungs at low $T$, followed by 
$L = 12$ and then $L = 10$ at higher temperatures; data for identical system 
sizes are connected by lines.}  
\label{ftwmtm}
\end{figure}

The other primary aspect of our results that can be explained quantitatively 
is the redistribution of spectral weight among the different discrete 
excitations as a function of the increasing temperature. Concentrating 
again on the fully frustrated case and the antisymmetric channel, the 
weights of the poles for any wave vector in the spectral functions show 
(Figs.~\ref{dsfkp211} and \ref{dsfkp2o3}) a clear growth with temperature 
of the intensity in a number of satellite peaks at the expense of the 
one-triplon peak. This information is shown in Fig.~\ref{ftwmtm} for the 
parameter choice $j' = 1.45$, where the three strongest satellites, at 
$\omega/\JL = 0.55$, $0.45$, and $2.45$, are respectively the transitions 
from the one-triplon band to the singlet, triplet, and quintet of the 
two-triplon bound state, and the fourth, at $\omega/\JL = 3.45$, contains 
the $j' + 2$ transitions within the set of $2 \rightarrow 3$ processes.

\begin{figure}[t!]
\centering\includegraphics[width=0.98\columnwidth]{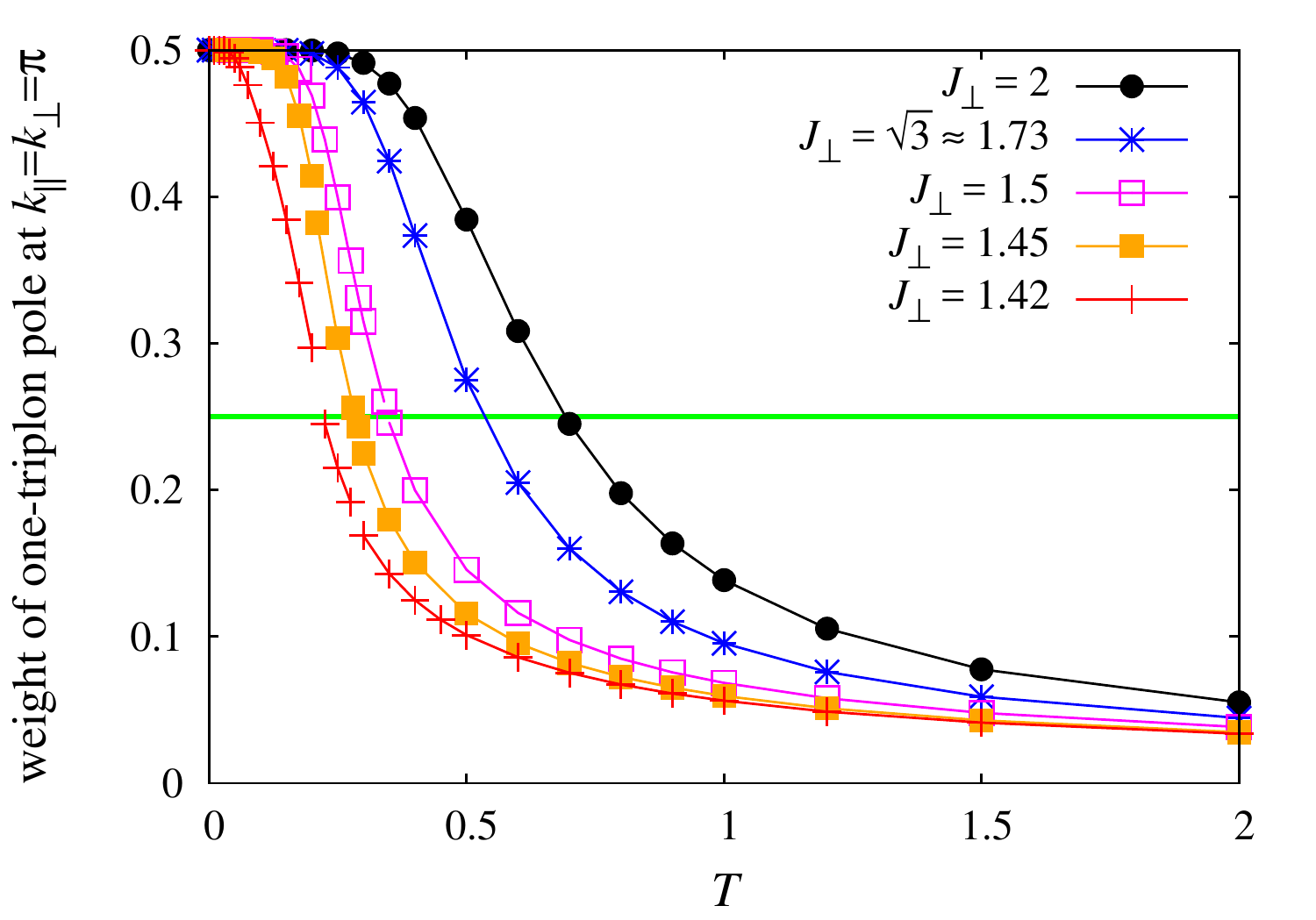}
\caption{Weight of the one-triplon pole in the dynamical structure factor
of the fully frustrated ladder as a function of temperature for five 
different coupling ratios approaching the quantum critical point ($j'_c = 
1.40148$). The system sizes used in our calculations are $L = 14$ at low 
temperatures for $j' = 1.42$ and $1.45$, $L = 12$ at intermediate temperatures 
for $j' = 1.42$ and 1.45 and at low temperatures for $j' = 1.5$, and $L = 10$ 
otherwise. The quantity $T_{1/2}$ is defined at a given coupling ratio, $j'$, 
as the temperature where the intensity falls to half of its $T = 0$ value 
(horizontal line).}  
\label{ftwj}
\end{figure}

Appendix \ref{appb} analyzes the spectral weights of the $1 \rightarrow 2$ 
processes. Because all transitions are independent of the wave vector, the 
result is an extremely straightforward form where the bound-state intensities 
are related by simple ratios multiplying a thermal factor and a single matrix 
element. However, several features complicate efforts to apply this analysis 
to the observed spectral weights, shown in the inset of Fig.~\ref{ftwmtm}. 
One is the problem noted above, concerning the attribution of intensities 
due to the different, degenerate $1 \rightarrow 2$ and $2 \rightarrow 3$ 
processes. In principle, the intensity of each line ($i$) should have a leading 
temperature dependence $I_i = a_i e^{-\JR/T} + b_i e^{-(2\JR - l_i \JL)/
T}$, with related coefficients $a_i$ and $b_i$, and $l_i$ a small 
(positive, negative, or zero) integer. It is clear that the three leading 
lines in the inset of Fig.~\ref{ftwmtm} cross each other at temperatures as 
low as $T = 0.2\,\JR$, and very obviously at $0.4\,\JR$, as well as having maxima at 
quite different peak values. These results imply both strong effects from 
multiple contributions at rather low temperatures and additional physics 
at higher temperatures, where the satellites lose intensity again due to 
further thermal excitation. We draw attention also to the fact that the fourth 
(weakest) line in the inset, which from above is the strongest dominated by 
$2 \rightarrow 3$ processes, clearly has a thermal evolution characterized by 
an initial-state energy (presumably of order $2\JR$) rather different from 
that ($\JR$) governing the three $1 \rightarrow 2$ processes. Because thermal 
effects extend so rapidly beyond the lowest order, the relevance of an 
analysis of the type shown in App.~\ref{appb} could be tested only at very low 
temperatures, perhaps around $0.1 \JR$. Finally, such a test could only be 
performed meaningfully at values of $j'$ not too close to $j'_c$ ($j' > 1.45$ 
for $L = 14$ ladders), because otherwise the low-lying, Haldane-type intruder 
states mentioned in Sec.~\ref{sec:FFL} also interfere with the overall 
intensities in our finite-size calculations.

\subsection{One-Triplon Spectral Weight}

\label{sec:AnalysisC}

We turn next to the question motivating our entire investigation, namely 
the factors responsible for the anomalously rapid loss of spectral weight 
from the one-triplon band visible in a material such as SrCu$_2$(BO$_3$)$_2$. 
We focus again on the fully frustrated system. For all coupling ratios, $j'$, 
there is a loss of weight from the one-triplon mode, which can be quantified 
from the weight of the one-triplon pole in Figs.~\ref{dsfkp211} to 
\ref{dsfkp2o3}. This is shown in Fig.~\ref{ftwj} for several values of 
$j'$; a discussion of finite-size effects in the context of this figure, 
specifically for the $j' = 1.42$ and $1.45$ curves, is presented in 
App.~\ref{app:FS}. The observed loss of intensity with temperature is not 
unusual, and its unconventional feature is only the way it happens, between 
discrete energy levels that remain sharp at all temperatures, rather than as 
a conventional thermal broadening (Figs.~\ref{dsf210} and \ref{dsfkp210}). 
Over much of the range of $j'$, the temperature at which a significant loss 
of one-triplon spectral weight occurs cannot be said to be anomalously low, 
as shown in Fig.~\ref{ftwj} for parameter values such as $j' = 2$. However, 
our single most important finding is that, close to $j'_c$, there is a regime 
where large numbers of multiparticle excitations, involving many-rung triplon 
clusters in strongly bound and low-lying states (Fig.~\ref{fig:LinfAllExec}), 
come to dominate the low-energy physics \cite{rus}. As Fig.~\ref{ftwj} shows 
clearly, these states do lead to rapid transfer of spectral weight out of the 
one-triplon band at anomalously low temperatures. 

\begin{figure}[t!]
\centering\includegraphics[width=0.98\columnwidth]{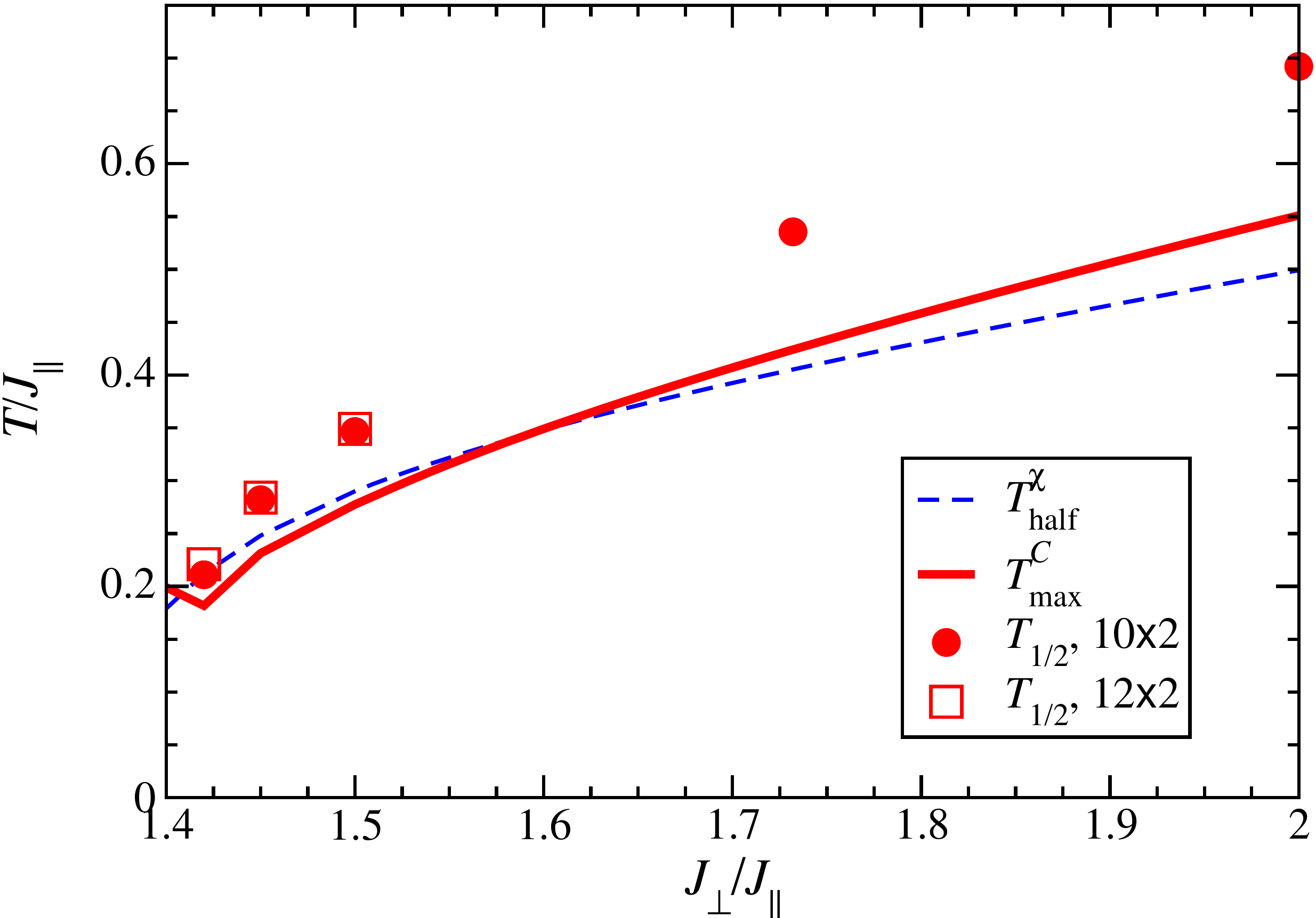}
\caption{Temperature, $T_{1/2} (j')$, at which the weight of the dynamical 
structure factor (Fig.~\protect{\ref{ftwj}}) falls to one half of its peak 
value, shown by the symbols for several values of the coupling ratio, $j'$.
Lines indicate the values of the quantities $T^{C}_{\rm max}$ and 
$T^{\chi}_{\rm half}$ (see text) in the thermodynamic limit, deduced from the 
numerical results of Ref.~\protect{\cite{rus}} for the magnetic specific 
heat and susceptibility.}  
\label{fth}
\end{figure}

To quantify the loss of spectral weight from the one-triplon signal, we 
consider the temperature, $T_{1/2}$, at which the intensity falls to one 
half of its zero-temperature value, given by the intersection of the
different curves with the horizontal line in Fig.~\ref{ftwj}. In 
Fig.~\ref{fth}, we show the values of $T_{1/2}$ as a function of $j'$ in 
the rung-singlet regime, noting that the phase transition, $j'_c \approx 
1.4015$, lies essentially on the left boundary of the figure. It is 
important to note that, although finite-size effects are expected to 
be relevant for small values of $|j' - j_c'|$ (Sec.~\ref{sec:FFL}, 
App.~\ref{app:FS}), our results for $j' = 1.42$, $1.45$, and $1.5$
from ladders  of $L = 12$, in addition to $L = 10$, demonstrate that 
$T_{1/2}$ can still be extracted reliably in this regime with the available 
system sizes. At values of $j'$ far from the transition, the one-triplon 
spectral weight remains close to unity until a temperature that is a 
significant fraction of $\JL$, as can be seen by considering the case $j'
 = 2$ in Fig.~\ref{ftwj}, where $T_{1/2} \simeq 0.692 \, \JL = 0.346 \, \JR$ 
(Fig.~\ref{fth}). However, as $j'$ approaches $j'_c$, $T_{1/2}$ falls to an 
anomalously low temperature, of order $0.22\, \JL \approx 0.15\, \JR$ when 
$j' = 1.42$. 

As a means to gain some initial insight into this emerging low energy scale, 
in Fig.~\ref{fth} we compare $T_{1/2}$ with the characteristic temperatures 
extracted from the thermodynamic response of the system. In Ref.~\cite{rus} 
we showed that the magnetic specific heat, $C(T)$, is characterized by a 
strong peak, which becomes narrower and moves to lower energies as $j' 
\rightarrow j'_c$, and in Fig.~\ref{fth} we show its position, $T_{\rm max}^C$. 
The magnetic susceptibility, $\chi (T)$, has a very broad peak, but can be 
characterized quite sensitively by its rapid onset, for which we use the 
temperature, $T_{\rm half}^\chi$, where $\chi (T)$ reaches half of its peak 
height. Because $C(T)$ is a consequence of excitations to all available 
levels in the spectrum, whereas $\chi(T)$ is a consequence of excitations 
to all available magnetic levels ($S \ge 1$), the fact that $T_{\rm max}^C$ 
and $T_{\rm half}^\chi$ track each other very closely makes clear that the 
descent of singlet and triplet bound-state branches as $j' \rightarrow j'_c$ 
[Fig.~\ref{fig:LinfAllExec}] is very similar not only in energy but also in 
density. To lowest order, the temperature scale obtained from a dynamical 
quantity, $T_{1/2}$, is effectively identical in the vicinity of the transition 
to the characteristic temperatures obtained from the static response. This 
confirms that the plethora of low-lying bound states accessible at finite 
temperatures near the quantum phase transition is explicitly the origin of 
the rapid loss of one-triplon spectral weight. We comment here that a 
connection between $C(T)$ and rapid spectral-weight transfer, based on 
the presence of well-localized bound-state modes of the system, was mooted 
for SrCu$_2$(BO$_3$)$_2$ on the basis of Raman-scattering studies in 
Ref.~\cite{rlea}. That $T_{1/2}$ rises above $T_{\rm max}^C$ and $T^\chi_{\rm half}$ 
in the regime far from $j'_c$ (Fig.~\ref{fth}) suggests that the 
spectral-weight transfer depends more critically than do $C(T)$ and 
$\chi(T)$ on a smaller number of processes (i.e., on a low density of 
levels) whose gaps become large.

The origin of the emerging low energy scale around $j'_c$ is somewhat complex, 
with two distinct types of contribution. In Ref.~\cite{rus} we discussed the 
quantity $E_{\rm bond}$, which is effectively the energy scale of a domain wall 
between the potentially long regions of singlet and triplet rungs that 
characterize the system in the vicinity of $j'_c$. However, the peak in the 
specific heat, $T_{\rm max}^C$, and the half-height temperature, $T^\chi_{\rm half}$, 
of the susceptibility are always irrational fractions of the characteristic 
energy scale, which remains on the order of the gap to the lowest excitations 
(Fig.~\ref{fig:LinfAllExec}). The second type of contribution is from the 
effective number of excited energy levels involved in draining weight away 
from the one-triplon branch, and this quantity remains difficult to define. 
A reasonable hypothesis for the rapid loss of spectral weight would be that 
pairs of triplons scatter strongly, with an effective scattering length 
significantly larger than the lattice constant \cite{rzrr}, and the area 
of this scattering region could be used to deduce a number of states. Our 
analysis demonstrates that this is not the correct picture, in that the 
origin of the strong quasiparticle scattering lies in the presence of many 
multi-triplon bound states rather than in extended two-triplon ones. Although 
it is clear from the data of Fig.~\ref{ftwj} that many states are involved 
in the loss of spectral weight, the deduction of an effective number of 
multi-triplon bound-state branches from these data remains an ill-defined task. 

As to the destination of the spectral weight lost from the one-triplon branch, 
this weight is transferred systematically to the primary multi-triplon 
excitations, whose intensities grow exponentially at low temperatures (as 
discussed above and shown in the inset of Fig.~\ref{ftwmtm}) and then more 
slowly beyond $T/\JL \approx 0.5$. Despite sometimes non-monotonic behavior, 
these satellite peak intensities do approach a constant ratio at high 
temperatures, as we will show in Sec.~\ref{sec:AnalysisHighT}. If the ladder 
couplings are altered away from perfect frustration, this picture is altered 
due to the presence of some ``continuum'' thermal effects. However, it is 
difficult to find a meaningful comparison of spectral-weight shifts with the 
fully unfrustrated system, where there are no special mode energies; in this 
case it would be necessary to characterize the shift of intensity by 
integrating the spectrum over certain frequency windows, chosen on the 
basis of the low-$T$ spectra, giving another quite ill-defined procedure 
that would lose validity at higher temperatures. 

In closing this subsection, we comment that Fig.~\ref{fth} allows a 
qualitative analogy with the situation in SrCu$_2$(BO$_3$)$_2$. This 
material is known to be located close to a first-order quantum phase 
transition out of its rung-singlet phase (to a plaquette valence-bond 
state), and to have an anomalously low value, $T_{1/2}/\Delta_0 = 7~{\rm
K}/35~{\rm K} = 0.2$ \cite{rzrr}, of the ratio between the 
half-height temperature of the one-triplon intensity and the size of 
the one-triplon gap. The conventional discussion of this material is 
phrased using $1/j'$, with the phase transition occurring at $1/j'_c = 
0.675$ \cite{rcm} and the best estimates of $1/j'$ (= 0.635 \cite{rmu,MiUe03}) 
falling within 5\% of this value. While it is tempting to place 
SrCu$_2$(BO$_3$)$_2$ directly on Fig.~\ref{fth}, where the ratio of 0.2
corresponds to $j' = 1.42$ and therefore lies very close to the transition, 
finding the limiting value of $T_{1/2}/\Delta_0$ for the Shastry-Sutherland 
system, analogous to that obtained here for the fully frustrated ladder, will 
require detailed calculations in a 2D model. In a very recent and exciting 
development \cite{rzea}, an applied pressure has been used to drive the 
SrCu$_2$(BO$_3$)$_2$ system through the quantum phase transition, i.e., 
to push $j'$ across $j'_c$, which raises the possibility of detailed 
thermodynamic and spectral-weight measurements as a function of the 
proximity, $\delta j' = |j' - j'_c|/j'_c$, to the transition.  

\subsection{High-Temperature Spectra}

\label{sec:AnalysisHighT}

\begin{figure}[t!]
\centering\includegraphics[width=0.98\columnwidth]{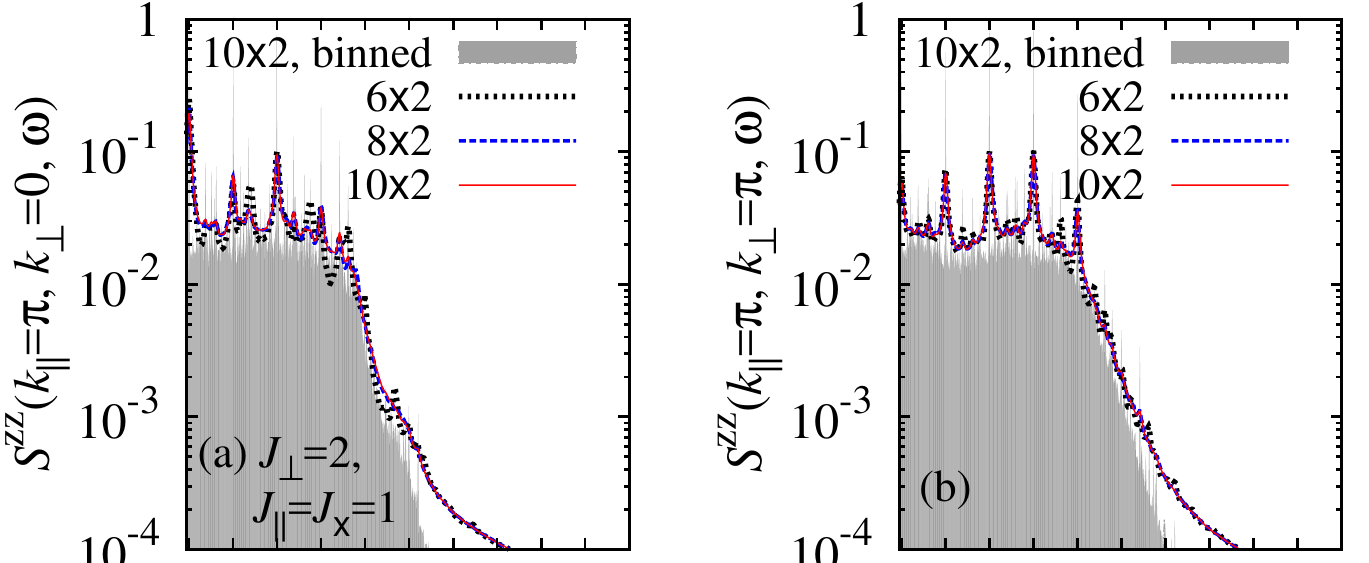}
\centering\includegraphics[width=0.98\columnwidth]{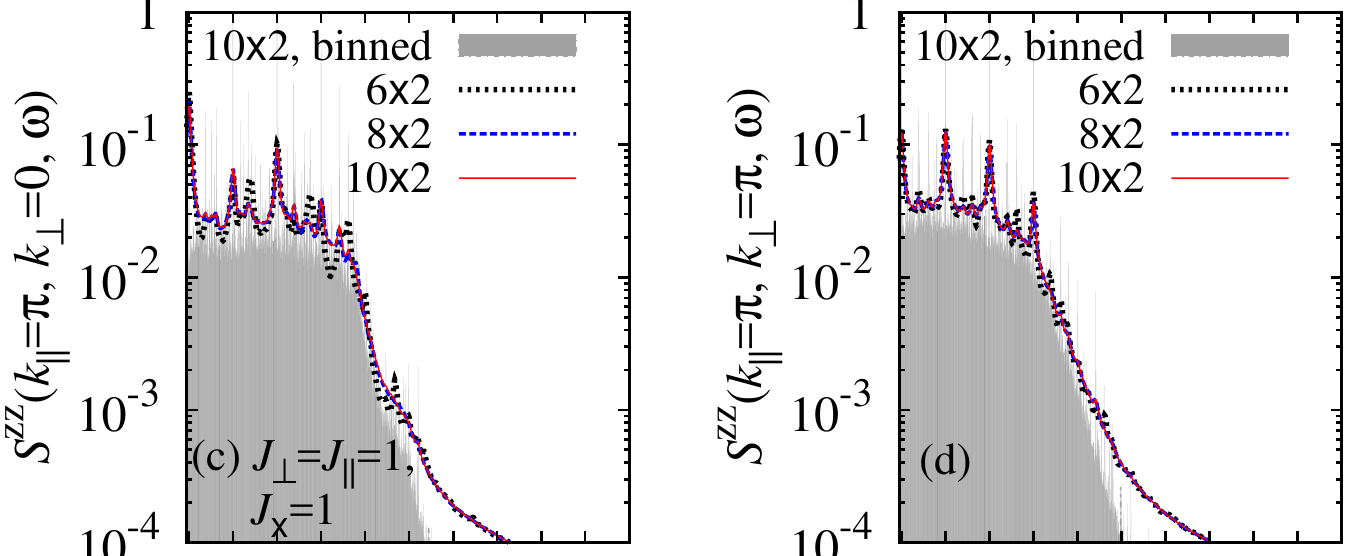}
\centering\includegraphics[width=0.98\columnwidth]{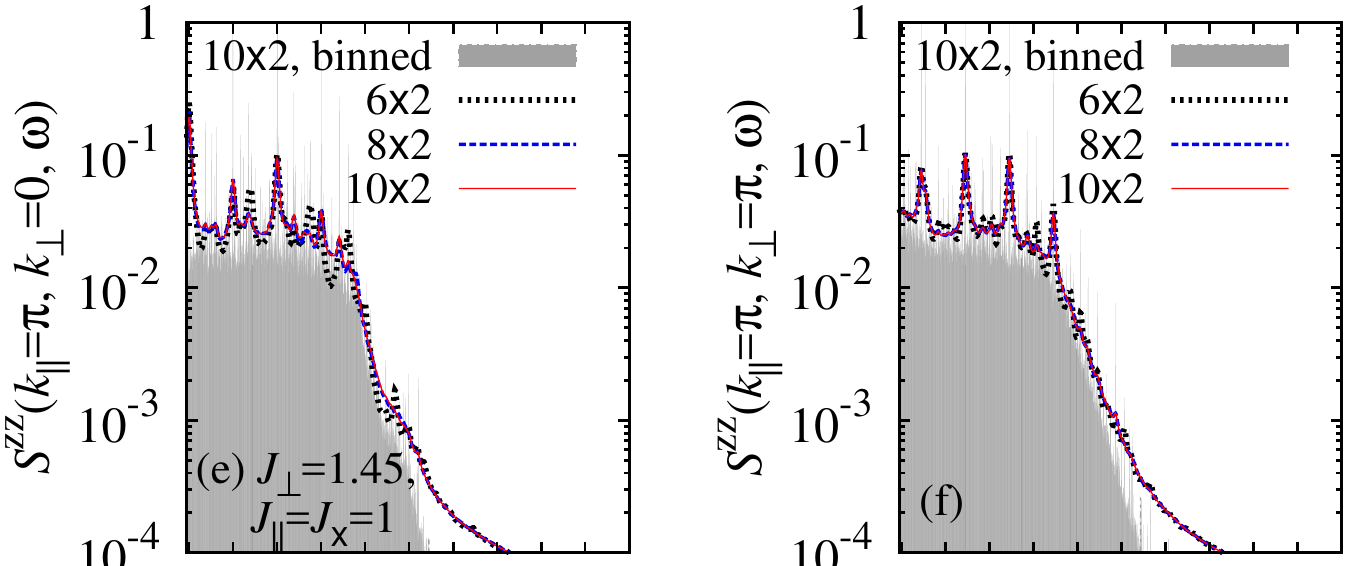}
\centering\includegraphics[width=0.98\columnwidth]{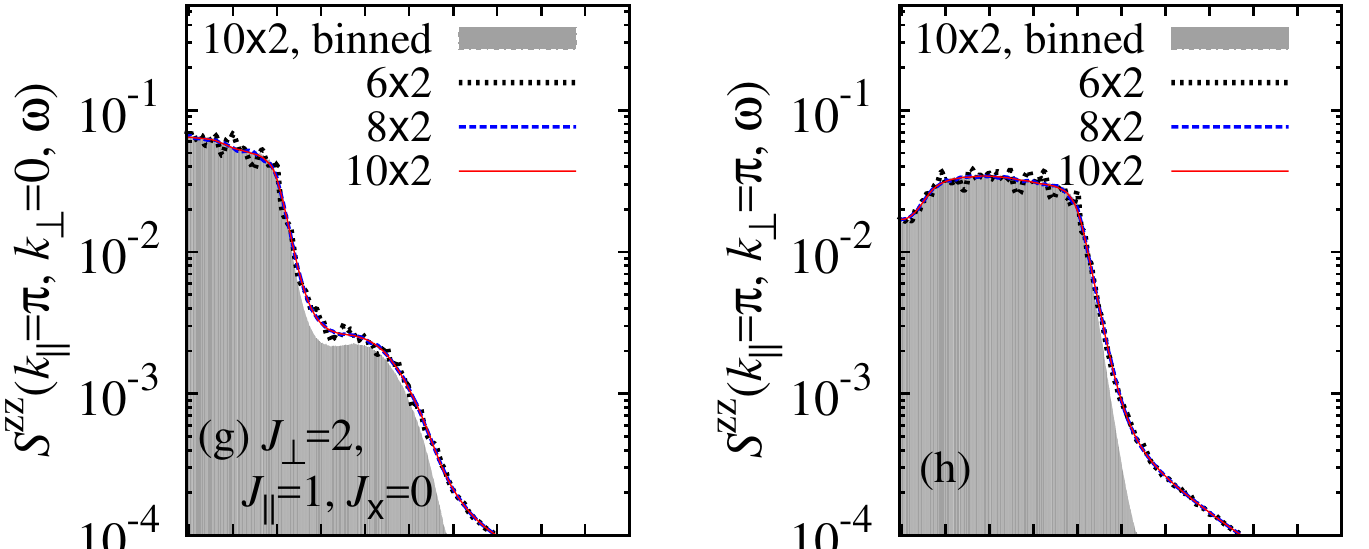}
\centering\includegraphics[width=0.98\columnwidth]{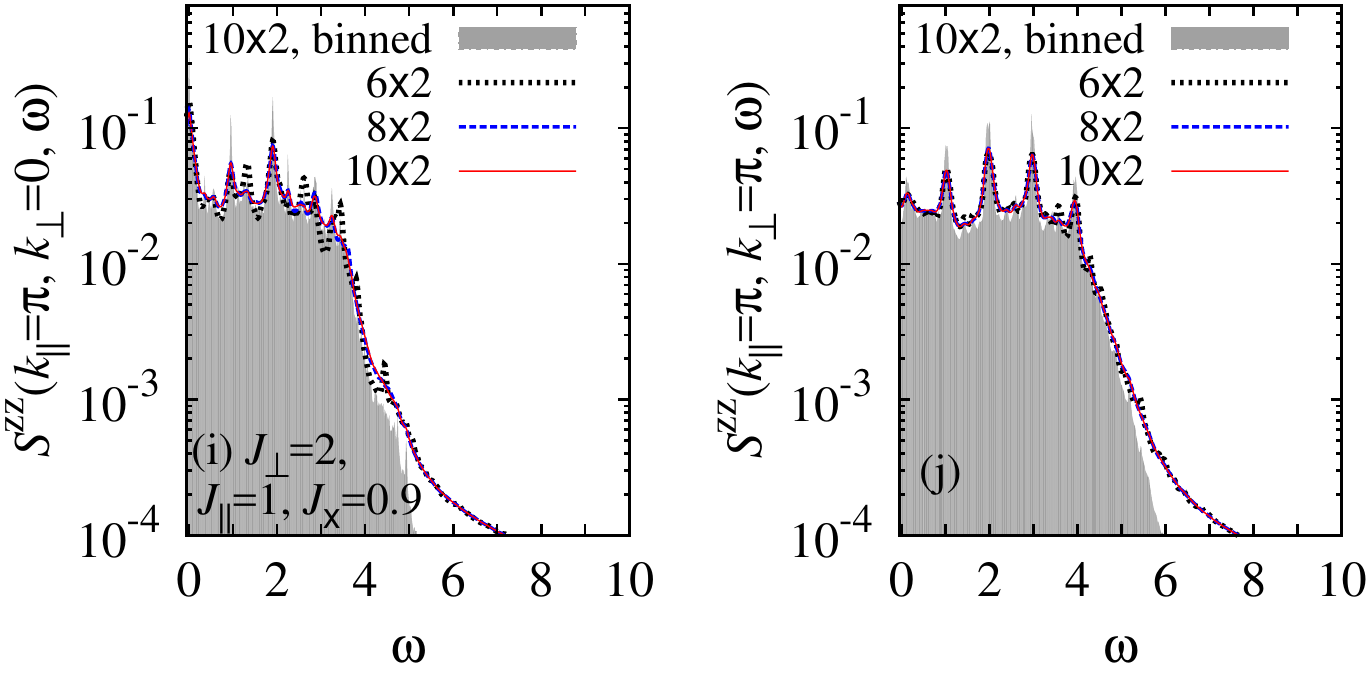}
\caption{Intensity at $k_{||} = \pi$ in (a,c,e,g,i) the symmetric and 
(b,d,f,h,j) the antisymmetric channel, shown as a function of energy at 
infinite temperature for the ladder of Fig.~\ref{fig:ladder} with coupling 
parameters $\JR = 2$ and $\JL = \JD = 1$ (a,b); $\JR = \JL = \JD = 1$ (c,d); 
$\JR = 1.45$ and $\JL = \JD = 1$ (e,f); $\JR = 2$, $\JL = 1$, and $\JD = 0$ 
(g,h); $\JR = 2$, $\JL = 1$, and $\JD = 0.9$ (i,j). The lines show ED data 
calculated for the specified system sizes with a Lorentzian broadening $\eta
 = \JL/20$; the grey shading shows binned data obtained from calculations 
on a ladder of $10 \times 2$ spins with $\eta = 0$.} 
\label{dsfkpinf}
\end{figure}

We conclude the analysis of our numerical results by discussing the situation 
at high temperatures. The results of Fig.~\ref{dsfinf} for the limit $T 
\rightarrow \infty$ are shown in Fig.~\ref{dsfkpinf} in the form of logarithmic 
intensities for the wave vector $k_\| = \pi$. For the fully frustrated ladders 
[Figs.~\ref{dsfkpinf}(a-f)], as noted in Sec.~\ref{sec:FTspec}, there are 
always special energies where the dynamical structure factor has a high 
spectral weight. This is a consequence of the discrete support, i.e., the 
$T = 0$ spectrum of non-dispersive bound-state energy levels arising from a 
hierarchy of $n$-triplon clusters, and of the matrix elements associated with 
transitions between these states [App.~\ref{appb}]. For the unfrustrated 
ladder [Figs.~\ref{dsfkpinf}(g,h)], this weight is spread out completely 
over the available states, which form continuous energy bands, but a plateau 
structure emerges in both the symmetric and antisymmetric channels. These 
plateaus may be ascribed to the scattering processes connecting sectors of 
different triplon number, $l$, specifically $0 \rightarrow 1$, $1 \rightarrow 
2$, $\dots$ in the antisymmetric channel and $0 \rightarrow 2$, $1 \rightarrow 
3$, $\dots$ in the symmetric channel. We note that these processes are not 
symmetrical in energy, but that the reverse processes start to contribute at 
higher temperatures, and in fact dominate the low-energy response once many 
thermally occupied states are available to give up their energy.

Figure \ref{dsfkpinf} may also be used to discuss the intrinsic line shape 
of the spectral features in spin ladders as a consequence of their frustration. 
To remove the effects of the Lorentzian broadening ($\eta = \JL/20$) used in 
Figs.~\ref{dsf211} to \ref{dsfkp21p9}, we have summed the weights of all the 
poles, meaning the coefficients of all the $\delta$-functions obtained in 
Eq.~(\ref{edsfah}) in the limit $\eta = 0$, in bins of sizes $\Delta \omega 
/ \JL = 0.006$, \ldots, 0.015 (depending on the specific ratio $j'$) and we 
present the resulting histograms for ladders of $L = 10$ rungs as the grey 
shaded regions in Fig.~\ref{dsfkpinf}. Focusing on the antisymmetric channel 
($k_\perp = \pi$), we observe that in fully frustrated ladders [$\JD = \JL$, 
Figs.~\ref{dsfkpinf}(b), \ref{dsfkpinf}(d), and \ref{dsfkpinf}(f)] the 
dominant modes of the dynamical spectral function (Sec.~\ref{sec:AnalysisA}) 
remain as $\delta$-functions up to $T = \infty$. This complete absence of 
thermal broadening is due to the complete flatness of the band, which is 
in one-to-one correspondence with the conservation of every rung spin, in 
the fully frustrated ladder. Figure \ref{dsfkpinf}(j) shows that a detuning 
of the frustration to $\JD = 0.9\,\JL$ smooths the spectral function: the 
previously sharp lines are now subject to thermal broadening, which gives 
them an intrinsic line width, on the order of $0.1\,\JL$, in approximate 
agreement with the band width in this case. Considering finally the
unfrustrated ladder [$\JD = 0$, Fig.~\ref{dsfkpinf}(h)], here the lines 
also disappear completely from the binned data, leaving only the same 
broad plateaus observed in the spectral function with Lorentzian broadening. 
We note also that the Lorentzian broadening gives rise to high-frequency 
tails in the spectral functions, which may be considered as artifacts by 
comparison with the much more rapid decay towards high frequencies found 
in the intrinsic line shape (again we remind the reader of the logarithmic 
scale on the intensity axes). Similar effects are also observed in the 
binned data at finite temperatures, $0 < T < \infty$ (not shown); the 
examples shown in Fig.~\ref{dsfkpinf} for $T = \infty$ represent the upper 
limit on intrinsic thermal broadening.

\subsection{Experiment}

\label{sec:AnalysisE}

Finally, we close our discussion of the thermal redistribution of spectral 
weight in the dynamical structure factor of frustrated systems by considering 
the experimental situation. To the best of our knowledge, no material has yet 
been identified with the effective magnetic Hamiltonian of the fully frustrated 
spin ladder. However, many low-dimensional frustrated quantum magnets are 
known and many have very narrow excitation bands as a consequence of their 
frustration. Thus we expect our results to serve as a valuable paradigm for 
understanding the thermodynamic \cite{rus} and dynamical response of frustrated 
quantum spin systems. 

Our primary motivation for this study was the anomalously rapid loss of 
spectral weight in the one-triplon excitation as a function of temperature, 
and we have identified the proximity to a quantum phase transition, with its 
accompanying plethora of multiparticle bound states, as a strong source of 
such behavior. However, it is not the only possible origin of anomalous 
effects on spectral weights, and a complete interpretation requires an 
account of other factors. The majority of low-dimensional quantum magnets 
have additional ``anisotropic'' terms in the magnetic Hamiltonian beyond 
the Heisenberg interaction. A good example is the spin-chain material 
Cu(C$_6$D$_5$COO)$_2\cdot3$D$_2$O (Cu-benzoate), where the anomalous opening 
of a gap in an applied magnetic field \cite{rdhrba} was explained by the 
presence of Dzyaloshinskii-Moriya (DM) interactions and $g$-tensor anisotropy 
\cite{roa,Affleck_1999}. Electron Spin Resonance (ESR) studies of Cu-benzoate
\cite{ranibsam,ranibsam2,rie}, KCuGaF$_6$ \cite{Umegaki_2009}, and 
Cu-pyrimidine dinitrate \cite{Zvyagin_2012_Review,rhea} report spectral 
weights that vanish rapidly with increasing temperature even in the absence 
of frustrating interactions, and these have been ascribed to the presence 
of breather and soliton modes in the effective quantum sine-Gordon model 
\cite{roa,Affleck_1999} for the material. Although no comparison of the 
field-induced gap with the temperature has yet shown that the thermally 
induced loss of ESR spectral weight is truly anomalous, the presence of 
additional modes arising due to spin anisotropy in the Hamiltonian does 
serve as an additional possible route to rapid thermal decay. Similar 
thermal effects are thought to be present in the strong-leg ladder material 
(C$_7$H$_{10}$N)$_2$CuBr$_4$ (DIMPY) \cite{Ozerov15}.

The archetypal perfectly frustrated quantum magnet, whose dynamical 
response at finite temperature motivated this work, is the 2D material 
SrCu$_2$(BO$_3$)$_2$. This system has complete frustration in its 
Shastry-Sutherland geometry, a very flat one-triplon excitation, and 
also possesses significant DM interactions, ensuring a degree of anisotropy 
in spin space. Here we have shown that fully frustrated spin interactions 
have a pronounced effect in creating multi-triplon bound states, whose 
presence causes a significant damping and redistribution of the one-triplon 
spectral weight even at low temperatures. One of our primary findings is 
that this effect is dramatically stronger near the quantum phase transition 
out of the rung-singlet state, where many multiparticle excitations with 
anomalously low energies come to dominate the response of the system. It 
is known that SrCu$_2$(BO$_3$)$_2$ lies close to this transition in the 
Shastry-Sutherland model, which occurs at the ratio $1/j' \equiv J_s/J_r 
\simeq 0.675$ \cite{rcm} of the square-lattice and rung couplings, and thus 
we strongly suspect that the emergence of many low-lying multi-triplon bound 
states is precisely the physics of this material. We point out that the 
measured decay rate of the one-triplon band intensity with temperature
\cite{rglhcbdqc,rzrr}, compared with calculations for the Shastry-Sutherland 
model exactly analogous to Fig.~\ref{ftwj}, could be used to estimate very 
accurately the proximity, $\delta j'$, of SrCu$_2$(BO$_3$)$_2$ to the critical 
point. Whether or not these generic properties of a fully frustrated, 
SU(2)-symmetric model are enhanced by DM interactions in causing the 
near-total destruction of the one-triplon mode at $T \approx \Delta/3$ in 
SrCu$_2$(BO$_3$)$_2$ will require detailed analysis of anisotropic models. 

\section{Summary}

\label{sec:Summary}

The dynamical response of low-dimensional quantum magnets presents a major 
challenge both to theoretical understanding and to the most advanced numerical 
methods. This is particularly true for highly frustrated spin systems, which 
are characterized by an almost flat one-triplon excitation band at zero 
temperature, where little insight is available either into the temperature 
dependence of this single-particle dispersion or into the multiparticle 
dynamics developing at finite temperature. We have investigated these 
questions using the example of the frustrated two-leg spin ladder and we 
present systematic numerical results from exact diagonalization for the 
dynamic structure factor as a function of temperature, coupling ratio, and 
degree of frustration. 

We find that the fully frustrated system has a discrete spectrum of 
excitations. As the temperature is increased, spectral weight is transferred 
out of the one-triplon band to levels at a wide range of energies, including 
those below the one-particle gap; these are the energy levels of highly 
localized bound states involving clusters of multiple neighboring triplons. 
Close to the quantum phase transition at $j'_c = 1.4015$, the bound states 
lie very low in energy, they involve very large numbers of triplets, and the 
weight transfer is anomalously rapid. These many-particle excitations are 
extended objects and, as a consequence of the discrete spectrum, they persist 
as sharp spectral features even to infinite temperatures. 

One may ask whether these two remarkable qualitative features, the rapid 
transfer of spectral weight at temperatures much lower than the one-triplon 
gap and the persistence of well-defined peaks at very high temperatures, are 
connected. In principle, the rapid transfer of spectral weight depends on 
having a large number of low-lying energy levels in the spectrum, whereas 
the well-defined peaks are a specific consequence of having completely flat 
(localized) bands. However, it remains an open question whether a truly 
high density of low-lying states could be achieved in a system with (weakly) 
dispersive bands or lacking an exact convergence of levels at a quantum 
phase transition, and the rate of spectral-weight transfer in such cases 
may be limited. 

Nevertheless, the first of these remarkable features offers an explanation 
for the qualitative physics underlying the observation by inelastic 
neutron scattering studies of the low-dimensional frustrated material 
SrCu$_2$(BO$_3$)$_2$ that the intensity of the one-triplon band is completely 
dispersed to states at all energies even at temperatures only 1/3 of its gap. 
The second of these features, namely the persistence of sharp excitation 
peaks at high temperatures in a fully frustrated system, remains to be 
investigated in experiment. 

\acknowledgments

We thank S.~Wessel for helpful discussions.
We are grateful to the HLRN Hannover for the allocation of CPU time.
This work was supported by the
Helmholtz Association via the Virtual Institute ``New states of matter and 
their excitations,'' by the Swiss NSF, by the NSF of China under Grant 
11174365, and by the National Basic Research Program of the Chinese MoST 
under Grant 2012CB921704.

\appendix
\section{Multi-Triplon Bound States}
\label{appa}

The dynamical structure factor [Eq.~(\ref{edsfah})] is obtained from the 
action of local spin-raising and -lowering operators. Analytical insight 
into its nature for the fully frustrated ladder begins from the expression 
of the bound states in the basis of triplons (one-rung triplet excitations). 
As noted in Sec.~\ref{sec:FFL}, in the case where the frustrating couplings 
$j' = \JR/\JL
 = \JR/\JD \ge j'_c \simeq 1.4015$, two triplons on neighboring rungs form an 
exact bound state. This bound state consists of a singlet, which we denote 
$|2s \rangle$, with energy $E_{2s}/\JL = 2(j'- 1)$, a triplet, denoted $|2t,m 
\rangle$, $m = -1$, 0, 1, with energy $E_{2t}/\JL = 2 j' - 1$, and a quintet, 
$|2q,m \rangle$, $m = -2$, $-1$, $\dots$, 2, with $E_{2q}/\JL = 2 j' + 1$. In 
the basis $|m_i,m_{i+1} \rangle$, $m_i = - 1$, 0, 1, of two triplons on 
neighboring rungs, the nine components of the bound state are straightforward 
linear combinations of the nine two-triplon states. Omitting the rung indices 
and denoting $m = -1$ by $\overline{1}$, these are 
\begin{eqnarray}
|2q,2 \rangle & = & |11 \rangle, \;\;\;\; |2q,1 \rangle \, = \, {\textstyle 
\frac{1}{\sqrt{2}}} (|10 \rangle + |01 \rangle), \nonumber \\
|2q,0 \rangle & = & {\textstyle \frac{1}{\sqrt{6}}} (|1 \overline{1} \rangle
 + 2 |00 \rangle + |\overline{1} 1 \rangle), \nonumber \\
|2t,1 \rangle & = & {\textstyle \frac{1}{\sqrt{2}}} (|10 \rangle - |01 
\rangle), \label{e2ts} \\
|2t,0 \rangle & = & {\textstyle \frac{1}{\sqrt{2}}} 
(|1 \overline{1} \rangle - |\overline{1} 1 \rangle), \nonumber \\
|2s \rangle & = & {\textstyle \frac{1}{\sqrt{3}}} (|1 \overline{1} \rangle
 - |00 \rangle + |\overline{1} 1 \rangle), \nonumber 
\end{eqnarray}
and symmetrically for the states $|2q,-1 \rangle$, $|2q,-2 \rangle$, and 
$|2t,-1 \rangle$. 

In the same way, the 27 components of states with three triplons on 
neighboring rungs also form a set of multiplets, or effective three-triplon 
bound states. Diagonalizing the three-rung Hamiltonian in the different $S_z$ 
sectors yields the energies $E_{3h}/\JL = 3 j' + 2$ (heptet, $|3h,m \rangle$), 
$E_{3qa}/\JL = 3 j' + 1$ (quintet, $|3qa,m \rangle$), $E_{3ta}/\JL = 3 j'$ 
(triplet, $|3ta,m \rangle$), $E_{3qb}/\JL = E_{3tb}/\JL = 3 j' - 1$ (degenerate 
quintet, $|3qb,m \rangle$, and triplet, $|3tb,m \rangle$), $E_{3s}/\JL = 3 j'
 - 2$ (singlet, $|3s \rangle$), and $E_{3tc}/\JL = 3 j' - 3$ (triplet, $|3tc,
m \rangle$). We draw attention to the fact that the lowest-lying state of the 
multiplet is a triplet, and that this is a generic property of all odd-length
multi-triplon clusters \cite{rus,NUZ97}. The wave functions of these states 
may be expressed as 
\begin{eqnarray} 
|3h,3 \rangle & = & |111 \rangle, \nonumber \\
|3h,2 \rangle & = & {\textstyle \frac{1}{\sqrt{3}}} (|110 \rangle + |101 
\rangle + |011 \rangle), \nonumber \\
|3h,1 \rangle & = & {\textstyle \frac{1}{\sqrt{15}}} (2 |100 \rangle + 2 |010 
\rangle + 2 |001 \rangle \label{e3tsh} \\ 
& & \;\;\;\;\;\;\;\; + |1 1 \overline{1} \rangle + |1 \overline{1} 1 \rangle
 + |\overline{1} 1 1 \rangle), \nonumber \\
|3h,0 \rangle & = & {\textstyle \frac{1}{\sqrt{10}}} (|1 0 \overline{1} 
\rangle + |0 \overline{1} 1 \rangle + |\overline{1} 1 0 \rangle \nonumber \\ 
& & \;\;\;\;\;\;\;\; + 2 |000 \rangle + |1 \overline{1} 0 \rangle + |0 1 
\overline{1} \rangle + |\overline{1} 0 1 \rangle), \nonumber 
\end{eqnarray}
for the heptet, with symmetrical expressions for $m = -1$, $-2$, and $-3$,
\begin{eqnarray} 
|3qa,2 \rangle & = & {\textstyle \frac{1}{\sqrt{2}}} (|110 \rangle - |011 
\rangle), \nonumber \\
|3qa,1 \rangle & = & {\textstyle \frac{1}{2}} (|100 \rangle + |1 1 \overline{1} 
\rangle - |\overline{1} 1 1 \rangle - |0 0 1 \rangle), \label{e3tsqa} \\
|3qa,0 \rangle & = & {\textstyle \frac{1}{2 \sqrt{3}}} (2 |1 0 \overline{1} 
\rangle - |0 \overline{1} 1 \rangle - |\overline{1} 1 0 \rangle \nonumber \\ 
& & \;\;\;\;\;\;\;\; + |1 \overline{1} 0 \rangle + |0 1 \overline{1} \rangle
 - 2 |\overline{1} 0 1 \rangle), \nonumber \\
|3qb,2 \rangle & = & {\textstyle \frac{1}{\sqrt{6}}} (|110 \rangle - 2 
|101 \rangle + |011 \rangle), \nonumber \\
|3qb,1 \rangle & = & {\textstyle \frac{1}{2 \sqrt{3}}} (|100 \rangle - 2 |010 
\rangle + |001 \rangle \label{e3tsqb} \\ 
& & \;\;\;\;\;\;\;\; - |1 1 \overline{1} \rangle + 2 | 1 \overline{1} 1 
\rangle - |\overline{1} 1 1 \rangle), \nonumber \\
|3qb,0 \rangle & = & {\textstyle \frac{1}{2}} (|1 \overline{1} 0 \rangle - 
|0 1 \overline{1} \rangle - |\overline{1} 1 0 \rangle + |0 \overline{1} 1 
\rangle), \nonumber 
\end{eqnarray}
for the quintets, 
\begin{eqnarray} 
|3ta,1 \rangle & = & {\textstyle \frac{1}{\sqrt{3}}} (|010 \rangle - |1 1 
\overline{1} \rangle - |\overline{1} 1 1 \rangle), \label{e3tsta} \\
|3ta,0 \rangle & = & {\textstyle \frac{1}{\sqrt{3}}} (|000 \rangle - |1 0 
\overline{1} \rangle - |\overline{1} 0 1 \rangle), \nonumber \\
|3tb,1 \rangle & = & {\textstyle \frac{1}{2}} (|100 \rangle - |001 
\rangle - |1 1 \overline{1} \rangle + |\overline{1} 1 1 \rangle), 
\label{e3tstb} \\
|3tb,0 \rangle & = & {\textstyle \frac{1}{2}} (|1 \overline{1} 0 \rangle
 - |0 1 \overline{1} \rangle - |0 \overline{1} 1 \rangle + |\overline{1} 1 0 
\rangle), \nonumber \\
|3tc,1 \rangle & = & {\textstyle \frac{1}{2 \sqrt{15}}} (- 3 |100 \rangle
 + 2 |010 \rangle - 3 |001 \rangle \nonumber \\ & & \;\;\;\;\;\;\;\; + |1 1 
\overline{1} \rangle + 6 |1 \overline{1} 1 \rangle + |\overline{1} 1 1 
\rangle), \nonumber \\
|3tc,0 \rangle & = & {\textstyle \frac{1}{2 \sqrt{15}}} (- 2 |1 0 \overline{1} 
\rangle + 3 |0 \overline{1} 1 \rangle + 3 |\overline{1} 1 0 \rangle 
\label{e3tstc} \\ & & 
\;\;\;\;\; - 4 |000 \rangle + 3 |1 \overline{1} 0 \rangle + 3 |0 1 \overline{1} 
\rangle - 2 |\overline{1} 0 1 \rangle), \nonumber
\end{eqnarray}
for the triplets, and 
\begin{eqnarray} 
|3s \rangle & = & {\textstyle \frac{1}{\sqrt{6}}} (|1 0 \overline{1} \rangle
 + |0 \overline{1} 1 \rangle + |\overline{1} 1 0 \rangle \label{e3tss} \\ 
& & \;\;\;\;\;\;\;\;  - |1 \overline{1} 0 \rangle - |0 1 \overline{1} \rangle
 - |\overline{1} 0 1 \rangle) \nonumber
\end{eqnarray}
for the singlet. 

One may continue the process for all higher-$n$ multiplets. Results for 
$n = 4$ are obtained most readily by numerical diagonalization of the 4-site 
Haldane chain with open boundary conditions (Sec.~\ref{sec:FFL}) \cite{rus}. 
We do not find strong contributions from 4-triplon bound states in the 
dynamical response and do not enter into further detail here. We note only 
that the two lowest-lying states of the multiplet are a total singlet with 
energy $E_{4s}^1/\JL = 4 j' - 4.64575$ and a triplet with $E_{4t}^1/\JL = 4 j' - 
4.13658$, with the singlet the lowest level as for all even-$n$ bound states.

\section{Dynamical Spectral Function}
\label{appb}

We consider the analytical calculation of the contributions from small-$n$ 
bound states to the dynamical spectral function in the antisymmetric channel. 
In its most general form,
\begin{equation}
S^{\alpha\beta} ({\bf q}, \omega, T) = \sum_{ij} p_i \langle i | S_{\bf q}^\alpha | 
j \rangle \langle j| S_{\bf q}^\beta | i \rangle \delta (\omega - E_j + 
E_i),
\label{edsf}
\end{equation}
where $|i \rangle$ and $|j \rangle$ denote the initial and final states of 
the scattering process. For a system with a Heisenberg Hamiltonian (preserving 
SU(2) symmetry), $S^{\alpha\beta} ({\bf q}, \omega, T)$ obeys the symmetries 
$S^{xx} = S^{yy} = S^{zz} = S^{+-} = S^{-+}$, and here it is most straightforward 
to obtain $S^{zz}$ (\ref{edsfah}) from $S^{+-}$. The form of Eq.~(\ref{edsfah})
is obtained by considering the delta function as $\pi \delta(E) = \lim_{\eta 
\rightarrow 0} {\rm Im} [E - i \eta]^{-1}$. The temperature dependence of the 
dynamical spectral function is contained within the energy $\delta$-function 
and in the probability function $p_i = e^{-E_i/T}/Z(T)$ for the occupation 
of the initial state. 

In this Appendix, we specialize to the case of the dynamical structure factor 
for inelastic neutron scattering. In the highly restricted single-rung basis,
a neutron spin-flip changes both the $S$ and $S_z$ quantum numbers by $\pm 1$, 
also changing the rung parity, and therefore the corresponding $S^{\alpha\beta}$ 
appears exclusively in the antisymmetric channel (Secs.~\ref{sec:FTspec} and 
\ref{sec:Analysis}). An 
incident neutron may either alter the triplon number, explicitly changing 
the state of the ladder between two different sectors with $l$ and $l+1$ 
triplons, or may cause excitations within single bound-state multiplets of 
$n \ge 2$. Because the latter require higher activation energies [$E_i$ in 
Eq.~(\ref{edsf})] than processes with zero or one initial triplons, they 
are significantly weaker and will not be considered explicitly here. As noted 
above, we also restrict our presentation to the fully frustrated ladder, 
where the fact that the band energies are completely independent of the 
wave vector, $k_\|$, means that the real-space matrix elements provide the 
full information required. Thus we may consider only the action of a neutron 
incident on a single ladder rung, whose scattering matrix element from the 
zero- to the one-triplon sector (0 $\rightarrow$ 1) is given by $M = 
\langle 1 |S^+| 0 \rangle$, to deduce the relative intensities of all 
higher processes ($l \rightarrow l + 1$). 

The calculation of $S^{+-} ({\bf q}, \omega, T)$ for these cases proceeds 
directly from the wave functions specified in Eqs.~(\ref{e2ts})--(\ref{e3tss}). 
The case 1 $\rightarrow$ 2 involves scattering of a neutron incident on a 
singlet rung, one of whose neighbors is a triplon. The final state of these 
two ladder rungs is then one of the bound states of two triplons specified 
in Eq.~(\ref{e2ts}). The probability of finding a suitable initial state 
with an isolated triplon on site $i+1$ is given by 
\begin{equation}
2 p_i (T) = 2 p_t (T) p_s (T) = 2 e^{- \JR/T} / Z,
\label{epl}
\end{equation}
where $p_t = e^{- \JR/T}$ is the probability for a rung triplon excitation, 
$p_s \equiv 1/Z$ is the (temperature-dependent) probability of the neighboring 
rung to be in its singlet (i.e., unexcited) state, and the factor of 2 is for 
the two possible processes of exciting the neighboring rung singlet. 

In the basis of rung quantum numbers, a singlet excited to the triplet state 
with $m_i = 1$, and with a triplet of arbitrary $m_{i+1}$ located at site 
$i+1$, results in the two-rung states $|1,1 \rangle$, $|1,0 \rangle$, and 
$|1,\overline{1} \rangle$. Clearly the first is identically the two-rung 
state $|2q,2 \rangle$, the second may be part of the state $|2q,1 \rangle$ or 
$|2t,1 \rangle$ (each, from Eq.~(\ref{e2ts}), with probability 1/2), and the 
third may be part of the states $|2q,0 \rangle$, $|2t,0 \rangle$, or $|2s,0 
\rangle$ (with respective probabilities 1/6, 1/2, and 1/3). In the basis of 
multiplet quantum numbers, one finds the conventional, less restrictive 
selection rules for neutron scattering, $\Delta S = \Delta S_z = 0, \pm 1$. 
The contributions of one- to two-triplon scattering processes to the 
dynamical structure factor are then 
\begin{eqnarray}
\!\!\!\! S^{+-}_{|1t \rangle \rightarrow |2q,2 \rangle} (q,\omega,T) & = & 2 p_t (T) p_s 
(T)|M|^2 \delta (\omega \! - \! J \! - \! J') \nonumber \\ 
\!\!\!\! S^{+-}_{|1t \rangle \rightarrow |2q,1 \rangle} (q,\omega,T) & = & p_t (T) p_s 
(T) |M|^2 \delta (\omega \! - \! J \! - \! J') \nonumber \\ 
\!\!\!\! S^{+-}_{|1t \rangle \rightarrow |2q,0 \rangle} (q,\omega,T) & = & p_t (T) p_s 
(T) |M|^2 \delta (\omega \! - \! J \! - \! J') \label{edsf12} \\ 
\!\!\!\! S^{+-}_{|1t \rangle \rightarrow |2t,1 \rangle} (q,\omega,T) & = & {\textstyle 
{\frac{1}{3}}} p_t (T) p_s (T) |M|^2 \delta (\omega \! - \! J \! + \! J') 
\nonumber \\ 
\!\!\!\! S^{+-}_{|1t \rangle \rightarrow |2t,0 \rangle} (q,\omega,T) & = & p_t (T) p_s 
(T) |M|^2 \delta (\omega \! - \! J \! + \! J') \nonumber \\ 
\!\!\!\! S^{+-}_{|1t \rangle \rightarrow |2s,0 \rangle} (q,\omega,T) & = & {\textstyle 
{\frac{2}{3}}} p_t (T) p_s (T) |M|^2 \delta (\omega \! - \! J \! + \! 2 J'), 
\nonumber
\end{eqnarray}
all independent of the neutron wave vector, $q$. We note in addition that 
these expressions contain, as specified in Eq.~(\ref{edsf}), no explicit 
dependence of the intensity on the final-state energy, and thus all the 
branches of each $n$-triplon bound state have similar probabilities in the 
same sector (differing only by the coefficients in their wave functions). 
We draw attention to the fact that, although the scattering energy, $\omega$, 
in Eqs.~(\ref{edsf12}) can become negative for $|1t \rangle \rightarrow 
|2s,0 \rangle$ processes in the range $j'_c < j' < 2$, the excitation energy 
for the initial state remains positive; this is indeed the dominant process 
observed in Secs.~\ref{sec:FTspec} and \ref{sec:Analysis} to gain spectral 
weight at finite temperatures. 

A similar exercise can be applied to deduce the probabilities of the 
two-triplon bound states and, with an additional excited triplet, their overlap 
with the three-triplon bound states [Eqs.~(\ref{e3tsh})--(\ref{e3tss})], to 
obtain the complete contribution to the dynamical structure factor from 2 
$\rightarrow$ 3 processes. Although some of these are visible in our results 
at intermediate temperatures, as discussed in Sec.~\ref{sec:Analysis}, their 
contributions remain small compared to 1 $\rightarrow$ 2 processes. Thus their 
energies are visible but their intensities are difficult to characterize 
quantitatively, and so we do not perform the spectral-weight calculation 
explicitly here. 

\section{Finite-Size Analysis of the One-Triplon Spectral Weight}
\label{app:FS}

\begin{figure}[t]
\centering\includegraphics[width=0.98\columnwidth]{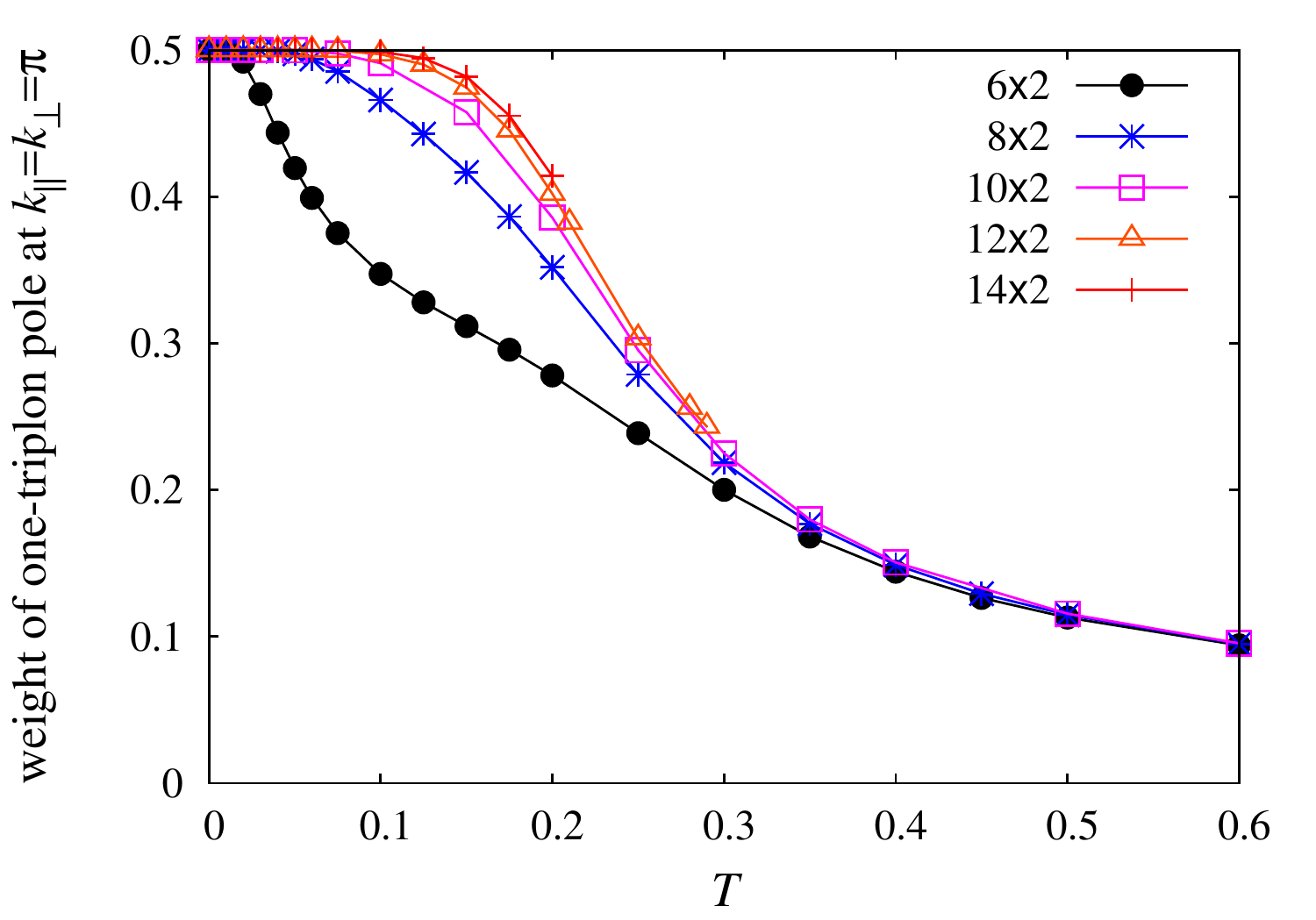}
\caption{Dependence on temperature of the weight of the one-triplon pole in 
the dynamical structure factor calculated for fully frustrated ladders of 
sizes $6 \le L \le 14$ rungs and with couplings $\JR = 1.45$, $\JL = \JD = 1$.}
\label{fig:J1_45}
\end{figure}

We consider the finite-size effects in our calculation of $S^{zz} (k_\|=\pi,
k_\perp=\pi,\omega)$ close to the critical point, $j_c'$, of the fully 
frustrated ladder ($\JD = \JL$). Among all of the features studied, the 
spectral weight of the one-triplon excitation, appearing at $\omega = \JR$, 
exhibits the most severe finite-size effects, and so we focus on this
quantity. At $T = 0$, the pole at $\omega = \JR$ in $S^{zz} (k_\|,k_\perp=\pi, 
\omega)$ has a weight of $1/2$, irrespective of the ladder length, whence 
finite-size effects appear only for $T > 0$. Also at high $T$, any dependence
of our results on the system size is weak and as a consequence we may restrict 
our focus further to the region of low but finite temperatures.

The primary problem in the calculation of the spectral weight is caused by 
the presence of the intruder states discussed in Sec.~\ref{sec:FFL} and 
illustrated in Fig.~\ref{fig:LinfAllExec}. For coupling ratios $j' = 
1.45$ and $1.42$, the spurious Haldane ground state lies below the first 
true excitation of the rung-singlet phase for ladders with $L \le 14$ rungs, 
as shown by the symbols in Fig.~\ref{fig:LinfAllExec}. Nevertheless, this 
state is nondegenerate, whereas the true low-lying excitations are at least 
$L$-fold degenerate, and thus we will find that systems with $L \le 14$ rungs 
are in fact sufficient to approximate the thermodynamic limit.

Figure \ref{fig:J1_45} shows the case $j' = 1.45$. As a gauge of the 
temperatures at which finite-size effects matter, the $L = 6$
ladder can be considered as converged to the thermodynamic limit for 
$T/\JL \gtrsim 0.5$ and the $L = 10$ ladder as a good approximation
to the infinite system for $T/\JL \gtrsim 0.35$. At lower temperatures, our
data for ladders of $L = 12$ and $14$ rungs show that finite-size
effects remain visible, but that corrections between $L = 10$ and 14 are 
small (of order 5\%) and converging. Overall, the largest available system 
presents a reasonable approximation to the thermodynamic limit for each 
temperature shown in Fig.~\ref{fig:J1_45}.

\begin{figure}[t]
\centering\includegraphics[width=0.98\columnwidth]{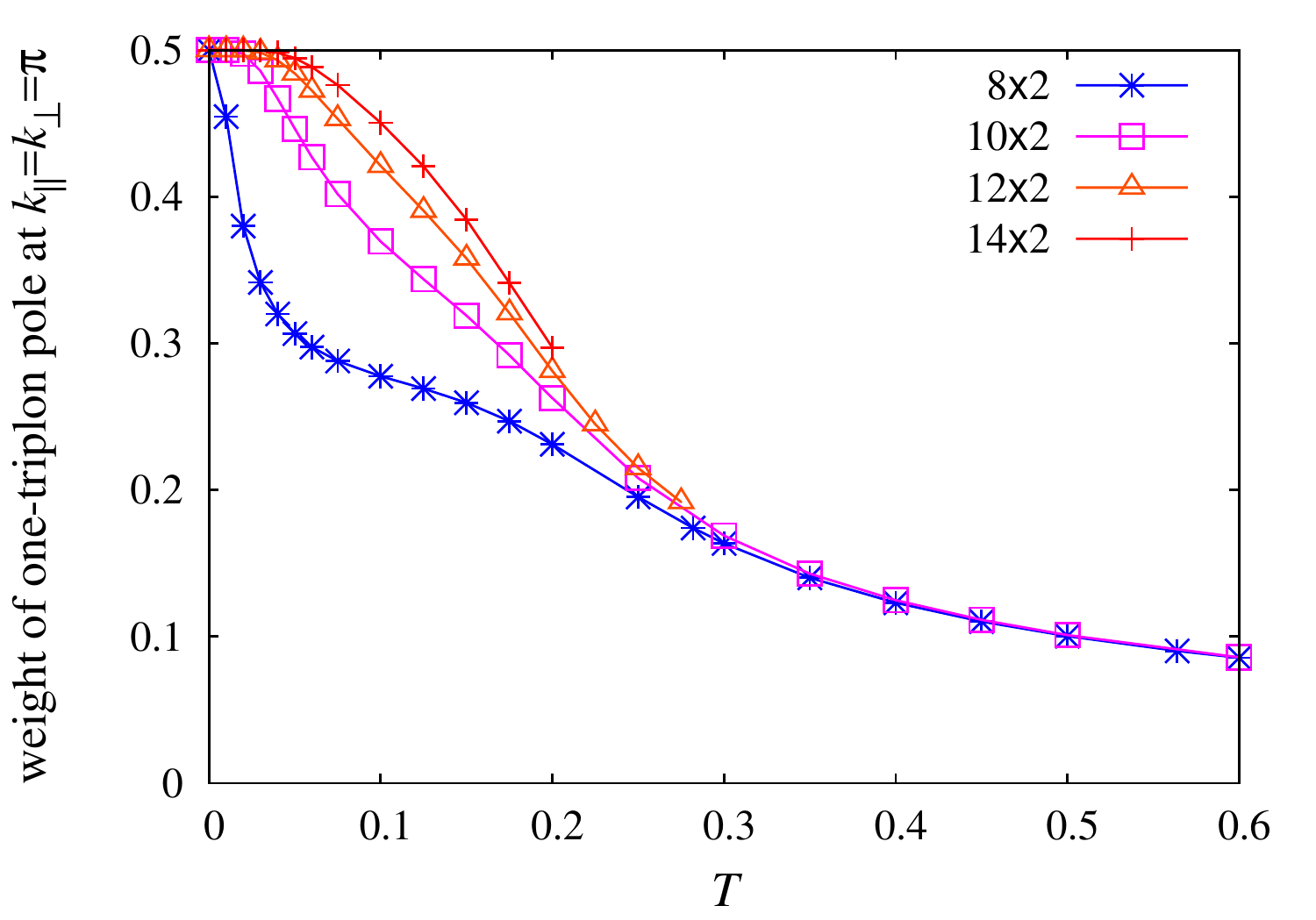}
\caption{Dependence on temperature of the weight of the one-triplon pole in 
the dynamical structure factor calculated for fully frustrated ladders of 
sizes $8 \le L \le 14$ rungs and with couplings $\JR = 1.42$, $\JL = \JD = 1$.}
\label{fig:J1_42}
\end{figure}

Figure \ref{fig:J1_42} shows the case $j' = 1.42$, where the coupling 
ratio approaches the critical point. Because the finite-size critical value 
for the $L = 6$ ladder is $j'_c = 1.436$, $j' = 1.42$ lies on the 
rung-triplet side of the transition for this system and the low-energy 
spectrum is completely different from the rung-singlet side; the $L = 6$ 
data must therefore be excluded from the analysis of $j' = 1.42$. In this 
case, the $L = 10$ ladder remains a good approximation to the 
infinite system for $T/\JL \gtrsim 0.3$. The situation at temperatures 
below $T/\JL \lesssim 0.25$ cannot be said to approach convergence for 
ladders up to $L = 14$. If one were to analyze the onset of the 
drop in spectral weight, which the $L = 14$ ladder places in the 
region $0.02 \lesssim T/\JL \lesssim 0.04$, this would in all probability 
underestimate the temperature scale that would be extracted from studies 
of longer ladders. However, for the temperature scale $T_{1/2}$ discussed 
in the main text, both the 12- and even 10-rung ladders appear to provide 
a satisfactory approximation to the thermodynamic limit, $T_{1/2} \approx 
0.225\,\JL$.

\end{document}